\documentclass[12pt]{article}
\usepackage{amsfonts,graphicx,amssymb,amsmath,amsthm,epsfig}

\allowdisplaybreaks

\begin{document}
\title{\bf Influence of Charge on Decoupled Anisotropic Spheres in $f(G,T)$ Gravity}
\author{M. Sharif$^1$ \thanks{msharif.math@pu.edu.pk} and K. Hassan$^2$ \thanks{komalhassan3@gmail.com}\\
$^1$ Department of Mathematics and Statistics, The University of Lahore,\\
1-KM Defence Road Lahore, Pakistan.\\
$^2$ Department of Mathematics, University of the Punjab,\\
Quaid-e-Azam Campus, Lahore-54590, Pakistan.}

\date{}

\maketitle
\begin{abstract}
In this paper, we develop two anisotropic solutions for static
self-gravitating spherical structure in the presence of
electromagnetic field through gravitational decoupling approach in
$f(G,T)$ theory, where $G$ and $T$ denote the Gauss-Bonnet term and
trace of the energy-momentum tensor, respectively. The extra source
with isotropic seed sector is responsible for generating anisotropy
in the spacetime. The system of field equations is decoupled into
two arrays by using minimal geometric deformation in the radial
component. The first set portrays the isotropic regime whereas the
second set represents the anisotropic system. The metric
coefficients of the Krori-Barua spacetime are employed to extract
solution of the first set while two constraints on the radial and
temporal components of the extra source yield the corresponding two
solutions. Finally, we investigate the influence of charge and
decoupling parameter on the physical viability and stability of the
obtained solutions. We conclude that the resulting solutions in this
modified theory indicate more feasible and stable structures.
\end{abstract}

\section{Introduction}

The immense and perplexing universe is composed of large-scale
structures such as clouds, stars, galaxies and clusters of galaxies.
General theory of relativity (GR) has played a crucial role to
examine the properties and mechanism of the cosmos. Dark energy and
dark matter with mysterious properties are assumed to be well
described by this theory to resolve the flat rotation curves of
galaxies \cite{1} along with late-time acceleration of the universe
\cite{2}. The Lambda cold dark matter model was developed to
elucidate the occurrence of dark energy by incorporating the
cosmological constant. In order to conform the values of
cosmological constant with the observational data and to elaborate
the evolution of the universe through various cosmic eras, its value
should be readapted. To address these problems, many researchers
have suggested to modify GR by changing the Einstein-Hilbert action.

The first and second-order terms in Lovelock gravity \cite{2a}, a
higher dimensional generalization of GR, represent GR and
Gauss-Bonnet (GB) invariant, respectively. Mathematically, in four
dimensions, the GB term is denoted as
$G=R^2-4R^{\psi\chi}R_{\psi\chi}+R^{\psi\chi\nu\mu}R_{\psi\chi\nu\mu}$
which is a combination of of the Ricci tensor $(R_{\psi\chi})$,
Riemann tensor $(R_{\psi\chi\nu\mu})$ and Ricci scalar $(R)$. Nojiri
and Odintsov \cite{3} formulated $f(G)$ gravity or modified GB
theory by including the generic function $f(G)$ in the
Einstein-Hilbert action. Modified theories with curvature-matter
coupling are considered as viable techniques that might elucidate
the remarkable phenomena of rapid cosmic expansion. Sharif and Ikram
\cite{8} proposed $f(G,T)$ gravity by including a generalized
$f(G,T)$ function in the Einstein-Hilbert action and discussed
energy conditions in the context of FRW universe. The non-zero
divergence of the energy-momentum tensor (EMT) creates an extra
force that leads test particles to trace the non-geodesic paths. The
same authors \cite{8a} studied wormholes by considering a variety of
matter configurations. Yousaf et al. \cite{8c} examined physical
properties of the stellar entities by decomposing the Riemann tensor
in this framework for charged/uncharged spherical system. Sharif and
Hassan \cite{8d} studied the complexity factor for a non-static
spherical and static/non-static cylindrical structures.

The inclusion of the electromagnetic field in celestial formations
has an interesting impact for studying and analyzing their
evolution. There has been a large body of literature to study the
influence of charge on stellar bodies in GR as well as modified
theories. Xingxiang \cite{8e} discussed static sphere constituting
charged perfect fluid. Das et al. \cite{8f} studied charged static
spherical solutions by matching the interior geometry with the
exterior Riessner-Nordstr\"{o}m metric. Sharif and Bhatti \cite{8g}
numerically solved the field equations for a shearfree charged
object and checked the viability through the energy conditions.
Murad \cite{8h} explored anisotropic charged celestial objects by
assuming a specific form of a metric potential. Different aspects
describing the internal structure of self-gravitating bodies have
been examined in the presence of an electric field \cite{8i}.

The presence of interacting substance in dense compact structures
reveals that they display different properties in various directions
which depicts the anisotropic nature of the compact objects
\cite{9}. Phase transition \cite{9a} and superfluid \cite{9b} are
considered the reasons to produce anisotropy in the system. Herrera
and Santos \cite{9c} investigated the origin of anisotropy and
inspected its influence in the evolution of self-gravitating bodies.
Harko and Mak \cite{10} determined the analytic solution of the
field equations by employing a particular anisotropic factor and
analyzed the static spherical anisotropic configurations. Paul and
Deb \cite{10c} investigated physical attributes of anisotropic star
models in hydrostatic equilibrium.

The analytic solutions to the field equations help us to understand
the intricate nature of the self-gravitating bodies. However, it is
often difficult to obtain solutions of the field equations due to
their non-linearity. The gravitational decoupling via minimal
geometric deformation (MGD) is the recently developed approach to
find viable solutions. In this approach, the radial function of the
line element is distorted by means of a linear transformation which
segregates the system of field equations into two sets. The first
set describes the seed sector, and the second corresponds to the
extra source. Both these sets are addressed independently and the
solution of the whole system is determined by applying the
superposition principle. Ovalle \cite{11} was the pioneer to develop
this scheme in the context of braneworld to calculate the exact
solutions of celestial objects. Later, Ovalle et al. \cite{13}
investigated the feasibility of a celestial object in GR by
extending the isotropic domain to anisotropic configuration.
Gabbanelli et al. \cite{14} used the Durgapal-Fuloria solution to
determine its anisotropic version using the same technique.

Sharif and Sadiq \cite{15} studied the impact of charge and
formulated two anisotropic solutions using Krori-Barua spacetime via
MGD method. Estrada and Tello-Ortiz \cite{17} used Heintzmann
solution to develop two consistent anisotropic solutions. Singh et
al. \cite{18} developed physically acceptable solutions through this
approach for class-I spacetime and determined the mass and radius of
the stars by plotting $M$-$R$ curve. Hensh and Stuchl{\'\i}k
\cite{19} worked on isotropic Tolman VII solution to calculate its
anisotropic version by using the decoupling method. Zubair and Azmat
\cite{20} constructed the anisotropic solution by deforming the
radial function of isotropic Tolman V solution. Maurya and his
collaborators \cite{20a} worked on the MGD technique to formulate
the anisotropic solutions from known isotropic domain in different
modified theories. Sharif and Saba \cite{16} constructed
charged/uncharged gravitational decoupled anisotropic solutions from
a known isotropic solution and examined the viability and stability
of the obtained solutions in the framework of $f(G)$ gravity. Many
researchers obtained anisotropic versions of the isotropic source
and checked the feasibility conditions of compact stars in the
formalism of different modified theories \cite{21}. We have recently
studied decoupled anisotropic spheres in $f(G,T)$ gravity
\cite{21a}.

This paper deals with the deformation of the radial component of
Krori-Barua metric through MGD scheme to extract charged anisotropic
solutions in $f(G,T)$ gravity. The format of the paper is as
follows. Section \textbf{2} addresses the key features of this
modified theory. In section \textbf{3}, the MGD procedure splits the
field equations into two arrays in which one describes the isotropic
source while the other represents the anisotropic configuration. We
obtain anisotropic solutions by using two constraints on radial as
well as temporal components of the extra source in section
\textbf{4}. The physical viability and stability of the constructed
solutions are examined in section \textbf{5}. In the last section,
we summarize our results.

\section{$f(G,T)$ Formalism}

The $f(G,T)$ field equations are acquired with the help of modified
action as
\begin{equation}\label{1}
\mathbb{A}_{f(G,T)}=\int\sqrt{-g}d^{4}x\bigg[\frac{\mathrm{R}+f(G,T)}{16\pi}+\pounds_{m}+\pounds_{E}+\xi\pounds_{\omega}\bigg],
\end{equation}
where determinant of the metric tensor ($g_{\psi\chi}$) and matter
Lagrangian density are represented by $g$ and $\pounds_{m}$,
respectively. The Lagrangian density of the electromagnetic field
and additional source are specified by $\pounds_{E}$ and
$\pounds_{\omega}$, respectively, whereas $\xi$ denotes the
decoupling parameter. The corresponding EMT of the sources are given
by the relation
\begin{align}\label{1a}
T_{\psi\chi}=g_{\psi\chi}\pounds_m-\frac{2\partial\pounds_m}{\partial
g^{\psi\chi}},\quad
\omega_{\psi\chi}=g_{\psi\chi}\pounds_{\omega}-\frac{2\partial\pounds_{\omega}}{\partial
g^{\psi\chi}}.
\end{align}
By varying the action \eqref{1} with respect to $g_{\psi\chi}$, we
obtain the modified field equations as
\begin{equation}\label{3}
G_{\psi\chi}=8\pi
T^{(tot)}_{\psi\chi}=8\pi(T^{(D)}_{\psi\chi}+T^{(M)}_{\psi\chi}+E_{\psi\chi}+\xi\omega_{\psi\chi}),
\end{equation}
where $G_{\psi\chi}=R_{\psi\chi}-\frac{1}{2}Rg_{\psi\chi}$ indicates
the Einstein tensor and $T^{(D)}_{\psi\chi}$ demonstrates the
correction terms caused by $f(G,T)$ theory as
\begin{eqnarray}\nonumber
T^{(D)}_{\psi\chi}&=&\frac{1}{8\pi}\bigg[\{(\mu+P)\upsilon_{\psi}\upsilon_{\chi}\}f_{T}(G,T)+\frac{1}{2}g_{\psi\chi}f(G,T)
+\big(4R^{\alpha\nu}R_{\psi\alpha \chi \nu}\\\nonumber&-&
2RR_{\psi\chi}-2R^{\alpha \nu \gamma} _{\psi}R_{\chi \alpha \nu
\gamma}\big)f_{G}(G,T)+4R_{\alpha\chi}R^{\alpha}_{\psi}
+(4g_{\psi\chi}R^{\alpha
\nu}\nabla_{\alpha}\nabla_{\nu}\\\nonumber&-&4R^{\alpha}_{\psi}\nabla_{\chi}\nabla_{\alpha}-4R_{\psi\alpha
\chi\nu}\nabla^{\alpha}\nabla^{\nu}-2g_{\psi\chi}R\nabla^{2}
+2R\nabla_{\psi}\nabla_{\chi}-4R^{\alpha}_{\chi}\nabla_{\psi}\nabla_{\alpha}\\\label{4}&+&4R_{\psi\chi}\nabla^{2})f_{G}(G,T)\bigg],
\end{eqnarray}
$\nabla^{2}=\nabla^{b}\nabla_{b}$ signifies the d' Alembert
operator. Further, $f_{G}(G,T)$ and $f_{T}(G,T)$ denote the partial
derivatives of an arbitrary function $f(G,T)$ with respect to $G$
and $T$, respectively. An additional source $\omega_{\psi\chi}$ is
found to be responsible for inducing anisotropy in the current
scenario which is associated with the seed sector via dimensionless
parameter $\xi$.

The EMT plays a significant role in determining the internal
configuration of the celestial bodies. In the current setup, the
perfect fluid source is presented by the EMT
\begin{equation}\label{5}
T^{(M)}_{\psi\chi} =(\mu+P)
\upsilon_{\psi}\upsilon_{\chi}+Pg_{\psi\chi},
\end{equation}
where $\mu$, $P$ and $\upsilon^{\psi}$ indicate the density,
pressure and four-velocity, respectively, satisfying
$\upsilon^{\psi}\upsilon_{\psi}=-1$. The tensor $E_{\psi\chi}$
describes EMT for the electromagnetic field as
\begin{equation}\label{3b}
E_{\psi\chi}=\frac{1}{4\pi}\left(F^{l}_{\psi}F_{\chi
l}-\frac{1}{4}F_{lm}F^{lm}g_{\psi\chi}\right),
\end{equation}
where $F_{\psi\chi}=\gamma_{\chi,\psi}-\gamma_{\psi,\chi}$ and
$\gamma_{\psi}$ indicate the Maxwell field tensor and four
potential, respectively. Here we take
$\gamma_{\psi}=\gamma(r)\delta^{0}_{\psi}$. The tensorial form of
the Maxwell field equations are expressed as
\begin{equation}\nonumber
F^{\psi\chi}_{~~;\chi}=4\pi J^{\psi} ,\quad F_{[\psi\chi;l]}=0,
\end{equation}
where $J^{\psi}=\sigma \upsilon^{\psi}$ is the four current density
while $\sigma$ denotes the charge density. The viability and
stability of the resulted anisotropic solutions will be checked by
assuming an explicit model of $f(G,T)$ gravity \cite{23a} as
\begin{equation}\label{60}
f(G,T)= \mathbf{f_1}(G)+\mathbf{f_2}(T),
\end{equation}
where $\mathbf{f_1}(G)$ and $\mathbf{f_2}(T)$  are independent
functions of $G$ and $T$, respectively. There can be many choices
regarding curvature and matter coupling, however, in order to
consider its role more effectively, we assume a quadratic $f(G,T)$
model. For this purpose, we choose $\mathbf{f_1}(G)=G^2$ and
$\mathbf{f_2}(T)=\beta T$, where $\beta$ refers to a free parameter.
The values of $G$ and its higher derivatives are provided in
Eqs.\eqref{63a}-\eqref{65} of Appendix \textbf{A}.

The interior region of the spherical compact object is given by
\begin{equation}\label{6}
ds^{2}=-e^{\phi}dt^{2}+e^{\lambda}dr^{2}+r^{2}(d\theta^{2}+{\sin^{2}\theta}{d\Phi^2}),
\end{equation}
where $\phi$ and $\lambda$ are functions of $r$ only. The Maxwell
field equations give
\begin{equation}\label{6b}
\gamma^{''}+\bigg(\frac{2}{r}-\frac{\phi^{'}+\lambda^{'}}{2}\bigg)\gamma^{'}=4
\pi \sigma e^{\frac{\phi}{2}+\lambda},
\end{equation}
prime means derivative with respect to $r$ and its integration leads
to
\begin{equation}\label{6c}
\gamma^{'}=\frac{\mathrm{s}e^{\frac{\phi+\lambda}{2}}}{r^2},
\end{equation}
$\mathrm{s}$ denotes the presence of charge in the interior of
self-gravitating body. The components of four-velocity in the
comoving frame take the form
\begin{equation}\label{6a}
\upsilon^{\psi}=\left(e^{\frac{-\phi}{2}},0,0,0\right).
\end{equation}
The corresponding field equations are
\begin{eqnarray}\label{8}
\frac{1}{r^{2}}+e^{-\lambda}(\frac{\lambda'}{r}-\frac{1}{r^{2}})
&=&8\pi(\bar{\mu}+\frac{\mathrm{s^2}}{8\pi
r^4}+T^{0(D)}_{0}-\xi\omega^{0}_{0}) ,
\\\label{9}
-\frac{1}{r^{2}}+e^{-\lambda}(\frac{1}{r^{2}}+\frac{\phi'}{r})
&=&8\pi(\bar{P}-\frac{\mathrm{s^2}}{8\pi
r^4}+T^{1(D)}_{1}+\xi\omega^{1}_{1}) ,
\\\label{10}
e^{-\lambda}(\frac{\phi''}{2}+\frac{\phi'^{2}}{4}-\frac{\lambda'\phi'}{4}
+\frac{\phi'}{2r}-\frac{\lambda'}{2r})&=&8\pi(\bar{P}+\frac{\mathrm{s^2}}{8\pi
r^4}+T^{2(D)}_{2}+\xi\omega^{2}_{2}),
\end{eqnarray}
where
\begin{align}\label{10a}
\bar{\mu}=\mu+\frac{\beta}{16\pi}(3\mu-P),\quad
\bar{P}=P+\frac{\beta}{16\pi}(-\mu+3P),
\end{align}
and the extra curvature terms $T^{0(D)}_{0},~T^{1(D)}_{1}$ and
$T^{2(D)}_{2}$ are mentioned in Eqs.\eqref{61}-\eqref{63} of
Appendix \textbf{A}.

Unlike GR, $f(G,T)$ theory yields the non-conserved form of EMT
which, in return, produces the additional force. The
non-conservation of the matter source is represented by the equation
\begin{eqnarray}\nonumber
\nabla^{\psi}T_{\psi\chi}&=&\frac{f_{T}(G,T)}{k^{2}-f_{T}(G,T)}
\bigg[-\frac{1}{2}g_{\psi\chi}\nabla^{\psi}T+(\Theta_{\psi\chi}+T_{\psi\chi})\nabla^{\psi}(\ln
f_{T}(G,T))\\\label{11}&+& \nabla^{\psi}\Theta_{\psi\chi}\bigg],
\end{eqnarray}
yielding
\begin{align}\label{12}
\frac{dP}{dr}+\frac{\lambda'}{2}(\mu+P)+\xi\frac{d\omega^{1}_{1}}{dr}-\frac{\mathrm{ss'}}{4\pi
r^4}
+\frac{\xi\lambda'}{2}(\omega^{1}_{1}-\omega^{0}_{0})+\frac{2\xi}{r}(\omega^{1}_{1}-\omega^{2}_{2})=\Omega,
\end{align}
where $\Omega$ includes the contribution of extra curvature terms
given in Eq.(\ref{66}) of Appendix \textbf{A}. The system of
non-linear differential equations \eqref{8}-\eqref{10} together with
\eqref{12} have eight unknowns, showing that our system is
under-determined (more unknowns than equations), therefore, we need
more constraints to solve the system. For this purpose, we employ
the systematic approach of MGD to close our system. We reformulate
the physical variables as
\begin{equation}\label{13}
\tilde{\mu}=\mu-\xi\omega^{0}_{0},\quad
\tilde{P_r}=P+\xi\omega^{1}_{1},\quad
\tilde{P_t}=P+\xi\omega^{2}_{2},
\end{equation}
where $\omega^{\psi}_{\chi}$ is the anisotropy producing factor for
the astrophysical objects.  The effective anisotropy is defined as
\begin{equation}\label{14}
\tilde{\Delta}=\tilde{P_t}-\tilde{P_r}=\xi(\omega^{2}_{2}-\omega^{1}_{1}),
\end{equation}
which will be zero for $\xi=0$.

\section{Gravitational Decoupling Via MGD}

Here, we apply the gravitational decoupling through MGD scheme to
solve the system \eqref{8}-\eqref{10} and evaluate the unknowns
(physical variables, metric potentials, charge and anisotropic
source). This approach splits the field equations in such a way that
the extra source $\omega^{\psi}_{\chi}$ is found to generate
anisotropy in the internal geometry. We start with the solution of
perfect matter configuration by the following line element
\begin{equation}\label{15}
ds^{2}=-e^{\vartheta(r)}dt^{2}+\frac{dr^{2}}{\tau(r)}+r^{2}(d\theta^{2}+{\sin^{2}\theta}{d\Phi^2}),
\end{equation}
where $\tau(r)=1-\frac{2m}{r}+\frac{\mathrm{s^2}}{r^2}$ and $m(r)$
corresponds to the Misner-Sharp mass of the compact object. We
distort the metric potentials to comprehend the influence of
anisotropy on the perfect matter by utilizing the linear
transformations as
\begin{equation}\label{16}
\vartheta\rightarrow\phi=\vartheta+\xi k,\quad \tau\rightarrow
e^{-\lambda(r)}=\tau+\xi h^\star,
\end{equation}
where $k$ and $h^\star$ are the deformations assigned to the
temporal and radial metric functions, respectively. In MGD, only
radial potential is translated, i.e., $k=0$ which means that the
temporal part remains unperturbed. The field equations
\eqref{8}-\eqref{10} are segregated into two sets by using the
deformed metric. By substituting $\xi=0$, the modified field
equations for the perfect fluid yield the first set as
\begin{align}\label{18}
8\pi(\mu+\frac{\beta}{16\pi}(3\mu-P)+\frac{\mathrm{s^2}}{8\pi
r^4}+T^{0(D)}_{0})
&=\frac{1}{r^{2}}-(\frac{\tau'}{r}+\frac{\tau}{r^{2}}),
\\\label{19}
8\pi(P+\frac{\beta}{16\pi}(-\mu+3P)-\frac{\mathrm{s^2}}{8\pi
r^4}+T^{1(D)}_{1})
&=-\frac{1}{r^{2}}+\frac{\tau}{r}(\frac{1}{r}+\phi'),
\\\nonumber
8\pi(P+\frac{\beta}{16\pi}(-\mu+3P)+\frac{\mathrm{s^2}}{8\pi
r^4}+T^{2(D)}_{2})&=\tau(\frac{\phi''}{2}+\frac{\phi'^{2}}{4}+\frac{\phi'}{2r})\\\label{20}&+\tau'(\frac{\phi'}{4}
+\frac{1}{2r}).
\end{align}

Solving the above equations simultaneously, the expressions for
$\mu$, $P$ and $\mathrm{s}^2$ become
\begin{align}\nonumber
\mu&=\frac{-1}{{16 \left(\beta ^2+12 \pi \beta +32 \pi ^2\right)
r^2}}\bigg[\beta  r^2 \tau ' \phi '-4 \beta+2 \beta  r^2 \tau \phi
''+\beta  r^2 \tau  \phi '^2\\\nonumber&+8 \pi  r^2 \tau ' \phi '+16
\pi r^2 \tau  \phi ''+8 \pi  r^2 \tau  \phi '^2+(12 \beta r^2 +64
\pi  r^2) T^{0(D)}_{0}+(8 \beta r^2 \\\nonumber&+32 \pi r^2
)T^{1(D)}_{1}-32 \pi(-4 \beta r^2 -32 \pi r^2 )T^{2(D)}_{2}+14 \beta
r \tau '-6 \beta r \tau  \phi '+4 \beta \tau \\\label{20a}&+80 \pi r
\tau '-16 \pi r \tau \phi '+32 \pi \tau \bigg],
\\\nonumber
P&=\frac{-1}{{16 (\beta +4 \pi ) (\beta +8 \pi ) r^2}}\bigg[4 \beta
-\beta r^2 \tau ' \phi '-2 \beta  r^2 \tau \phi ''-\beta r^2 \tau
\phi '^2-16 \pi  r^2 \tau \phi ''\\\nonumber&-8 \pi r^2 \tau ' \phi
'-8 \pi r^2 \tau \phi '^2+4 \beta r^2 T^{0(D)}_{0}+(8 \beta r^2 +32
\pi r^2 )T^{1(D)}_{1}+2 \beta r \tau '-4 \beta \tau\\\label{20b}&+(4
\beta r^2 +32 \pi r^2 )T^{2(D)}_{2}-10 \beta r \tau \phi ' -16 \pi r
\tau '-48 \pi r \tau \phi '-32 \pi \tau +32 \pi\bigg],
\\\nonumber
\mathrm{s}^{2}&=\frac{1}{8} \big(r^4 \tau ' \phi '+2 r^4 \tau \phi
''+r^4 \tau \phi '^2+4 r^4 T^{1(D)}_{1}-4 r^4 T^{2(D)}_{2}+2 r^3
\tau '-2 r^3 \tau  \phi '\\\label{20c}&-4 r^2 \tau +4 r^2\big).
\end{align}
The anisotropy generated by the new source is studied by the second
set
\begin{eqnarray}\label{21}
8\pi\omega^{0}_{0} &=& \frac{h^{\star'}}{r}+\frac{h^{\star}}{r^{2}},
\\\label{22}
8\pi\omega^{1}_{1} &=&\frac{h{^\star}}{r}(\frac{1}{r}+\phi'),
\\\label{23}
8\pi\omega^{2}_{2}&=&h^{\star}(\frac{\phi''}{2}+\frac{\phi'^{2}}{4}+\frac{\phi'}{2r})+h^{\star'}(\frac{\phi'}{4}
+\frac{1}{2r}).
\end{eqnarray}
One can observe that the above system of field equations seems
similar for the charged spherical anisotropic matter source through
the metric
\begin{equation}\label{24}
ds^{2}=-e^{\vartheta(r)}dt^{2}+\frac{dr^{2}}{h^{\star}}+r^{2}(d\theta^{2}+{\sin^{2}\theta}{d\Phi^2}).
\end{equation}
It can also be noted that the term $\frac{1}{r^2}$ is the only
varying quantity between Eqs.\eqref{18}-\eqref{19} and
\eqref{21}-\eqref{22}. In order to make this system equivalent to
the standard field equations for charged anisotropic stellar object,
we specify the physical variables as
$\bar{\mu}+\frac{\mathrm{s}^2}{8\pi
r^4}=\omega^{\star0}_{0}=\omega^{0}_{0}+\frac{1}{ r^2},~
\bar{P_r}-\frac{\mathrm{s}^2}{8\pi
r^4}=\omega^{\star1}_{1}=\omega^{1}_{1}+\frac{1}{ r^2}$, and
$\bar{P_t}+\frac{\mathrm{s}^2}{8\pi
r^4}=\omega^{\star2}_{2}=\omega^{2}_{2}=\omega^{\star3}_{3}=\omega^{3}_{3}$.

Junction conditions play a crucial role in understanding the
fundamental characteristics of the astrophysical objects at the
boundary $(\Sigma)$. The first and second fundamental forms of
junction conditions assure the smooth matching of exterior and
interior geometries at the junction. The choice of an outer region
is examined on the basis that its properties (static, irrotational,
charged) at the hypersurface match with the interior regime. The
matter source is restricted only within the stellar object, and a
boundary is marked with the outer Reissner-Nordstr\"{o}m (presence
of charge in the exterior region) metric to separate both the
structures. In the scenario of $f(G,T)$ theory, the inclusion of
higher curvature terms in $G$ will be significantly restrained and
there is no contribution of $T$ in charged spacetime. Hence, the
external geometry as in GR can be chosen for modified theories.
Furthermore, the Reissner-Nordstr\"{o}m metric has been used in the
literature for $f(G)$ and $f(G,T)$ gravity theories with the same
spacetime \cite{23b}. We choose the interior region as
\begin{equation}\label{24}
ds^{2}=-e^{\phi(r)}dt^{2}+\frac{1}{(1-\frac{2\check{m}}{r}+\frac{\mathrm{s^2}}{r^2})}dr^{2}+r^{2}(d\theta^{2}+{\sin^{2}\theta}{d\Phi^2}),
\end{equation}
where $\check{m}=m(r)-\frac{\xi r }{2}h^{\star}(r)$ indicates the
inner geometric mass. The matching of the first fundamental form
$\big([ds^2=0]_\Sigma\big)$ at the boundary $(r=\textsf{R})$ of the
stellar object yields
\begin{equation}\label{26}
\phi_{-}(\texttt{R})=\phi_{+}(\texttt{R}),\quad
e^{-\lambda_{+}(\texttt{R})}=e^{-\lambda_{-}(\texttt{R})}=1-\frac{2\mathbb{M}_o}{\texttt{R}}+\frac{\mathbb{S}^2_o}{\texttt{R}^2}+\xi
h^{\star}(\texttt{R}),
\end{equation}
where $\tau=e^{-\lambda}-\xi h^{\star}$ has been utilized. Here,
plus and minus signs in the metric potentials indicate the exterior
and interior regions, respectively. Moreover,
$\mathbb{M}_o=m(\texttt{R})$,
$\mathbb{S}^2_o=\mathrm{s}^2(\texttt{R})$ and
$h^{\star}(\texttt{R})$ represent the total mass, charge and
deformation function at the boundary of the star. The matching of
the second fundamental form
$\big[(T^{(tot)}_{\psi\chi}\textsl{W}^{\chi})_\Sigma=0$,
($\textsl{W}^{\chi}=\big(0,e^{-\frac{\lambda}{2}},0,0\big)$)\big] at
the hypersurface yields
\begin{equation}\label{27}
\bar{P}(\texttt{R})-\frac{\mathbb{S}^2_o}{8\pi\texttt{R}^4}+\xi\big(\omega^{1}_{1}(\texttt{R})\big)_{-}+\big(T^{1(D)}_{1}(\texttt{R})\big)_{-}=
\xi\big(\omega^{1}_{1}(\texttt{R})\big)_{+}+\big(T^{1(D)}_{1}(\texttt{R})\big)_{+}.
\end{equation}
Using Eq.\eqref{26} in the above equation , we have
\begin{equation}\label{28}
\bar{P}(\texttt{R})-\frac{\mathbb{S}^2_o}{8\pi\texttt{R}^4}+\xi\big(\omega^{1}_{1}(\texttt{R})\big)_{-}=
\xi\big(\omega^{1}_{1}(\texttt{R})\big)_{+},
\end{equation}
which can be also be expressed as
\begin{equation}\label{29}
\bar{P}(\texttt{R})-\frac{\mathbb{S}^2_o}{8\pi\texttt{R}^4}+\frac{\xi
h{^\star}
(\texttt{R})}{8\pi}(\frac{\phi'}{\texttt{R}}+\frac{1}{\texttt{R}^{2}})=\frac{\xi
a{^\star} (\texttt{R})}{8\pi
\texttt{R}}(\frac{1}{\texttt{R}}+\frac{2\mathbb{M}-\frac{2\mathcal{S}^2}{\texttt{R}}}{\texttt{R}(\texttt{R}-2\mathbb{M})+\mathcal{S}^2}).
\end{equation}
Here $\mathbb{M}$ and $\mathcal{S}$ indicate the mass and charge of
the exterior geometry, respectively, while $a{^\star}$ corresponds
to the outer radial geometric deformation. The external geometric
structure representing the impact of anisotropic matter
configuration is expressed by the Reissner-Nordstr\"{o}m spacetime
\begin{equation}\label{30}
ds^{2}=-\bigg(1-\frac{2\mathbb{M}}{r}+\frac{\mathcal{S}^2}{r^2}\bigg)dt^{2}+\frac{1}{\big(1-\frac{2\mathbb{M}}{r}+\frac{\mathcal{S}^2}{r^2}+\xi
a{^\star}\big)}dr^{2}+r^{2}(d\theta^{2}+{\sin^{2}\theta}{d\Phi^2}).
\end{equation}
The necessary and sufficient requirements are provided by
Eqs.\eqref{26} and \eqref{29} to remove any discontinuity or
irregularity at the boundary. The assumption $(a{^\star}=0)$
converts Eq.\eqref{29} to the standard Reissner-Nordstr\"{o}m case
as
\begin{equation}\label{31}
\tilde{P}(\texttt{R})-\frac{\mathbb{S}^2_o}{8\pi\texttt{R}^4}=\bar{P}(\texttt{R})-\frac{\mathbb{S}^2_o}{8\pi\texttt{R}^4}+\frac{\xi
h{^\star}
(\texttt{R})}{8\pi}(\frac{\phi'}{\texttt{R}}+\frac{1}{\texttt{R}^{2}})=0.
\end{equation}

\section{Anisotropic Solutions}

In this section, we evaluate the anisotropic charged spherical
solutions by using the isotropic (seed) solution called Krori-Barua
metric \cite{22} which has a singularity-free nature. This is given
by
\begin{equation}\label{32}
e^{\phi(r)}=e^{\mathcal{B}r^{2}+\mathcal{C}},\quad
e^{\lambda(r)}=\tau^{-1}=e^{\mathcal{A}r^{2}}.
\end{equation}
Using this solution in the field equations \eqref{18}-\eqref{20}, we
obtain
\begin{align}\nonumber
\mu&=\frac{1}{4\big(\beta^2+12\pi\beta+32\pi^2\big)
r^2}\bigg[e^{-\mathcal{A} r^2}\big(-8\pi\big(-\mathcal{A}r^2
\big(\mathcal{B} r^2+5\big)+e^{\mathcal{A}r^2}\big(r^2 (2\\\nonumber
&\times T^{0(D)}_{0}+T^{1(D)}_{1}-T^{2(D)}_{2})-1\big)+\mathcal{B}^2
r^4+1\big)-\beta\big(-\mathcal{A} r^2 \big(\mathcal{B}
r^2+7\big)+e^{\mathcal{A} r^2} \\\label{33a} &\times\big(r^2 (3
T^{0(D)}_{0}+2 T^{1(D)}_{1}-T^{2(D)}_{2})-1\big) + \big(\mathcal{B}
r^2-1\big)^2\big)\big)\bigg],\\\nonumber P&= \frac{1}{4 (\beta +4
\pi ) (\beta +8 \pi ) r^2}\bigg[e^{-\mathcal{A} r^2} \big(-\beta
\big(\mathcal{A} \mathcal{B} r^4+e^{\mathcal{A} r^2} \big(r^2
(T^{0(D)}_{0}+2
T^{1(D)}_{1}+T^{2(D)}_{2})\\\nonumber&+1\big)-\mathcal{A} r^2-B^2
r^4-6 \mathcal{B} r^2-1\big)-8 \pi \big(\mathcal{A}\mathcal{B}
r^4+e^{\mathcal{A} r^2} \big(r^2
(T^{1(D)}_{1}+T^{2(D)}_{2})+1\big)\\\label{33b}&+\mathcal{A}
r^2-\mathcal{B}^2 r^4-4 B r^2-1\big)\big)\bigg],\\\nonumber
\mathrm{s}^2&=\frac{r^2 e^{-\mathcal{A} r^2}}{2}
\big(\big(\mathcal{B} r^2+1\big) \big(-\mathcal{A} r^2+\mathcal{B}
r^2-1\big)+e^{\mathcal{A} r^2} \big(r^2
(T^{1(D)}_{1}-T^{2(D)}_{2})+1\big)\big).\\\label{33a}
\end{align}
The matching conditions can be used to evaluate the values of
unknown constants $\mathcal{B}, \mathcal{C}$ and $\mathcal{A}$. The
smooth matching between the external and internal geometries over
the hypersurface assists in determining these constants. The
continuum of the metric potentials between the outer and interior
structures yields
\begin{align}\label{36}
\mathcal{B}&=\frac{\mathbb{M}_o
\texttt{R}-\mathbb{S}_o^2}{\texttt{R}^4 \left(-\frac{2
\mathbb{M}_o}{\texttt{R}}+\frac{\mathbb{S}_o^2}{\texttt{R}^2}+1\right)},\\\label{36a}
\mathcal{C}&=\frac{\texttt{R}^2 \left(-\frac{2
\mathbb{M}_o}{\texttt{R}}+\frac{\mathbb{S}_o^2}{\texttt{R}^2}+1\right)
\ln \left(-\frac{2
\mathbb{M}_o}{\texttt{R}}+\frac{\mathbb{S}_o^2}{\texttt{R}^2}+1\right)-\mathbb{M}_o
\texttt{R}+\mathbb{S}_o^2}{\texttt{R}^2 \left(-\frac{2
\mathbb{M}_o}{\texttt{R}}+\frac{\mathbb{S}_o^2}{\texttt{R}^2}+1\right)},\\\label{36b}
\mathcal{A}&=\frac{1}{\texttt{R}^2}\ln\bigg(\frac{1}{1-\frac{2\mathbb{M}_o}{\texttt{R}}+\frac{\mathbb{S}_o^2}{\texttt{R}^2}}\bigg).
\end{align}
The anisotropic solutions of the internal compact structure is
developed by using the radial and temporal metric components given
in Eq.\eqref{32}. Some new constraints can be employed to determine
the solution of Eqs.\eqref{21}-\eqref{23} in which the anisotropic
sector and geometric deformation function $h^\star$ are related. We
consider the physical behavior of the compact star 4U 1820-30
\cite{22a} with radius and mass as $9.1\pm0.4$km and
$1.58\pm0.06M_\odot$, respectively, which help to calculate the
values of constant.

In the following, we study two anisotropic solutions.

\subsection{Solution I}

Here, we determine the deformation function $h^\star$ and
constituents of extra source $(\omega^{\psi}_{\chi})$ by taking an
additional constraint at the radial part of the additional source.
One can note that compatibility between inner source and outer
Riessner-Nordstr\"{o}m spacetime holds if
$\bar{P}(\texttt{R})-\frac{\mathbb{S}^2_o}{8\pi\texttt{R}^4}+T^{1(D)}_{1}(\texttt{R})\sim\xi(\omega^{1}_{1}(\texttt{R}))_{-}$.
This constraint is satisfied \cite{13} when
\begin{equation}\label{37}
\bar{P}-\frac{\mathrm{s}^2}{8\pi r^4}+T^{1(D)}_{1}=\omega^{1}_{1},
\end{equation}
yielding the deformation function $h^\star$ (using Eqs.\eqref{19}
and \eqref{22}) as
\begin{equation}\label{37a}
h^\star=\tau-\frac{1}{1+\phi'r}.
\end{equation}
The matching of the first fundamental form turns out to be
\begin{eqnarray}\label{39}
\texttt{R}^2e^{\mathcal{B}\texttt{R}^2+\mathcal{C}}&=&\texttt{R}^2-2\mathbb{M}
\texttt{R}+\mathbb{S}^2,\\\label{40}
\tau(1+\beta)-\frac{\beta}{1+2\mathcal{C}\texttt{R}^2}&=&1-\frac{2\mathbb{M}}{\texttt{R}}+\frac{\mathbb{S}^2}{\texttt{R}^2}.
\end{eqnarray}
In a similar way, the second fundamental form, i.e.,
$\bar{P}(\texttt{R})-\frac{\mathbb{S}^2_o}{8\pi\texttt{R}^4}+T^{1(D)}_{1}(\texttt{R})-\xi(\omega^{1}_{1}(\texttt{R}))_{-}=0$
yields the constant $\mathcal{A}$ through Eq.\eqref{37} as
\begin{equation}\label{41}
\mathcal{A}=\frac{\ln(1+2\mathcal{B} \texttt{R}^2)}{\texttt{R}^2}.
\end{equation}
Now, we extract the value of mass from  Eqs.\eqref{26} and
\eqref{40}, and using in Eq.\eqref{39} provides the constant
$\mathcal{C}$ as
\begin{equation}\label{43}
\mathcal{C}=\ln\bigg[1-\frac{2\mathbb{M}_o}{\texttt{R}}+\frac{\mathbb{S}^2_o}{\texttt{R}^2}
+\beta\bigg\{1-\frac{2\mathbb{M}_o}{\texttt{R}}+\frac{\mathbb{S}^2_o}{\texttt{R}^2}-\frac{1}{1+2\mathcal{B}
\texttt{R}^2}\bigg\}\bigg]-\mathcal{B}\texttt{R}^2.
\end{equation}
Equations \eqref{41} and \eqref{43} are useful in matching the
internal and external structures of the compact object. Applying the
above mentioned constraints, we obtain $\tilde{\mu}, \tilde{P_r}$,
$\tilde{P_t}$ and $\mathrm{s}^2$ as
\begin{align}\nonumber
\tilde{\mu}&=\frac{e^{-\mathcal{\mathcal{A}} r^2}}{32 (\beta +4 \pi
) (\beta +8 \pi )}\bigg[\frac{1}{\pi  \big(2 \mathcal{B}
r^3+r\big)^2}\big\{\xi \big(8 \pi \big(r^4 \big(-2
\mathcal{A}^2\big(\mathcal{B} r^2+1\big) \\\nonumber &\times \big(2
\mathcal{B} r^2+1\big)+\mathcal{A} \mathcal{B} \big(4 \mathcal{B}
r^2 \big(\mathcal{B} r^2+8\big)+27\big)-\mathcal{B}^2 \big(14
\mathcal{B} r^2+45\big)\big)\\\nonumber &+e^{\mathcal{A} r^2}
\big(r^2 \big(2 \mathcal{B} \big(5 r^2
(T^{1(D)}_{1}+T^{2(D)}_{2})+3\big)+3
(T^{1(D)}_{1}+T^{2(D)}_{2})\big)+1\big)\\\nonumber &-1+r^2 (5
\mathcal{A}-18 \mathcal{B})\big)+\beta \big(2 \mathcal{B} r^6
\big(\mathcal{A}^2+20 \mathcal{A} \mathcal{B}-7
\mathcal{B}^2\big)+r^4 \big(2 \mathcal{A}^2\\\nonumber &+11
\mathcal{A} \mathcal{B}-65 \mathcal{B}^2\big)+4 \mathcal{A}
\mathcal{B}^2 r^8 (\mathcal{B}-\mathcal{A})+e^{\mathcal{A} r^2}
\mathcal{B}^2\big)+4 \mathcal{\mathcal{A}} \mathcal{B}^2 r^8
(\mathcal{B}-\mathcal{\mathcal{A}})\\\nonumber &+e^{\mathcal{A}
r^2}[ \big(r^2 \big(2 \mathcal{B} \big(5 r^2 (T^{0(D)}_{0}+2
T^{1(D)}_{1}+T^{2(D)}_{2})+3\big)+3 (T^{0(D)}_{0}+2
T^{1(D)}_{1}\\\nonumber &+T^{2(D)}_{2})\big)+1\big)-r^2
(\mathcal{A}+24
\mathcal{B})-1\big)\big)\big\}+\frac{8}{r^2}\big\{\beta
\big(\mathcal{B} r^4 (\mathcal{\mathcal{A}}-\mathcal{B})+r^2(7
\mathcal{\mathcal{A}}\\\nonumber &+2 \mathcal{B})+e^{\mathcal{A}
r^2} \big(r^2 (-3 T^{0(D)}_{0}-2
T^{1(D)}_{1}+T^{2(D)}_{2})+1\big)-1\big)+8 \pi \big(\mathcal{B}
r^4\\\label{43a} &\times
(\mathcal{\mathcal{A}}-\mathcal{B})+e^{\mathcal{\mathcal{A}} r^2}
\big(r^2 (-2 T^{0(D)}_{0}-T^{1(D)}_{1}+T^{2(D)}_{2})+1\big)+5
\mathcal{\mathcal{A}} r^2-1\big)\big\}\bigg],\\\nonumber
\tilde{P_r}&=\frac{1}{32 \pi (\beta +4 \pi ) (\beta +8 \pi )
r^2}[(\xi +8 \pi ) e^{-\mathcal{\mathcal{A}} r^2} \big(\beta  r^2
\big(\mathcal{B} r^2
(\mathcal{B}-\mathcal{\mathcal{A}})+\mathcal{A}\\\nonumber&+6
\mathcal{B}\big)+8 \pi \big(\mathcal{B} r^4
(\mathcal{B}-\mathcal{A})-r^2 (\mathcal{A}-4
\mathcal{B})-e^{\mathcal{A} r^2} \big(r^2
(T^{1(D)}_{1}+T^{2(D)}_{2})+1\big)\\\label{43b}&+1\big)-\beta
e^{\mathcal{\mathcal{A}} r^2} \big(r^2 (T^{0(D)}_{0}+2
T^{1(D)}_{1}+T^{2(D)}_{2})+1\big)+\beta \big)],
\\\nonumber
\tilde{P_t}&=\frac{1}{4 (\beta +4 \pi ) (\beta +8 \pi ) \big(2
\mathcal{B} r^2+1\big)^2}\bigg[\xi e^{-\mathcal{A} r^2}
\big(\mathcal{A}^2 r^2 \big(2 \mathcal{B}^2 r^4+3 \mathcal{B}
r^2+1\big)
\\\nonumber&\big(\beta \big(\mathcal{B} r^2-1\big)+8 \pi \big(\mathcal{B}
r^2+1\big)\big)-2 \mathcal{B}^3 r^4 \big(\beta r^2+1\big)\big)-2
\mathcal{B}^3 r^4 \big(\beta \big(e^{\mathcal{\mathcal{A}} r^2}
\big(r^2 (T^{0(D)}_{0}\\\nonumber&+2
T^{1(D)}_{1}+T^{2(D)}_{2})+1\big)-33\big)+8 \pi
\big(e^{\mathcal{\mathcal{A}} r^2}
\big(r^2(T^{1(D)}_{1}+T^{2(D)}_{2})+1\big)-24\big)\big)\\\nonumber&-\mathcal{B}^2
r^2 \big(\beta \big(e^{\mathcal{\mathcal{A}} r^2} \big(9 r^2
(T^{0(D)}_{0}+2 T^{1(D)}_{1}+T^{2(D)}_{2})+7\big)-51\big)+8 \pi
\big(e^{\mathcal{A} r^2} \big(9 r^2\\\nonumber&\times
(T^{1(D)}_{1}+T^{2(D)}_{2})+7\big)-37\big)\big)-2 \mathcal{A}
\big(\mathcal{B} \beta r^2 \big(2 \mathcal{B}^3 r^6+12 \mathcal{B}^2
r^4+11 \mathcal{B} r^2\\\nonumber&+2\big)+8 \pi \big(2 \mathcal{B}^4
r^8+12 \mathcal{B}^3 r^6+17 \mathcal{B}^2 r^4+8 \mathcal{B}
r^2+1\big)\big)-\mathcal{B} \big(\beta \big(e^{\mathcal{\mathcal{A}}
r^2} \big(7 r^2 (T^{0(D)}_{0}\\\nonumber &+2
T^{1(D)}_{1}+T^{2(D)}_{2})+4\big)-10\big)+8 \pi
\big(e^{\mathcal{\mathcal{A}} r^2} \big(7 r^2
(T^{1(D)}_{1}+T^{2(D)}_{2})+4\big)-8\big)\big)\\\nonumber&-e^{\mathcal{\mathcal{A}}
r^2} (\beta (T^{0(D)}_{0}+2 T^{1(D)}_{1}+T^{2(D)}_{2})+8 \pi
(T^{1(D)}_{1}+T^{2(D)}_{2}))+2 \mathcal{B}^5 (\beta +8 \pi )
r^8\\\nonumber&+\mathcal{B}^4 (23 \beta +152 \pi )
r^6\big)\bigg]+\frac{1}{4 (\beta +4 \pi ) (\beta +8 \pi )
r^2}\{(e^{-\mathcal{\mathcal{A}} r^2} \big(-\beta
\big(\mathcal{\mathcal{A}} \mathcal{B} r^4+e^{\mathcal{\mathcal{A}}
r^2} \\\nonumber&\times\big(r^2 (T^{0(D)}_{0}+2
T^{1(D)}_{1}+T^{2(D)}_{2})+1\big)-\mathcal{\mathcal{A}}
r^2-\mathcal{B}^2 r^4-6 \mathcal{B} r^2-1\big)-8 \pi
\big(\mathcal{\mathcal{A}} \mathcal{B}
r^4\\\label{43c}&+e^{\mathcal{A} r^2} \big(r^2
(T^{1(D)}_{1}+T^{2(D)}_{2})+1\big)+\mathcal{A} r^2-\mathcal{B}^2
r^4-4 \mathcal{B} r^2-1\big)\big))\},
\end{align}
\begin{align}\nonumber
\mathrm{s}^2&=r^2 e^{-\mathcal{A} r^2} \bigg[\frac{1}{4 (\beta +4
\pi ) (\beta +8 \pi ) \big(2 \mathcal{B} r^2+1\big)^2}\{\xi \big(-8
\pi \beta r^2 \big(\mathcal{B} \big(-2 r^4
\big(\mathcal{A}^2\\\nonumber&+11 \mathcal{A} \mathcal{B}-33
\mathcal{B}^2\big)+2 \mathcal{B}^2 r^8
(\mathcal{A}-\mathcal{B})^2+\mathcal{B} r^6 (\mathcal{A}-23
\mathcal{B}) (\mathcal{A}-\mathcal{B})+r^2 (51
\mathcal{B}\\\nonumber&-4 \mathcal{A})+10\big)-\mathcal{A}^2
r^2\big)+64 \pi ^2 r^2 \big(-\mathcal{A}^2 r^2 \big(2 \mathcal{B}
r^2+1\big) \big(\mathcal{B} r^2+1\big)^2+2 \mathcal{A}
\big(\mathcal{B} r^2 \\\nonumber&\times\big(2 \mathcal{B} r^2
\big(\mathcal{B} r^2+5\big)+7\big)+1\big) \big(\mathcal{B}
r^2+1\big)-\mathcal{B} \big(\mathcal{B} r^2 \big(\mathcal{B} r^2
\big(\mathcal{B} r^2 \big(2 \mathcal{B}
r^2+19\big)\\\nonumber&+48\big)+37\big)+8\big)\big)+\beta \big(2
\mathcal{B} r^2+1\big)^2 \big(\mathcal{B} r^4
(\mathcal{B}-\mathcal{A})+r^2 (\mathcal{A}+6
\mathcal{B})+1\big)\\\nonumber&+8 \pi \big(2 \mathcal{B}
r^2+1\big)^2 \big(\mathcal{B} r^4 (\mathcal{B}-\mathcal{A})-r^2
(\mathcal{A}-4 \mathcal{B})+1\big)\big)+e^{\mathcal{A} r^2} \big(64
\pi ^2 \big(r^2 \big(2 \mathcal{B}^3 \xi  r^4
\\\nonumber&\times\big(r^2
(T^{1(D)}_{1}+T^{2(D)}_{2})+1\big)+\mathcal{B}^2 r^2 \big(\xi \big(9
r^2 (T^{1(D)}_{1}+T^{2(D)}_{2})+7\big)+24 \beta r^2
(T^{1(D)}_{1}\\\nonumber&-T^{2(D)}_{2})+8\big)+\mathcal{B} \big(\xi
\big(7 r^2 (T^{1(D)}_{1}+T^{2(D)}_{2})+4\big)+24 \beta r^2
(T^{1(D)}_{1}-T^{2(D)}_{2})+8\big)\\\nonumber&+\xi
(T^{1(D)}_{1}+T^{2(D)}_{2})+6 \beta
(T^{1(D)}_{1}-T^{2(D)}_{2})\big)+2\big)+1024 \pi ^3 \big(2
\mathcal{B} r^3+r\big)^2
(T^{1(D)}_{1}\\\nonumber&-T^{2(D)}_{2})+\beta \big(2 \mathcal{B}
r^2+1\big)^2 \big(4 \beta +\xi \big(r^2 (-(T^{0(D)}_{0}+2
T^{1(D)}_{1}+T^{2(D)}_{2}))-1\big)\big)\\\nonumber&+8 \pi \big(\xi
\beta r^2 \big(\mathcal{B} \big(r^2 \big(2 \mathcal{B} r^2+7\big)
\big(\mathcal{B} \big(r^2 (T^{0(D)}_{0}+2
T^{1(D)}_{1}+T^{2(D)}_{2})+1\big)+T^{0(D)}_{0}\\\nonumber&+2
T^{1(D)}_{1}+T^{2(D)}_{2}\big)+4\big)+T^{0(D)}_{0}+2
T^{1(D)}_{1}+T^{2(D)}_{2}\big)-\xi  \big(2 \mathcal{B} r^2+1\big)^2
\big(r^2 (T^{1(D)}_{1}\\\nonumber&+T^{2(D)}_{2})+1\big)+2 \beta
\big(2 \mathcal{B} r^2+1\big)^2 \big(2 \beta  r^2
(T^{1(D)}_{1}-T^{2(D)}_{2})+3\big)\big)\big)\}+\mathcal{B} r^4
(\mathcal{B}-\mathcal{A})\\\label{43d}&-\mathcal{A} r^2-1\bigg].
\end{align}

\subsection{Solution II}

Now, we employ density like constraint
$(\bar{\mu}+\frac{\mathrm{s}^2}{8\pi
r^4}+T^{0(D)}_{0}=\omega^{0}_{0})$ to obtain the second anisotropic
solution. Equations \eqref{18} and \eqref{21} along with this
constraint yield
\begin{equation}\label{48}
\frac{h^{\star'}}{r}+\frac{h^{\star}}{r^{2}}-\frac{1}{r^{2}}
-\frac{e^{-\mathcal{\mathcal{A}} r^2}}{r^2}+2\mathcal{\mathcal{A}}
e^{-\mathcal{\mathcal{A}} r^2}=0,
\end{equation}
whose integration gives the solution
\begin{equation}\label{49}
h^{\star}=\frac{F_1}{r} +1-e^{-{\mathcal{A}} r^2},
\end{equation}
where $F_1$ is the constant of integration. We choose $F_1=0$ so
that our resulting solution becomes free from any singularity at the
core of the astrophysical object. In the current set up, the
matching conditions are obtained by following the same procedure as
in solution I
\begin{align}\label{53}
&2\texttt{R}
(\mathbb{M}-\mathbb{M}_o)+\mathbb{S}^2_o-\mathbb{S}^2+\beta
\texttt{R}^2 (1-e^{-\mathcal{\mathcal{A}}
\texttt{R}^2})=0,\\\label{53a} &C+\mathcal{B}
\texttt{R}^2=\ln\bigg[1-\frac{2\mathbb{M}_o}{\texttt{R}}+\frac{\mathbb{S}^2_o}{\texttt{R}^2}-\frac{\mathbb{S}^2}{\texttt{R}^2}+\beta(1-e^{-\mathcal{A}
\texttt{R}^2})\bigg].
\end{align}
The expressions for the state variables and charge are as follows
\begin{align}\nonumber
\tilde{\mu}&=\frac{1}{32 \pi  (\beta +4 \pi ) (\beta +8 \pi )
r^2}\bigg[(8 \pi -\xi ) e^{-{\mathcal{A}} r^2} \big(\beta
\big(\mathcal{B} r^4 (\mathcal{\mathcal{A}}-\mathcal{B})+r^2 (7
\mathcal{A}\\\nonumber&+2 \mathcal{B})+e^{\mathcal{A} r^2} \big(r^2
(-3 T^{0(D)}_{0}-2 T^{1(D)}_{1}+T^{2(D)}_{2})+1\big)-1\big)+8 \pi
\big(\mathcal{B} r^4
\\\label{53b}&\times(\mathcal{A}-\mathcal{B})+e^{\mathcal{A} r^2} \big(r^2
(-2 T^{0(D)}_{0}-T^{1(D)}_{1}+T^{2(D)}_{2})+1\big)+5
\mathcal{\mathcal{A}} r^2-1\big)\big)\bigg] ,\\\nonumber
\tilde{P_r}&= \frac{1}{768 (\beta +4 \pi ) (\beta +8 \pi )
r^3}\bigg[e^{-\mathcal{A} r^2} \bigg(\frac{1}{\pi
\mathcal{A}^{5/2}}\{\xi \big(2 \mathcal{B} r^2+1\big) \big(3
\sqrt{\pi } (\mathcal{A}+\mathcal{B})\\\nonumber &\times
e^{\mathcal{A} r^2} (2 \mathcal{A} (5 \beta +24 \pi )-3 \mathcal{B}
(\beta +8 \pi ))\text {Erf}\big(\sqrt{\mathcal{A}} r\big)-2
\sqrt{\mathcal{A}} r \big(2 \mathcal{A}^2 \big(2 e^{\mathcal{A} r^2}
\\\nonumber& \big(2 \mathcal{\mathcal{A}}^2 \big(2 e^{\mathcal{A} r^2}
\big(\beta \big(r^2 (3 T^{0(D)}_{0}+2
T^{1(D)}_{1}-T^{2(D)}_{2})-3\big)+8 \pi \big(r^2 (2
T^{0(D)}_{0}+T^{1(D)}_{1}\\\nonumber &-T^{2(D)}_{2})-3\big)\big)+3
\beta \big(\mathcal{B} r^2+7\big)+24 \pi  \big(\mathcal{B}
r^2+5\big)\big)+3 \mathcal{A} \mathcal{B} \big(\beta \big(7-2
\mathcal{B} r^2\big)\\\nonumber &+8 \pi \big(3-2 \mathcal{B}
r^2\big)\big)-9 \mathcal{B}^2 (\beta +8 \pi )\big)\big)\}+\{192 r
\big(\beta r^2 \big(\mathcal{B} r^2
(\mathcal{B}-\mathcal{A})+\mathcal{A}+6
\mathcal{B}\big)\\\nonumber&+8 \pi \big(\mathcal{B} r^4
(\mathcal{B}-\mathcal{\mathcal{A}})-r^2 (\mathcal{A}-4
\mathcal{B})-e^{\mathcal{A} r^2} \big(r^2
(T^{1(D)}_{1}+T^{2(D)}_{2})+1\big)+1\big)\\\label{54}&-\beta
e^{\mathcal{A} r^2} \big(r^2 (T^{0(D)}_{0}+2
T^{1(D)}_{1}+T^{2(D)}_{2})+1\big)+\beta \big)\big\}\bigg)\bigg],
\\\nonumber
\tilde{P_t}&=\frac{1}{192 (\beta +4 \pi ) (\beta +8 \pi )
r^3}\bigg[e^{-\mathcal{\mathcal{A}} r^2}
\bigg(\frac{1}{\mathcal{\mathcal{A}}^{5/2}}\{3 \sqrt{\pi } \xi
(\mathcal{A}+\mathcal{B}) e^{\mathcal{A} r^2} \big(\mathcal{B} r^2
\big(2 \mathcal{B} r^2\\\nonumber &+3\big)-1\big) (2 \mathcal{A} (5
\beta +24 \pi )-3 \mathcal{B} (\beta +8 \pi ))
\text{Erf}\big(\sqrt{\mathcal{A}}
r\big)\}+\frac{1}{\mathcal{A}^2}\{2
r\big)\}+\frac{1}{{\mathcal{A}}^2}\\\nonumber &\times\{2 r \big(12
{\mathcal{A}}^3 r^2\big(\beta \big(\mathcal{B} r^2 \big(\mathcal{B}
\xi r^2+8 \xi -2\big)+7 \xi +2\big)+8 \pi \big(\mathcal{B}
r^2+1\big) \big(\mathcal{B} \xi r^2\\\nonumber &+5 \xi
-2\big)\big)+2 \mathcal{A}^2 \big(-4 \beta e^{\mathcal{A} r^2}
\big(r^2 \big(\xi \big(\mathcal{B} r^2 \big(\mathcal{B}
r^2+3\big)+1\big) (3 T^{0(D)}_{0}+2
T^{1(D)}_{1}\\\nonumber&-T^{2(D)}_{2})+3 \big(-\mathcal{B} \xi
\big(\mathcal{B} r^2+2\big)+2 T^{1(D)}_{1}+T^{2(D)}_{2}\big)+3
T^{0(D)}_{0}\big)+3\big)+8 \pi \\\nonumber&\times\big(-4
e^{\mathcal{A} r^2} \big(r^2 \big(\xi \big(\mathcal{B} \big(r^2
\big(\mathcal{B} \big(r^2 (2
T^{0(D)}_{0}+T^{1(D)}_{1}-T^{2(D)}_{2})-3\big)+3 (2
T^{0(D)}_{0}\\\nonumber&+T^{1(D)}_{1}-T^{2(D)}_{2})\big)-6\big)+2
T^{0(D)}_{0}+T^{1(D)}_{1}-T^{2(D)}_{2}\big)+3
(T^{1(D)}_{1}+T^{2(D)}_{2})\big)\\\nonumber&+3\big)-3 \mathcal{B}
r^2 \big(\mathcal{B} r^2 \big(4 \mathcal{B} \xi r^2+15 \xi
-4\big)+16 (\xi -1)\big)+9 \xi +12\big)+3 \beta \big(\mathcal{B} r^2
\\\nonumber&\times\big(\mathcal{B} r^2 \big(-4 \mathcal{B} \xi r^2-15 \xi +4\big)-18
\xi +24\big)+5 \xi +4\big)\big)+3 {\mathcal{A}} \mathcal{B} \xi
\big(\mathcal{B} r^2 \big(2 \mathcal{B}
r^2\\\nonumber&+3\big)-1\big) \big(\beta \big(2 \mathcal{B}
r^2-7\big)+8 \pi \big(2 \mathcal{B} r^2-3\big)\big)+9 \mathcal{B}^2
\xi (\beta +8 \pi ) \big(\mathcal{B} r^2 \big(2 \mathcal{B}
r^2+3\big)\\\label{55}&-1\big)\big)\}\bigg)\bigg],
\end{align}
\begin{align}\nonumber
\mathrm{s}^2&=\frac{1}{96 \mathcal{A}^{5/2} (\beta +4 \pi ) (\beta
+8 \pi )}\bigg[r^2 e^{-\mathcal{A} r^2} \big(-2 \sqrt{\mathcal{A}}
\big(48 \mathcal{A}^3 r^2 \big(\mathcal{B} r^2+1\big) \big(\pi \beta
\big(\mathcal{B} \xi  r^2\\\nonumber&+7 \xi +12\big)+8 \pi ^2
\big(\mathcal{B} \xi r^2+5 \xi +4\big)+\beta ^2\big)+2 \mathcal{A}^2
\big(3 \big(\beta \big(-8 \mathcal{B}^2 \beta r^4+\xi
\big(\mathcal{B} r^2\\\nonumber&+7\big) \big(2 \mathcal{B}
r^2+1\big)+8 \beta \big)-4 \pi \beta \big(\mathcal{B} r^2
\big(\mathcal{B} r^2 \big(4 \mathcal{B} \xi r^2+15 \xi +24\big)+18
\xi \big)-5 \xi \\\nonumber&-24\big)+8 \pi \xi \big(\mathcal{B}
r^2+5\big) \big(2 \mathcal{B} r^2+1\big)-32 \pi ^2 \big(\mathcal{B}
r^2 \big(\mathcal{B} r^2 \big(4 \mathcal{B} \xi r^2+15 \xi
+8\big)\\\nonumber&+16 \xi \big)-3 \xi -8\big)\big)-2 e^{\mathcal{A}
r^2} \big(\beta \big(\xi \big(-2 \mathcal{B} r^2-1\big) \big(r^2 (3
T^{0(D)}_{0}+2 T^{1(D)}_{1}-T^{2(D)}_{2})\\\nonumber&-3\big)+12
\beta \big)+64 \pi ^2 \big(r^2 \big(\xi  \big(\mathcal{B} \big(r^2
\big(\mathcal{B} \big(r^2 (2
T^{0(D)}_{0}+T^{1(D)}_{1}-T^{2(D)}_{2})-3\big)+3
\\\nonumber&\times(2
T^{0(D)}_{0}+T^{1(D)}_{1}-T^{2(D)}_{2})\big)-6\big)+2
T^{0(D)}_{0}+T^{1(D)}_{1}-T^{2(D)}_{2}\big)+18 \beta
(T^{1(D)}_{1}\\\nonumber&-T^{2(D)}_{2})\big)+6\big)+8 \pi  \big(\xi
\beta r^2 \big(\mathcal{B} \big(r^2 \big(\mathcal{B} \big(r^2 (3
T^{0(D)}_{0}+2 T^{1(D)}_{1}-T^{2(D)}_{2})-3\big)\\\nonumber&+9
T^{0(D)}_{0}+6 T^{1(D)}_{1}-3 T^{2(D)}_{2}\big)-6\big)+3
T^{0(D)}_{0}+2 T^{1(D)}_{1}-T^{2(D)}_{2}\big)+\xi  \big(-2
\mathcal{B} r^2\\\nonumber&-1\big) \big(r^2 (2
T^{0(D)}_{0}+T^{1(D)}_{1}-T^{2(D)}_{2})-3\big)+6 \beta \big(2 \beta
r^2 (T^{1(D)}_{1}-T^{2(D)}_{2})+3\big)\big)\\\nonumber&+3072 \pi ^3
r^2 (T^{1(D)}_{1}-T^{2(D)}_{2})\big)\big)+3 \mathcal{A} \mathcal{B}
\xi \big(8 \pi  \mathcal{B}^2 r^4+2 (6 \pi -1) \mathcal{B} r^2-4 \pi
-1\big)\\\nonumber&\times \big(\beta \big(2 \mathcal{B} r^2-7\big)+8
\pi \big(2 \mathcal{B} r^2-3\big)\big)+9 \mathcal{B}^2 \xi (\beta +8
\pi ) \big(8 \pi  \mathcal{B}^2 r^4+2 (6 \pi -1) \mathcal{B}
r^2\\\nonumber&-4 \pi -1\big)\big)-\frac{1}{r}\{3 \sqrt{\pi } \xi
(\mathcal{A}+\mathcal{B}) e^{\mathcal{A} r^2} \big(8 \pi
\mathcal{B}^2 r^4+2 (6 \pi -1) \mathcal{B} r^2-4 \pi
-1\big)\\\label{55c}&\times (2 \mathcal{A} (5 \beta +24 \pi )-3
\mathcal{B} (\beta +8 \pi )) \text{Erf}\big(\sqrt{\mathcal{A}}
r\big)\}\big)\bigg].
\end{align}

\section{Physical Aspects}

In this section, we examine physical viability and stability of the
resulting solutions. The anisotropy for the solution I becomes
\begin{eqnarray}\nonumber
\tilde{\Delta}&=&\frac{\xi e^{-\mathcal{A} r^2}}{32 (\beta +4 \pi )
(\beta +8 \pi )}\bigg[ \frac{8}{\big(2 \mathcal{B} r^2+1\big)^2}
\{\big(\mathcal{A}^2 r^2 \big(\mathcal{B} r^2+1\big) \big(2
\mathcal{B} r^2+1\big) \big(\beta \big(\mathcal{B}
r^2\\\nonumber&-&1\big)+8 \pi \big(\mathcal{B} r^2+1\big)\big)-2
\mathcal{A} \mathcal{B} \beta r^2 \big(\mathcal{B} r^2 \big(2
\mathcal{B} r^2 \big(\mathcal{B}
r^2+6\big)+11\big)+2\big)-e^{\mathcal{A} r^2} \big
(\beta\\\nonumber&\times& \big(\mathcal{B} \big(r^2 \big(2
\mathcal{B} r^2+7\big) \big(\mathcal{B} \big(r^2 (T^{0(D)}_{0}+2
T^{1(D)}_{1}+T^{2(D)}_{2})+1\big)+T^{0(D)}_{0}+2
T^{1(D)}_{1}\\\nonumber&+&T^{2(D)}_{2}\big)+4\big)+T^{0(D)}_{0}+2
T^{1(D)}_{1}+T^{2(D)}_{2}\big)+8 \pi  \big(\mathcal{B} \big(r^2
\big(2 \mathcal{B} r^2+7\big) \big(\mathcal{B} \big(r^2
(T^{1(D)}_{1}\\\nonumber&+&T^{2(D)}_{2})+1\big)+T^{1(D)}_{1}+T^{2(D)}_{2}\big)+4\big)+T^{1(D)}_{1}+T^{2(D)}_{2}\big)\big)-16
\pi  \mathcal{A} \big(\mathcal{B} r^2+1\big)\\\nonumber&\times&
\big(\mathcal{B} r^2 \big(2 \mathcal{B} r^2 \big(\mathcal{B}
r^2+5\big)+7\big)+1\big)+2 \mathcal{B}^5 \beta r^8+16 \pi
\mathcal{B}^5 r^8+23 \mathcal{B}^4 \beta  r^6+152
\\\nonumber&\times&\pi  \mathcal{B}^4 r^6+66 \mathcal{B}^3 \beta
r^4+384 \pi \mathcal{B}^3 r^4+51 \mathcal{B}^2 \beta  r^2+296 \pi
\mathcal{B}^2 r^2+10 \mathcal{B} \beta +64 \pi
\mathcal{B}\big)\}\\\nonumber&+&\frac{1}{\pi r^2}\{\beta
\big(\mathcal{B} r^4 (\mathcal{A}-\mathcal{B})-r^2 (\mathcal{A}+6
\mathcal{B})+e^{\mathcal{A} r^2} \big(r^2 (T^{0(D)}_{0}+2
T^{1(D)}_{1}+T^{2(D)}_{2})+1\big)\\\nonumber&-&1\big)+8 \pi
\big(\mathcal{B} r^4 (\mathcal{A}-\mathcal{B})+r^2 (\mathcal{A}-4
\mathcal{B})+e^{\mathcal{A} r^2} \big(r^2
(T^{1(D)}_{1}+T^{2(D)}_{2})+1\big)-1\big)\}\bigg],\\\label{55a}
\end{eqnarray}
whereas for solution II, we have
\begin{align}\nonumber
\tilde{\Delta}&=\frac{e^{-\mathcal{A} r^2}}{768 (\beta +4 \pi )
(\beta +8 \pi ) r^3}\bigg[\frac{1}{\sqrt{\pi } \mathcal{A}^{5/2}}\{3
\xi (\mathcal{A}+\mathcal{B}) e^{\mathcal{A} r^2} \big(8 \pi
\mathcal{B}^2 r^4+2 (6 \pi -1) \mathcal{B} r^2\\\nonumber&-4 \pi
-1\big) (2 \mathcal{A} (5 \beta +24 \pi )-3 \mathcal{B} (\beta +8
\pi )) \text{Erf}\big(\sqrt{\mathcal{A}} r\big)\}+\frac{2 \xi r}{\pi
\mathcal{A}^2}\{ \big(48 \pi \mathcal{A}^3 r^2 \big(\mathcal{B}
r^2\\\nonumber&+1\big) \big(\beta \big(\mathcal{B} r^2+7\big)+8 \pi
\big(\mathcal{B} r^2+5\big)\big)+2 \mathcal{A}^2 \big(\beta \big(2
\mathcal{B} r^2+1\big) \big(e^{\mathcal{A} r^2} \big(r^2 (6
T^{0(D)}_{0}+4 T^{1(D)}_{1}\\\nonumber&-2 T^{2(D)}_{2})-6\big)+3
\big(\mathcal{B} r^2+7\big)\big)+4 \pi \big(-4 e^{\mathcal{A} r^2}
\big(\beta r^2 \big(\mathcal{B} \big(r^2 \big(\mathcal{B} \big(r^2
(3 T^{0(D)}_{0}+2 T^{1(D)}_{1}\\\nonumber&-T^{2(D)}_{2})-3\big)+9
T^{0(D)}_{0}+6 T^{1(D)}_{1}-3 T^{2(D)}_{2}\big)-6\big)+3
T^{0(D)}_{0}+2 T^{1(D)}_{1}-T^{2(D)}_{2}\big)\\\nonumber&+\big(-2
\mathcal{B} r^2-1\big) \big(r^2 (2
T^{0(D)}_{0}+T^{1(D)}_{1}-T^{2(D)}_{2})-3\big)\big)-3 \mathcal{B}
r^2 \big(\mathcal{B} r^2 \big(4 \mathcal{B} \beta r^2+15 \beta
\\\nonumber&-4\big)+18 \beta -22\big)+15 (\beta +2)\big)+32 \pi ^2 \big(r^2
\big(-4 e^{\mathcal{A} r^2} \big(\mathcal{B} \big(r^2
\big(\mathcal{B} \big(r^2 (2
T^{0(D)}_{0}+T^{1(D)}_{1}\\\nonumber&-T^{2(D)}_{2})-3\big)+3 (2
T^{0(D)}_{0}+T^{1(D)}_{1}-T^{2(D)}_{2})\big)-6\big)+2
T^{0(D)}_{0}+T^{1(D)}_{1}-T^{2(D)}_{2}\big)\\\nonumber&-3
\mathcal{B} \big(\mathcal{B} r^2 \big(4 \mathcal{B}
r^2+15\big)+16\big)\big)+9\big)\big)+3 \mathcal{A} \mathcal{B}
\big(8 \pi \mathcal{B}^2 r^4+2 (6 \pi -1) \mathcal{B} r^2-4 \pi
\\\nonumber&-1\big)\big(\beta \big(2 \mathcal{B} r^2-7\big)+8 \pi \big(2
\mathcal{B} r^2-3\big)\big)+9 \mathcal{B}^2 (\beta +8 \pi ) \big(8
\pi \mathcal{B}^2 r^4+2 (6 \pi -1) \mathcal{B} r^2\\\label{55b}&-4
\pi -1\big)\big)\}\bigg].
\end{align}
It is well-known that the anisotropic factor shows positive behavior
if $\tilde{P_t}>\tilde{P_r}$, indicating that the anisotropic force
is acting outward. When $\tilde{P_r}>\tilde{P_t}$, the anisotropy is
negative and squeezes the matter within the star (together with
gravitational force).

In order to observe the physical analysis of the obtained solutions,
we consider the model \eqref{60} with parameter $\beta$ as 0.01 and
charge parameter $(\mathrm{s})$ as 0.1 and 0.9. The value of the
constant $\mathcal{A}$ is taken from \eqref{41} while $\mathcal{B}$
as well as $\mathcal{C}$ are fixed from \eqref{36} and \eqref{36a}.
The effective matter determinants
$(\tilde{\mu},\tilde{P_r},\tilde{P_t})$ must be positive, maximum
and finite in the inner structure of the charged celestial object or
we can say that these parameters should decrease as $r$ increases.
Figure \textbf{1} shows that $\tilde{\mu}$, $\tilde{P_r}$ and
$\tilde{P_t}$ for solution I decrease monotonically towards the
boundary as $r$ increases. Also, it is observed that the effective
density of the system decreases with charge showing that the stellar
object becomes less dense for larger charge while the
radial/tangential pressures become zero at the boundary. The last
plot of Figure \textbf{1} demonstrates that anisotropy decreases
with $r$ but increases with $\xi$ in the current set up. Thus, one
can observe that anisotropy is positive, meaning that it will act in
the outward direction of the star. The higher values of the
decoupling parameter $\xi$ assures more anisotropic system.
\begin{figure}\center
\epsfig{file=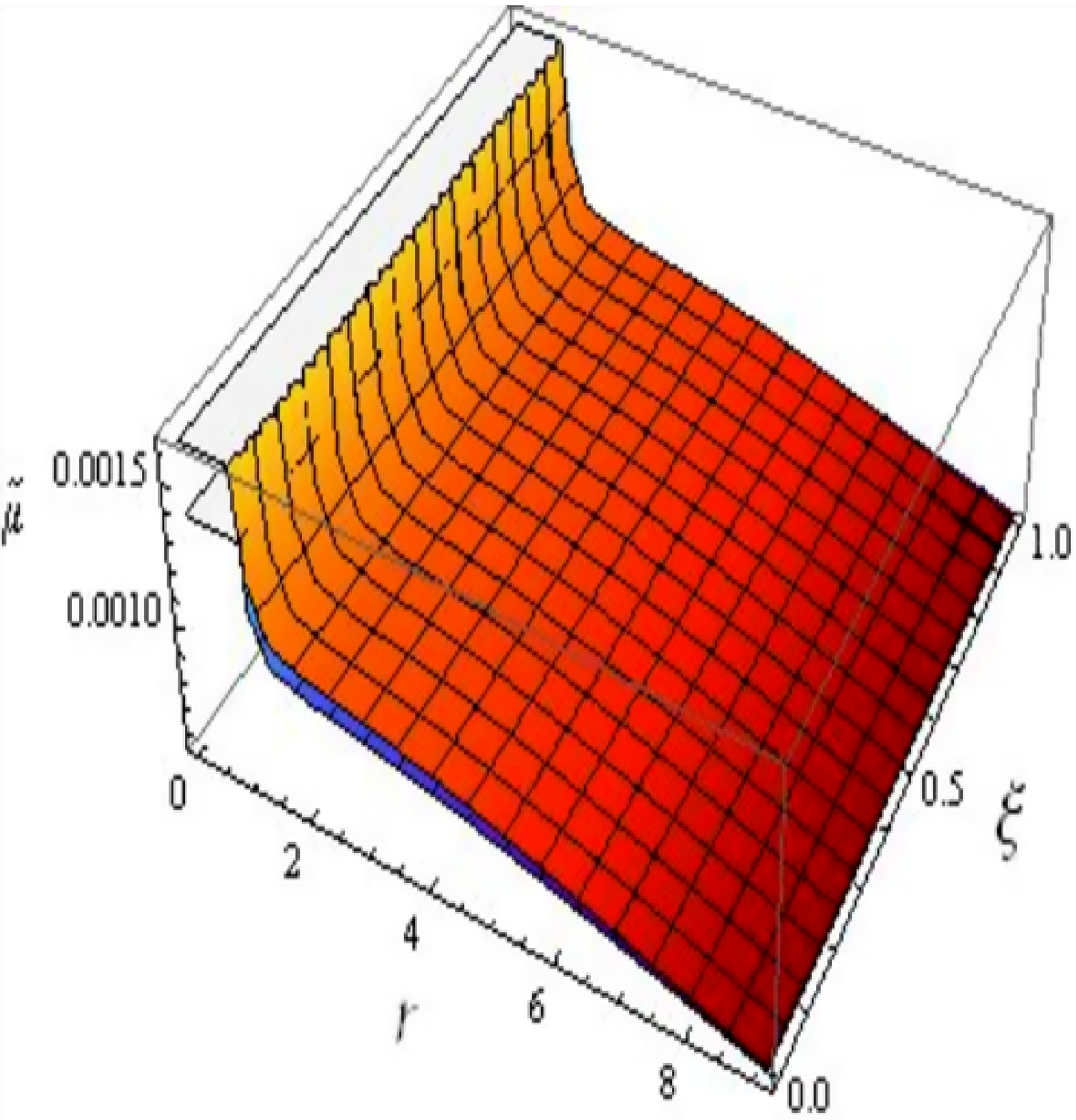,width=0.4\linewidth}\epsfig{file=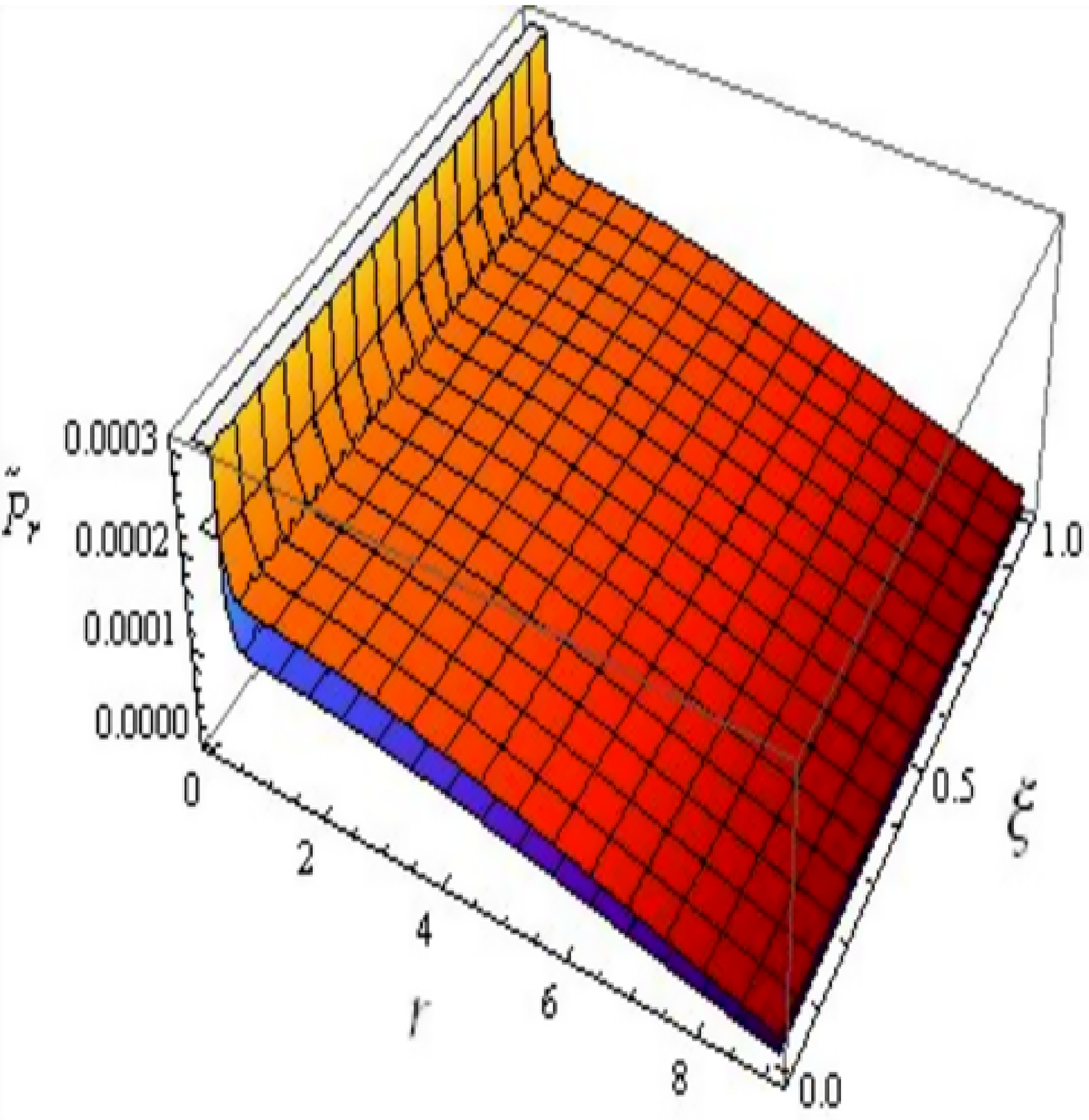,width=0.4\linewidth}
\epsfig{file=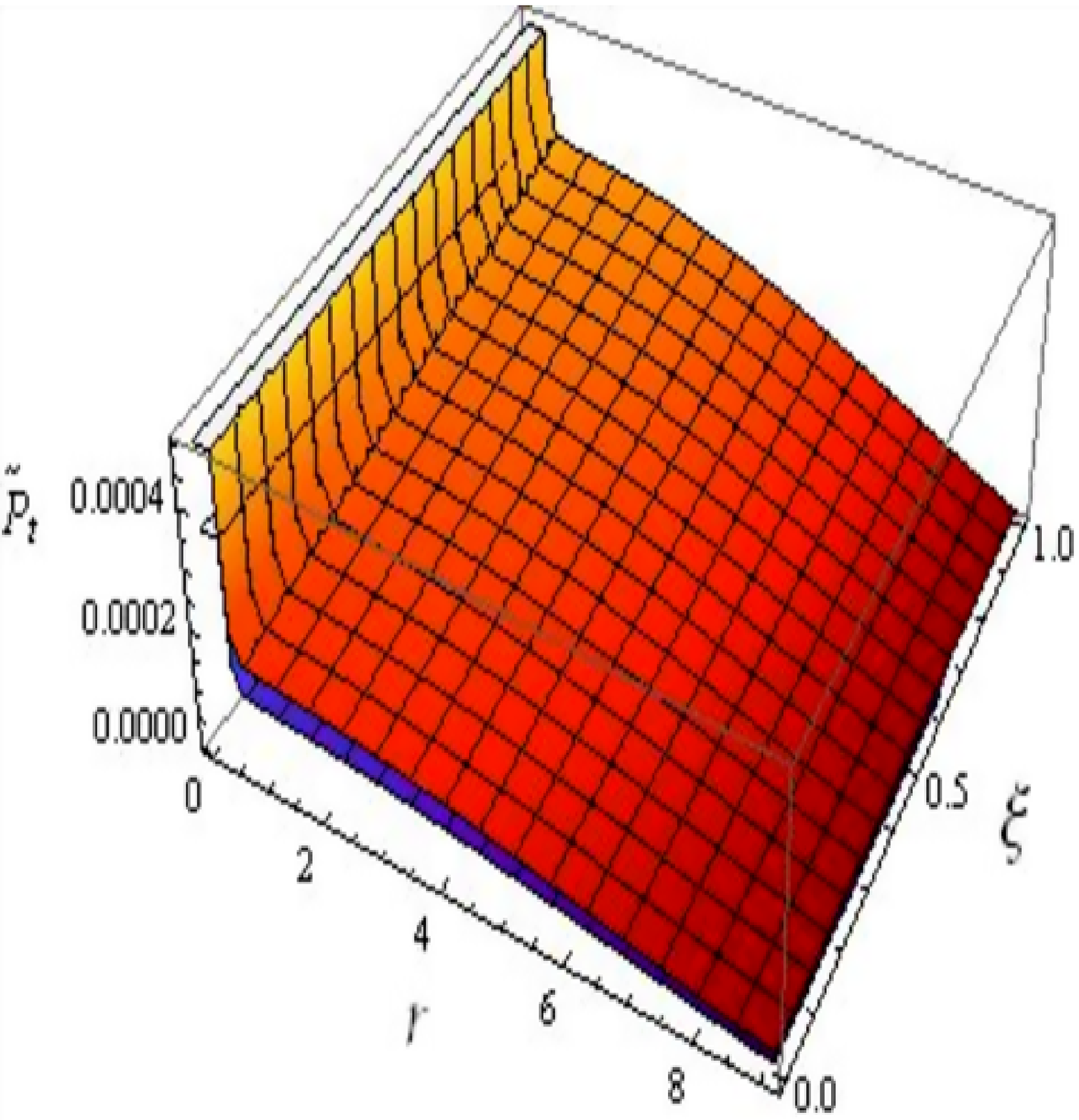,width=0.4\linewidth}\epsfig{file=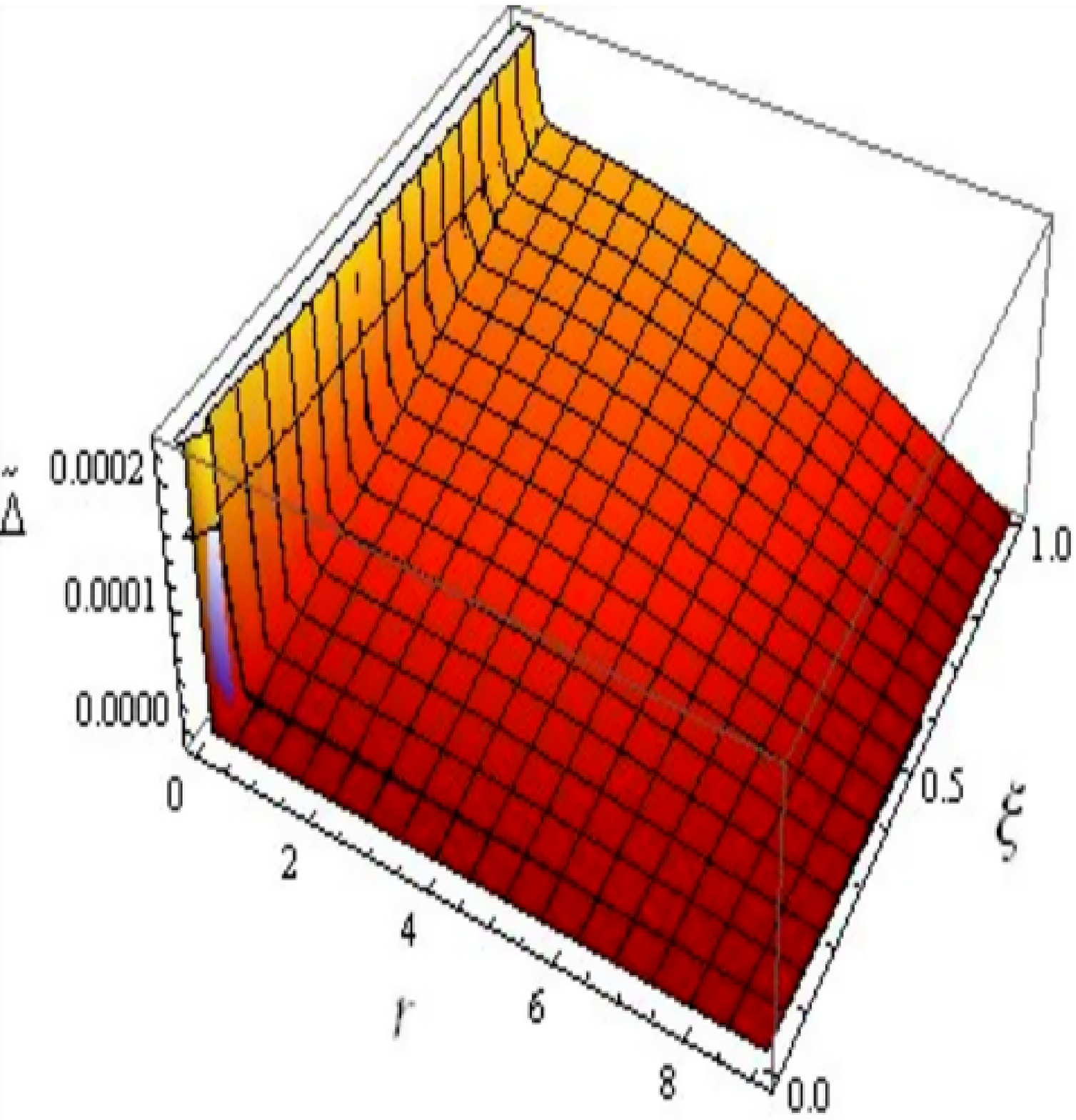,width=0.4\linewidth}
\caption{Plots of $\tilde{\mu},\tilde{P_r},\tilde{P_t}$ and
$\tilde{\Delta}$ versus $r$ and $\xi$ with $\mathbb{S}_o=0.1$
(Orange), $\mathbb{S}_o=0.9$ (Blue) for the solution I.}
\end{figure}

The compactness parameter (ratio between star's mass and radius) is
considered as an important aspect of self-gravitating stellar
objects. Buchdahl \cite{22b} matched the Schwarzschild exterior
vacuum region with the interior spherical geometry to evaluate the
upper limit of compactness parameter as
$\frac{m}{\texttt{R}}<\frac{4}{9}$, where
$m=\frac{\texttt{R}}{2}(1-e^{-\lambda})$. However, the Buchdahl
limit (upper bound) has been modified due to the involvement of
charge in the matter source. The upper bound of mass-radius ratio is
modified by Andreasson \cite{22bc} for a charged spherical
astrophysical object and is defined as
$\zeta(r)=\frac{m}{\texttt{R}}<\frac{2}{9}+\frac{3\mathbb{S}_o^2+2\texttt{R}\sqrt{\texttt{R}^2+3\mathbb{S}_o^2}}{9\texttt{R}^2}$.
The mass of the static spherical celestial object can be determined
by
\begin{equation}\label{58}
m=4\pi\int^{\texttt{R}}_{0}\big(\tilde{\mu}+\frac{\mathrm{s^2}}{8\pi
r^4}\big)r^2dr.
\end{equation}
The numerical technique is employed to determine the mass of
anisotropic structure along with the initial condition $m(0)=0$. The
wavelength of the electromagnetic radiations produced by a celestial
object, having a strong gravitational force, increases. This
increment is calculated by the redshift factor whose mathematical
expression is $Z(r)=\frac{1}{\sqrt{1-2\zeta}}-1$. For perfect matter
source, Buchdahl restricted this parameter to $Z(r)<2$ at the star's
surface while for the anisotropic configurations, this value is
observed as 5.211 \cite{22c}. Another important quantity of
self-gravitating objects is the equation of state (EoS) parameter,
defined as
\begin{equation}\label{58a}
\tilde{\textbf{w}}_{r}=\frac{\tilde{P_r}}{\tilde{\mu}}, \quad
\tilde{\textbf{w}}_{t}=\frac{\tilde{P_t}}{\tilde{\mu}}.
\end{equation}
For the effectiveness of the stellar matter source, both the EoS
parameters (radial and tangential) must lie in [0,1] \cite{23}.

Stability is another important factor in the analysis of compact
objects. There are various methods to check the stability of the
celestial objects like Herrera cracking approach and causality
condition. According to Herrera cracking approach \cite{24}, the
velocity of the considered system must satisfy
$|\nu^{2}_{t}-\nu^{2}_r|<1$, where $\nu^{2}_r$ and $\nu^{2}_{t}$ are
the squared speed sound in radial and tangential directions given as
$\nu^{2}_{r}=\frac{dP_{r}}{d\mu}$ and
$\nu^{2}_{t}=\frac{dP_{t}}{d\mu}$, respectively. The causality
condition states that the speed of sound should be less than the
speed of light and its components must lie in the interval 0 and 1,
i.e., $0<\nu^2_{t}<1$ and $0<\nu^2_{r}<1$.
\begin{figure}\center
\epsfig{file=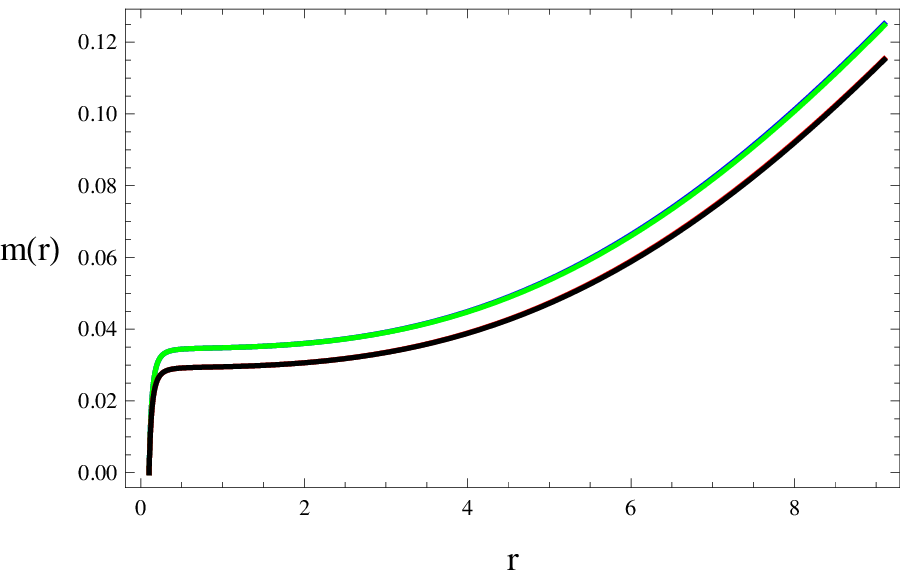,width=0.4\linewidth}\epsfig{file=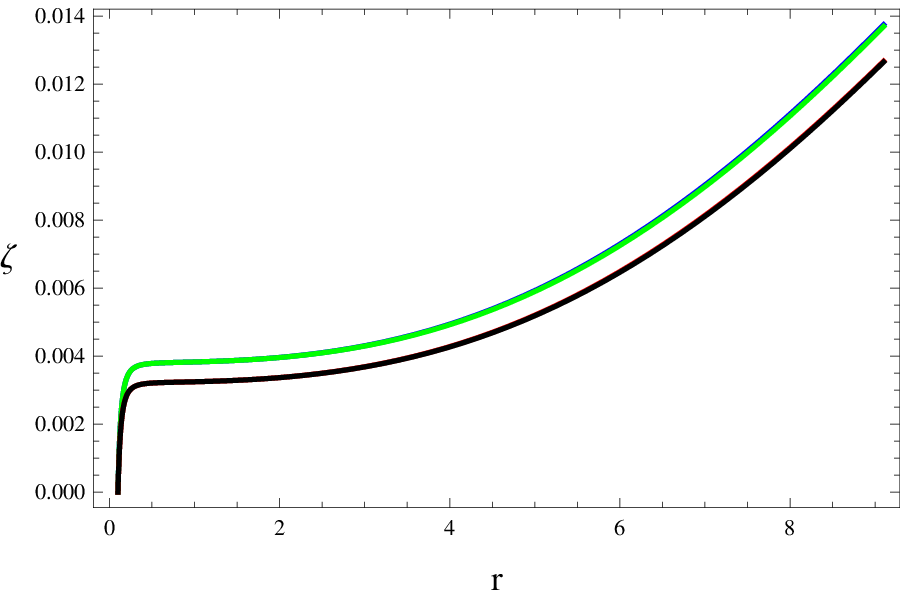,width=0.4\linewidth}
\epsfig{file=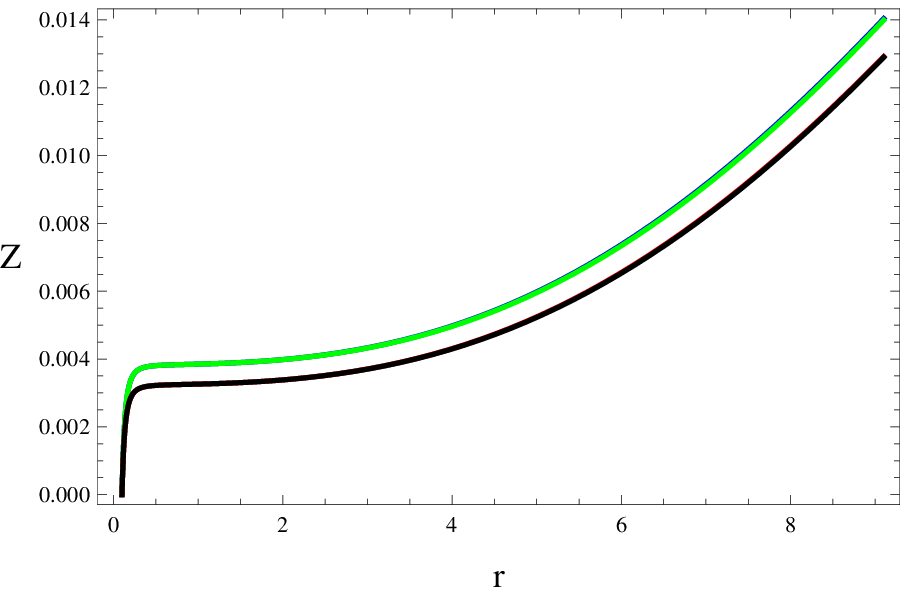,width=0.4\linewidth} \caption{Plots of mass,
compactness and redshift versus $r$ corresponding to
$\mathbb{S}_o=0.1$, $\xi=0.01$ (Blue), $\xi=0.9$ (Green) and
$\mathbb{S}_o=0.9$, $\xi=0.01$ (Red), $\xi=0.9$ (Black) for solution
I.}
\end{figure}
\begin{figure}\center
\epsfig{file=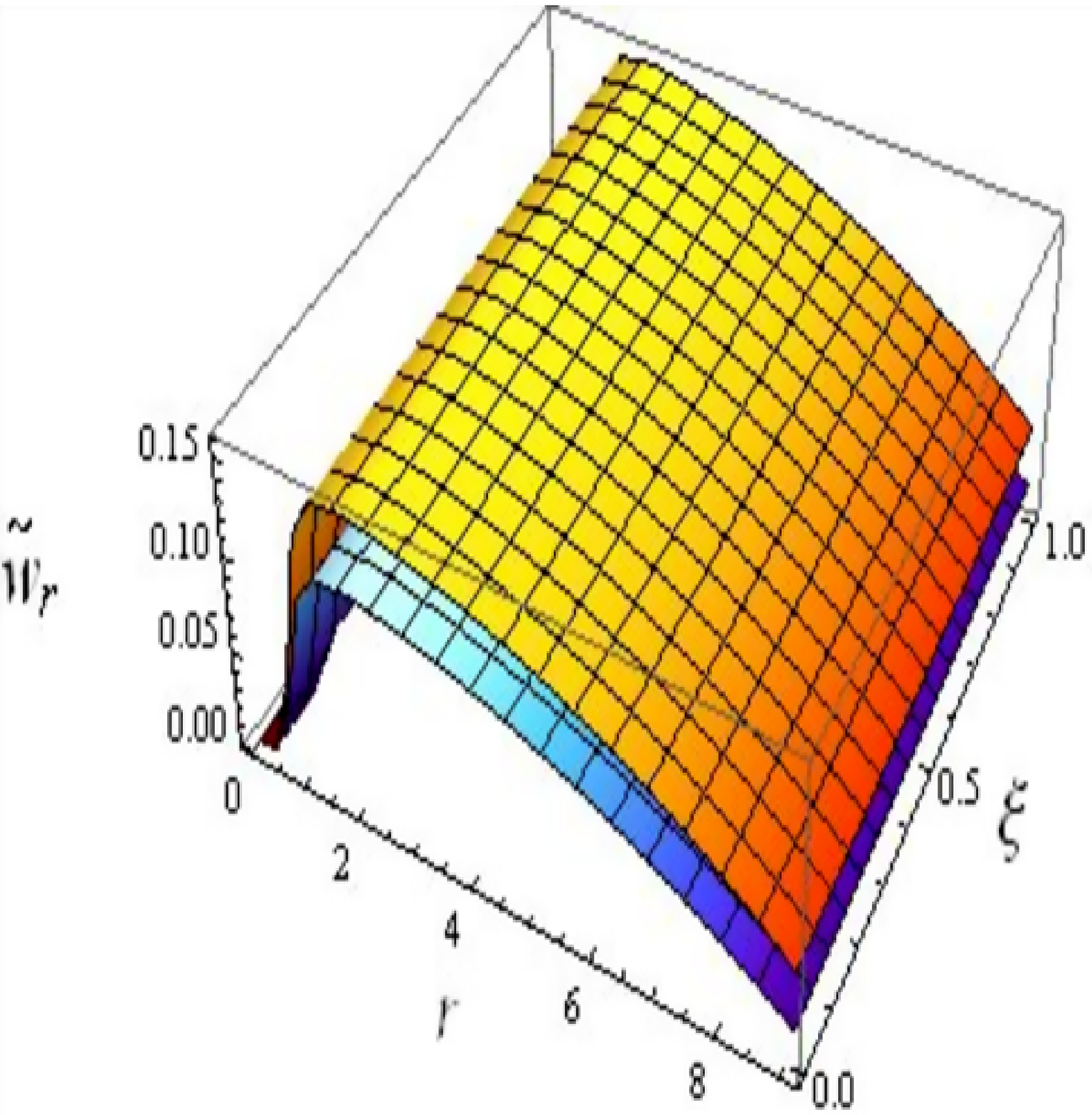,width=0.4\linewidth}
\epsfig{file=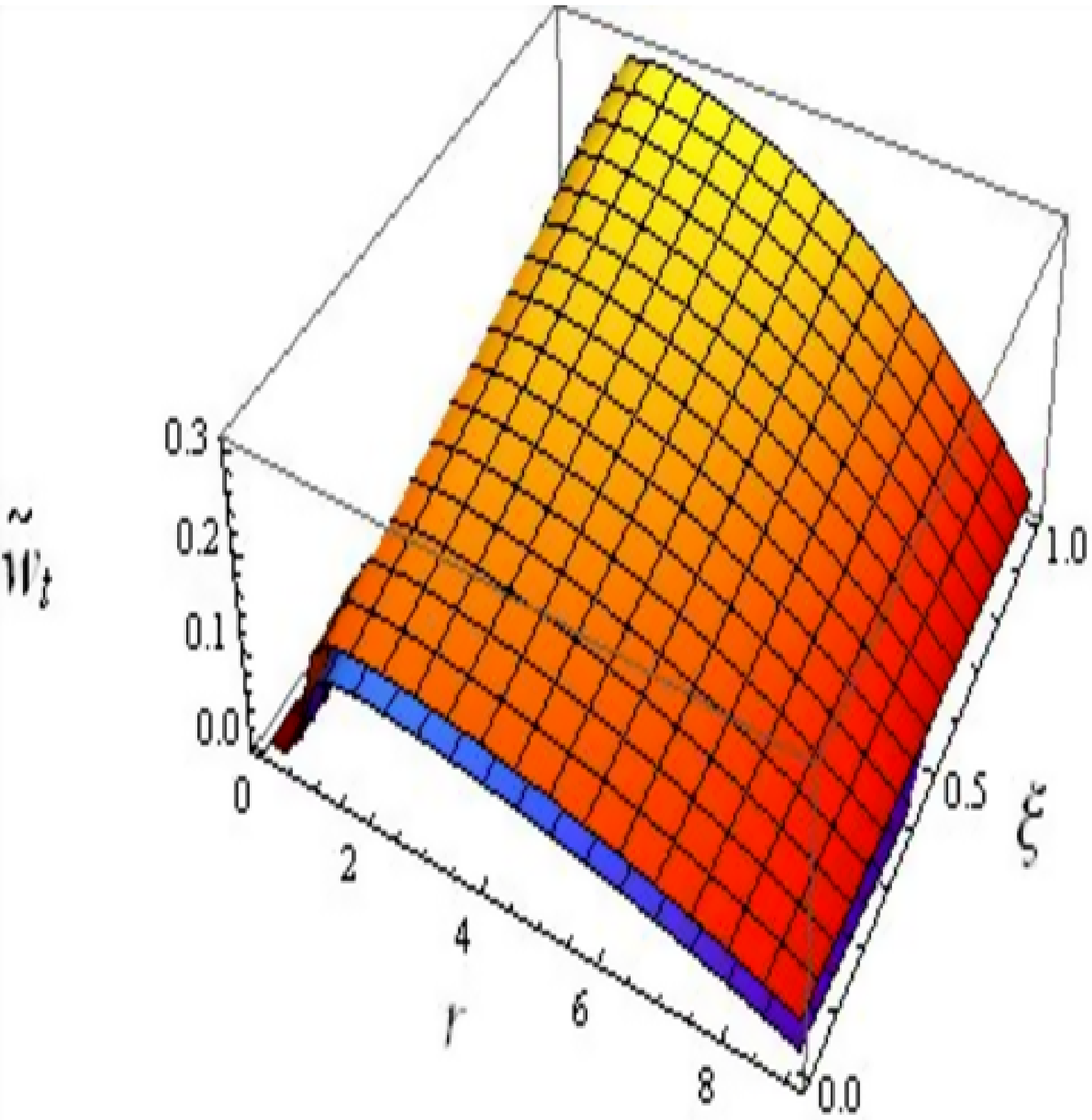,width=0.4\linewidth} \caption{Plots of EoS
parameters versus $r$ and $\xi$ with $\mathbb{S}_o=0.1$ (Orange),
$\mathbb{S}_o=0.9$ (Blue) for the solution I.}
\end{figure}

The fluid distribution of a stellar body is characterized by EMT on
which some limitations are enforced called as energy conditions. The
viability of the obtained solutions as well as the existence of
normal matter are checked using these conditions. It is also
important that the parameters governing the internal structure must
satisfy these conditions. These limitations are divided as dominant,
weak, null and strong energy conditions. In $f(G,T)$ theory, these
constraints are described as
\begin{align}\nonumber
&\tilde{\mu}+\frac{\mathrm{s}^2}{8\pi r^4}\geq0, \quad
\tilde{\mu}+\tilde{P_r}\geq0, \\\nonumber
&\tilde{\mu}+\tilde{P_t}+\frac{\mathrm{s}^2}{4\pi r^4}\geq0, \quad
\tilde{\mu}-\tilde{P_r}+\frac{\mathrm{s}^2}{4\pi
r^4}\geq0,\\\label{59} &\tilde{\mu}-\tilde{P_t}\geq0, \quad
\tilde{\mu}+\tilde{P_r}+2\tilde{P_t}+\frac{\mathrm{s}^2}{4\pi
r^4}\geq0.
\end{align}
\begin{figure}\center
\epsfig{file=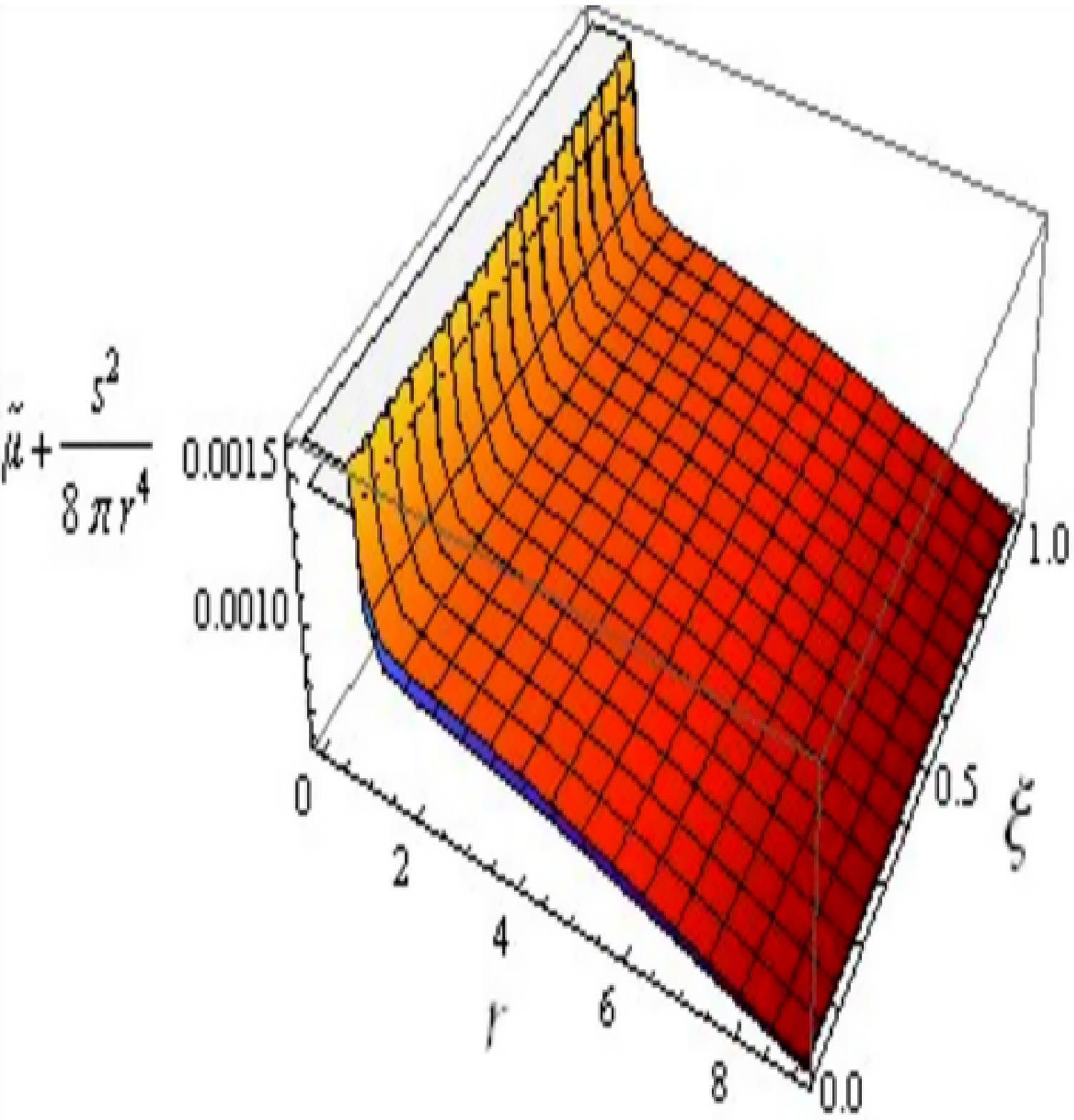,width=0.4\linewidth}\epsfig{file=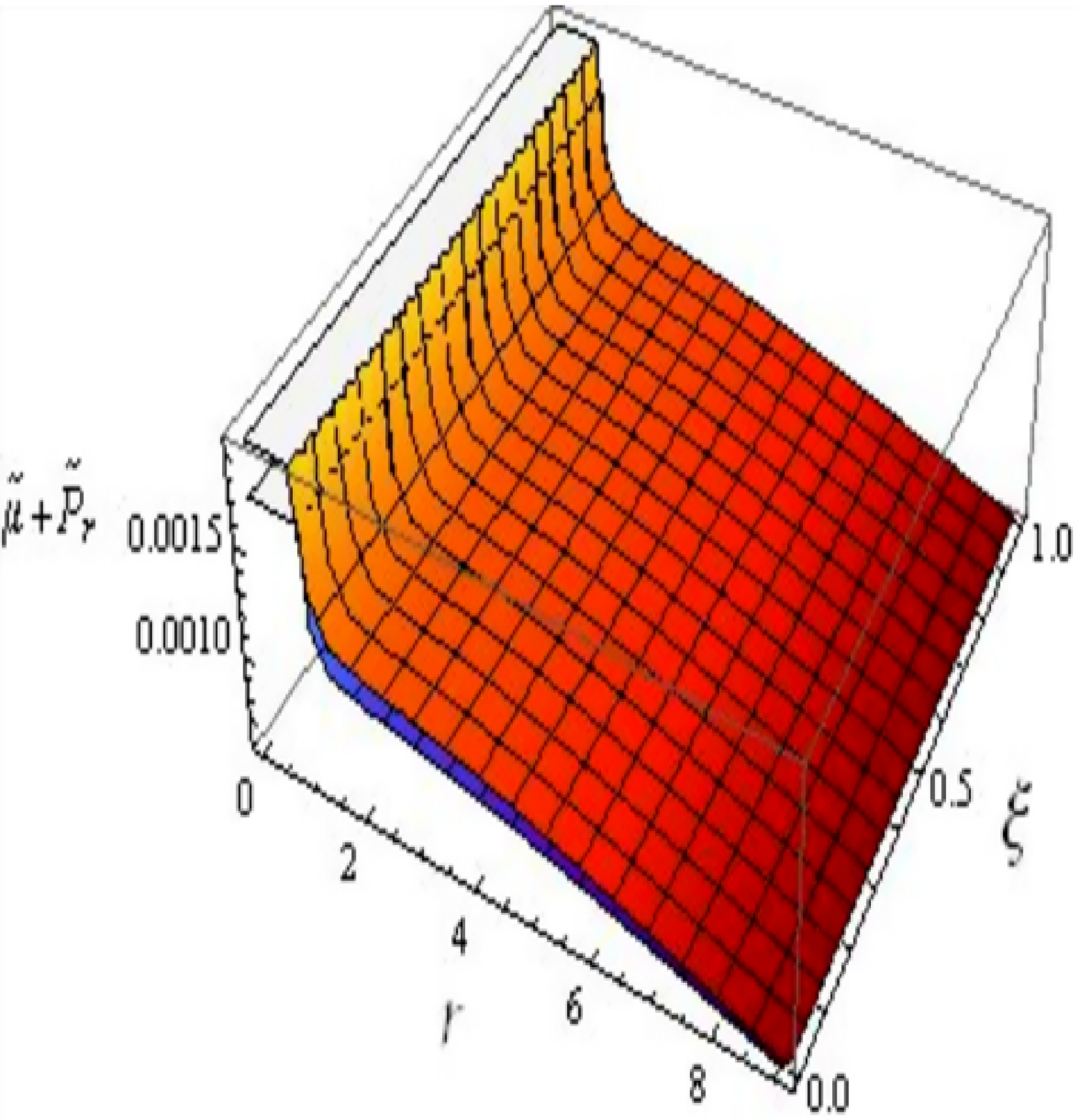,width=0.4\linewidth}
\epsfig{file=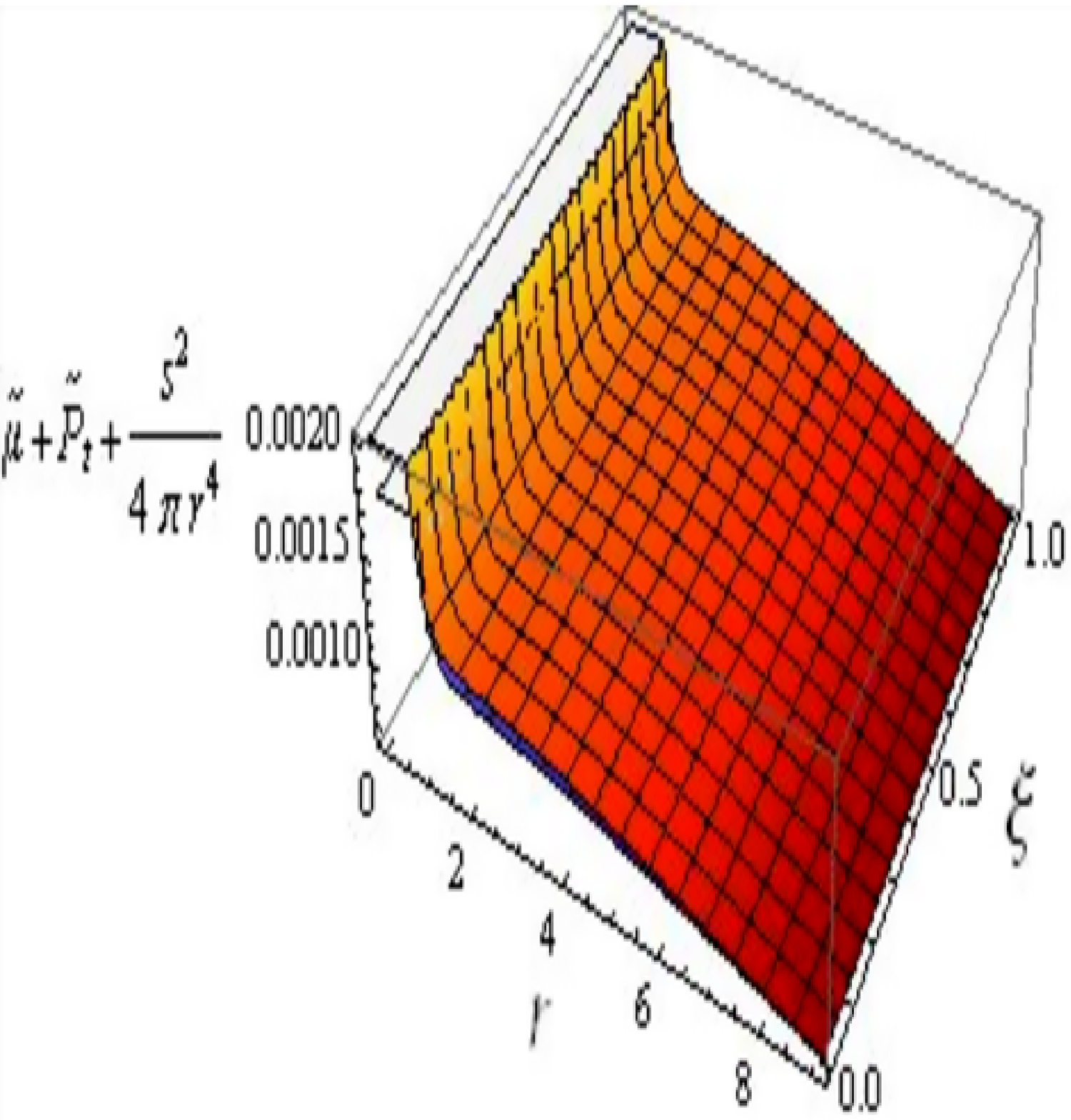,width=0.4\linewidth}\epsfig{file=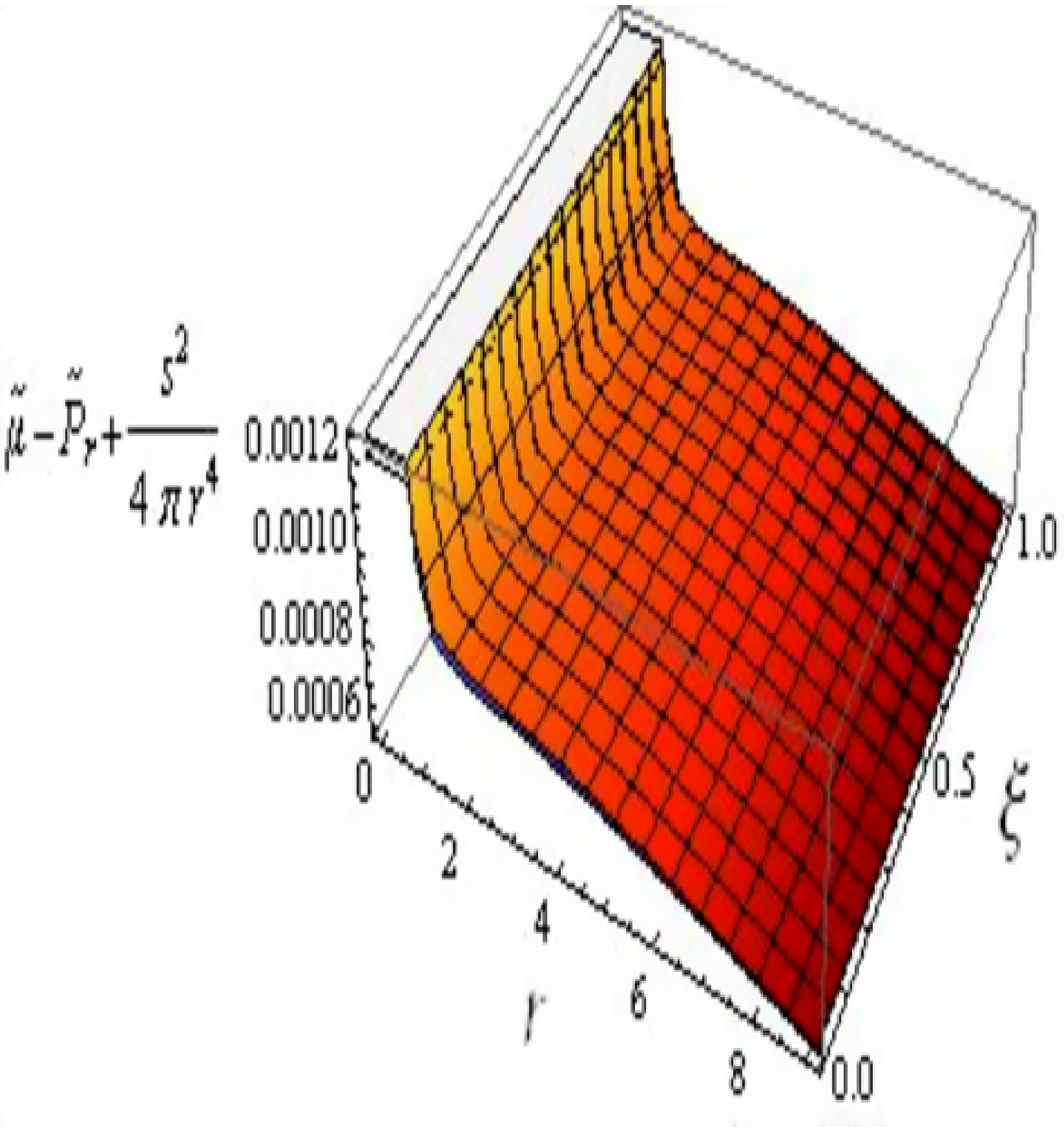,width=0.4\linewidth}
\epsfig{file=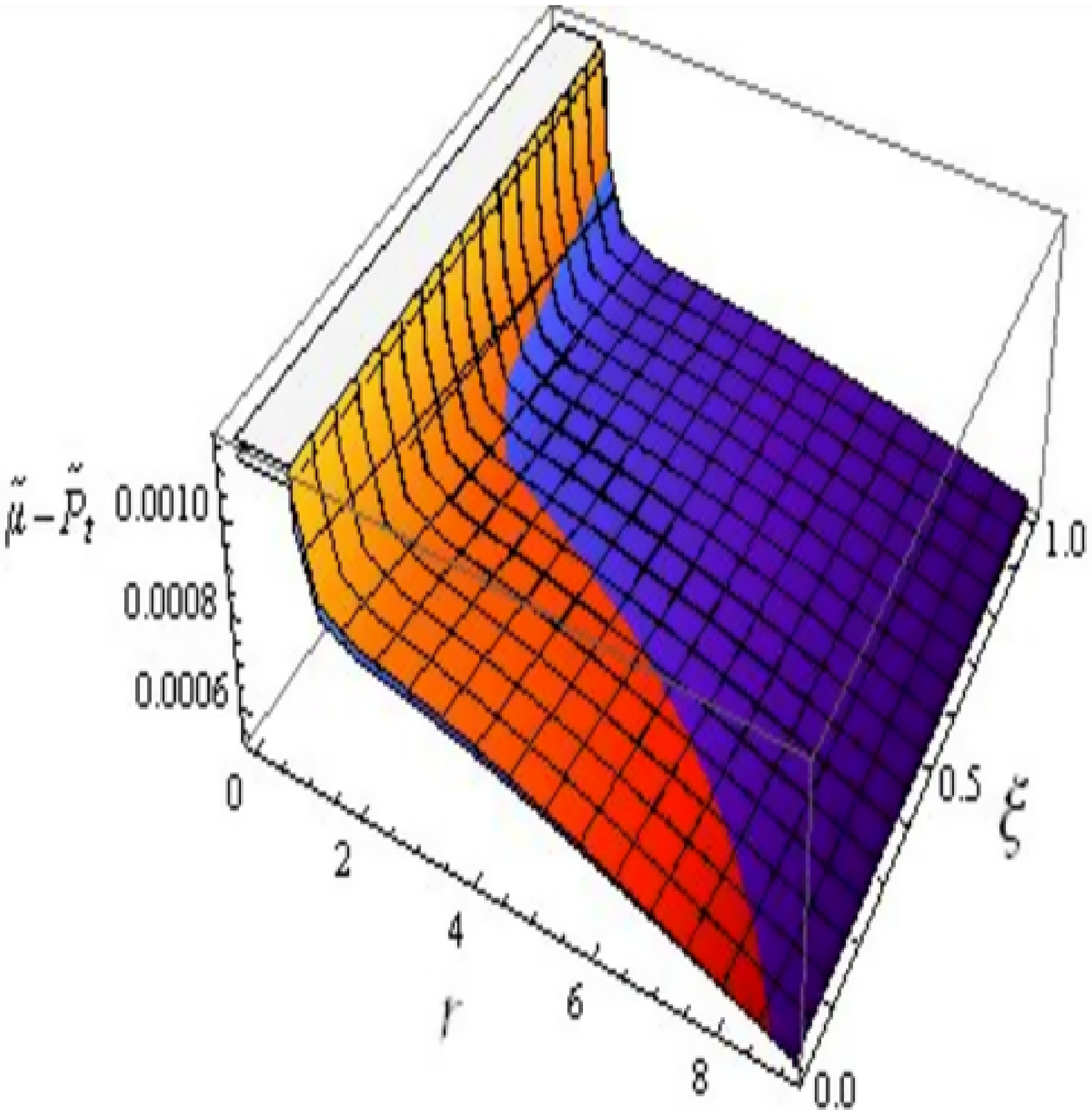,width=0.4\linewidth}\epsfig{file=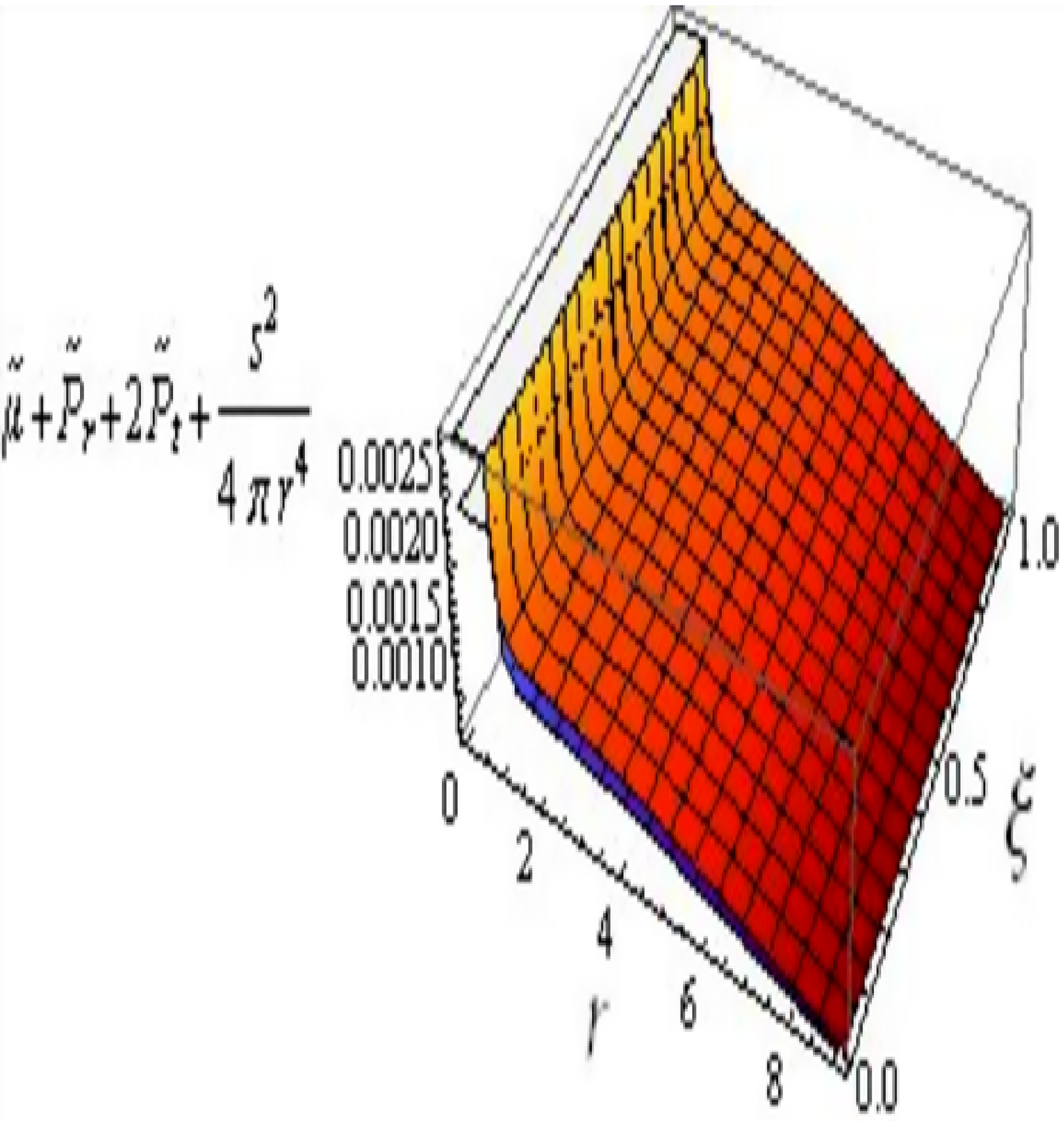,width=0.4\linewidth}
\caption{Plots of energy conditions versus $r$ and $\xi$ with
$\mathbb{S}_o=0.1$ (Orange), $\mathbb{S}_o=0.9$ (Blue) for the
solution I.}
\end{figure}
\begin{figure}\center
\epsfig{file=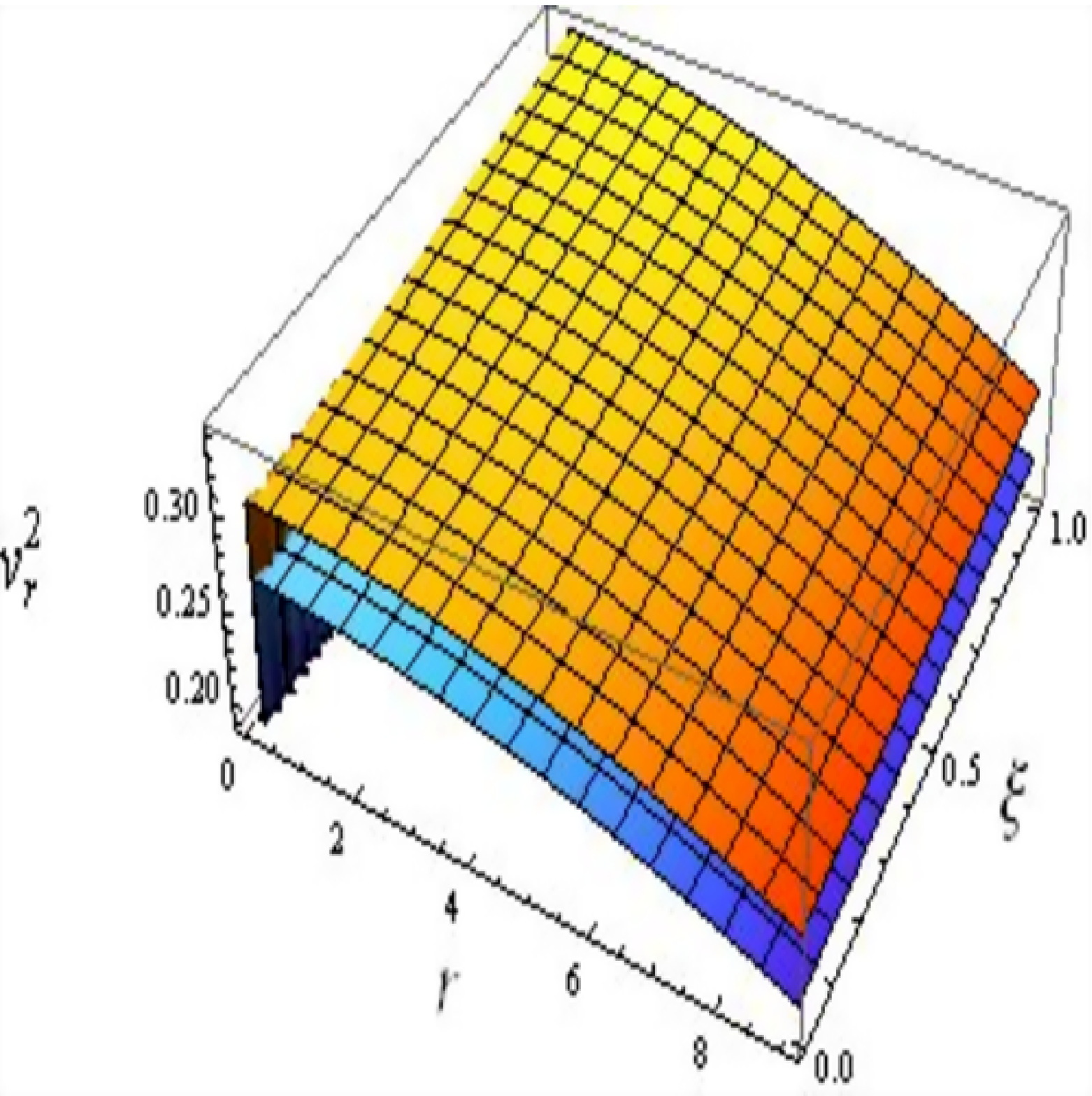,width=0.4\linewidth}\epsfig{file=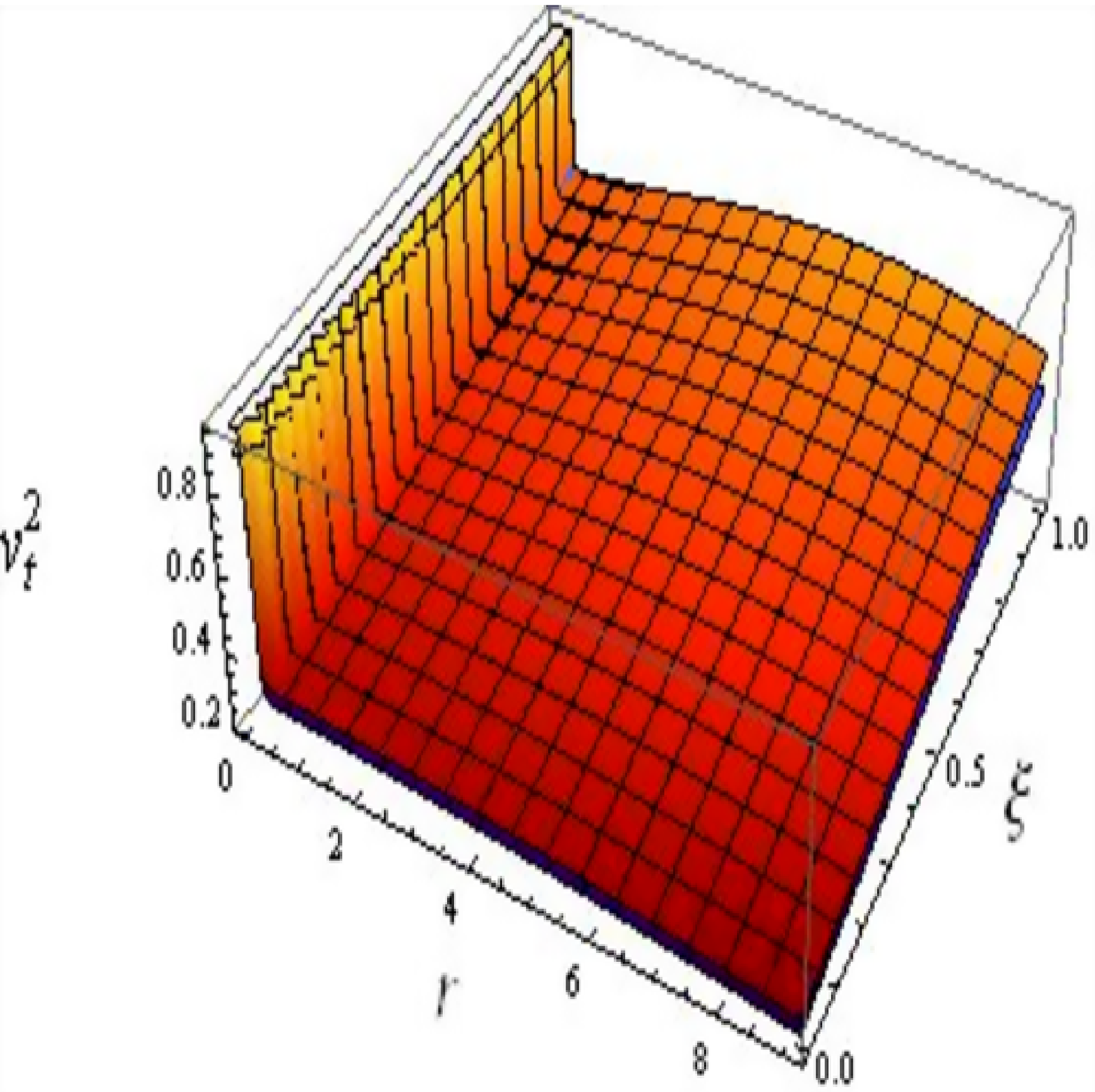,width=0.4\linewidth}
\epsfig{file=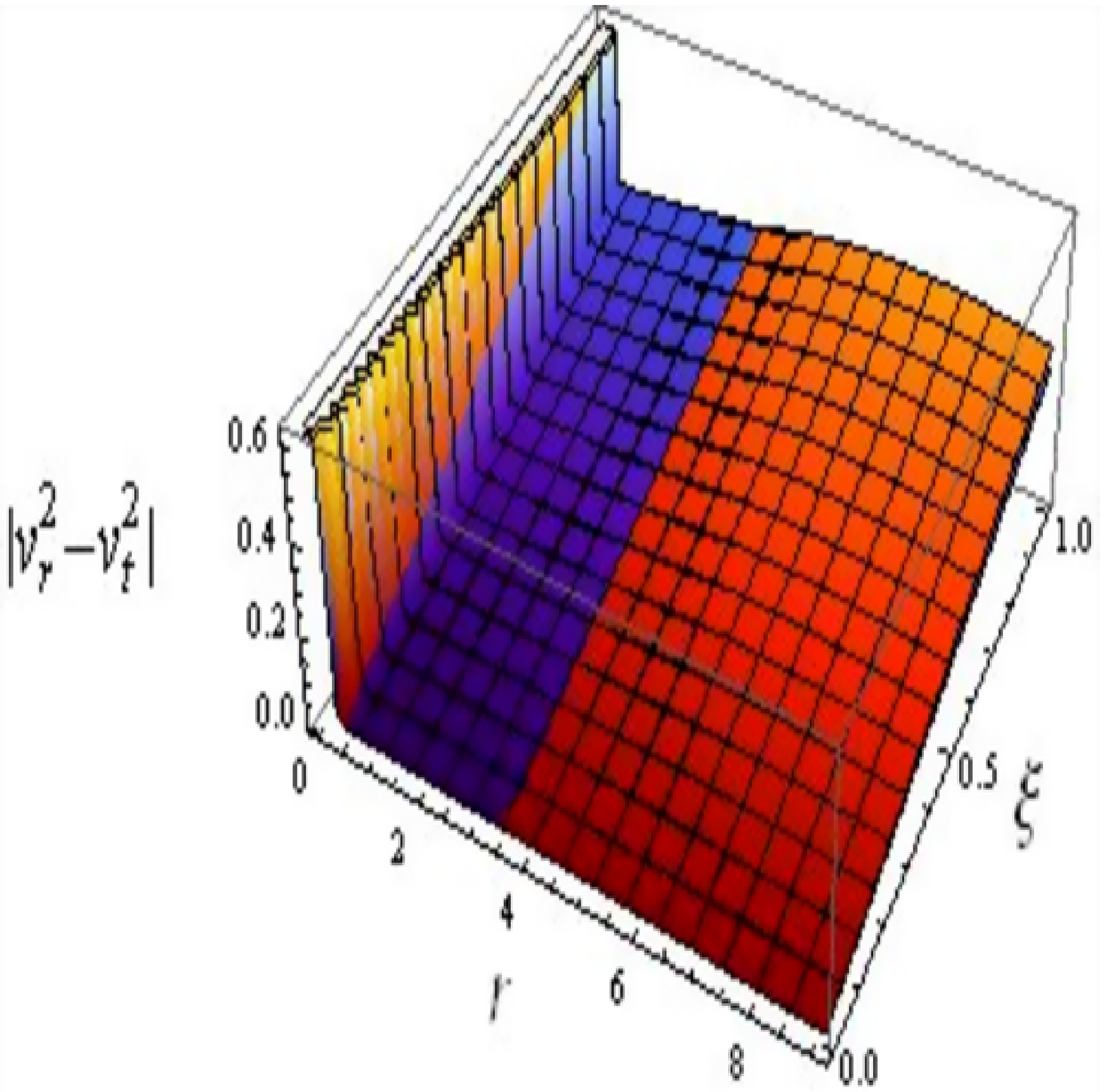,width=0.4\linewidth} \caption{Plots of
causality condition and Herrera cracking approach versus $r$ and
$\xi$ with $\mathbb{S}_o=0.1$ (Orange), $\mathbb{S}_o=0.9$ (Blue)
for the solution I.}
\end{figure}
\begin{figure}\center
\epsfig{file=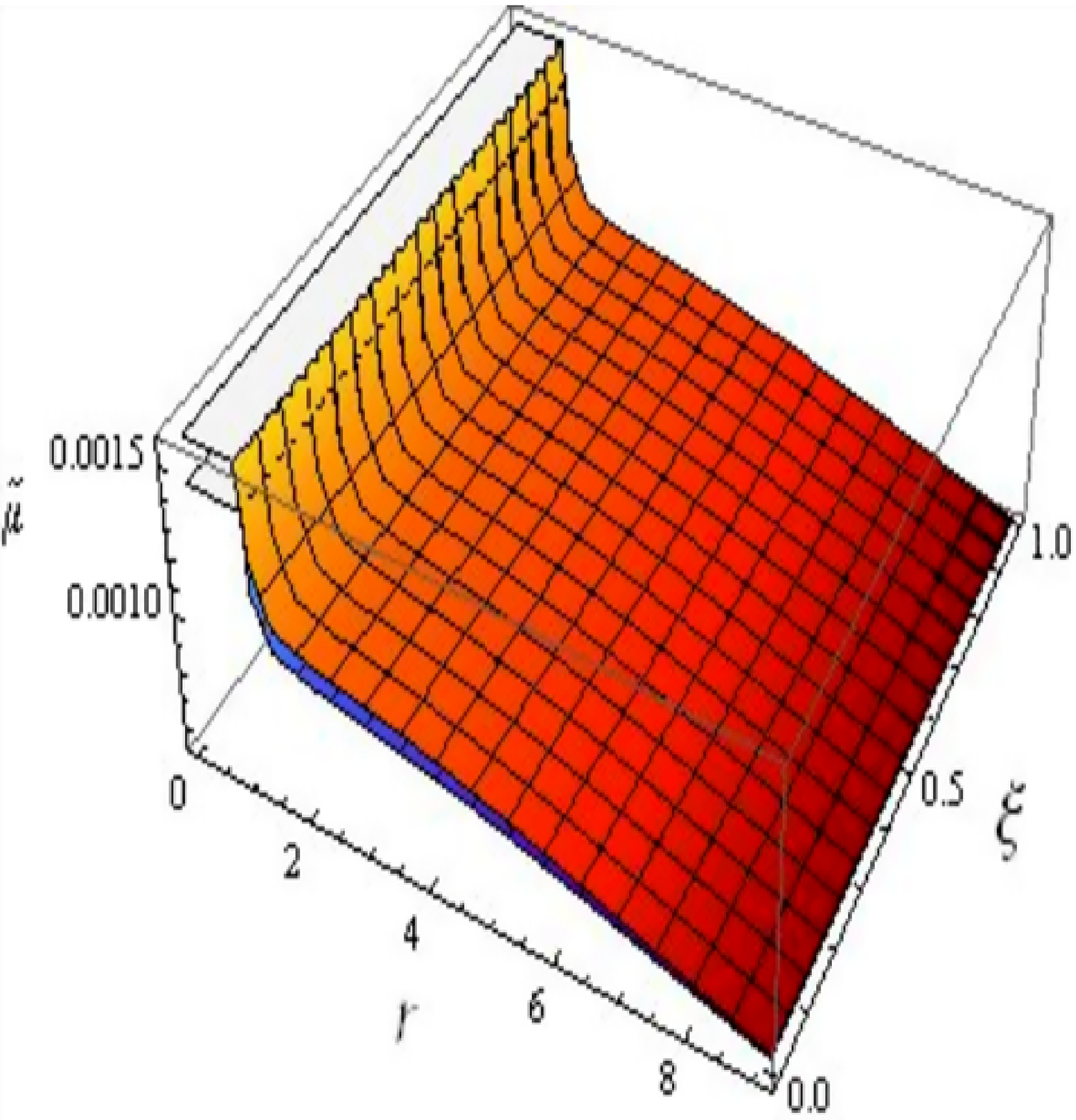,width=0.4\linewidth}\epsfig{file=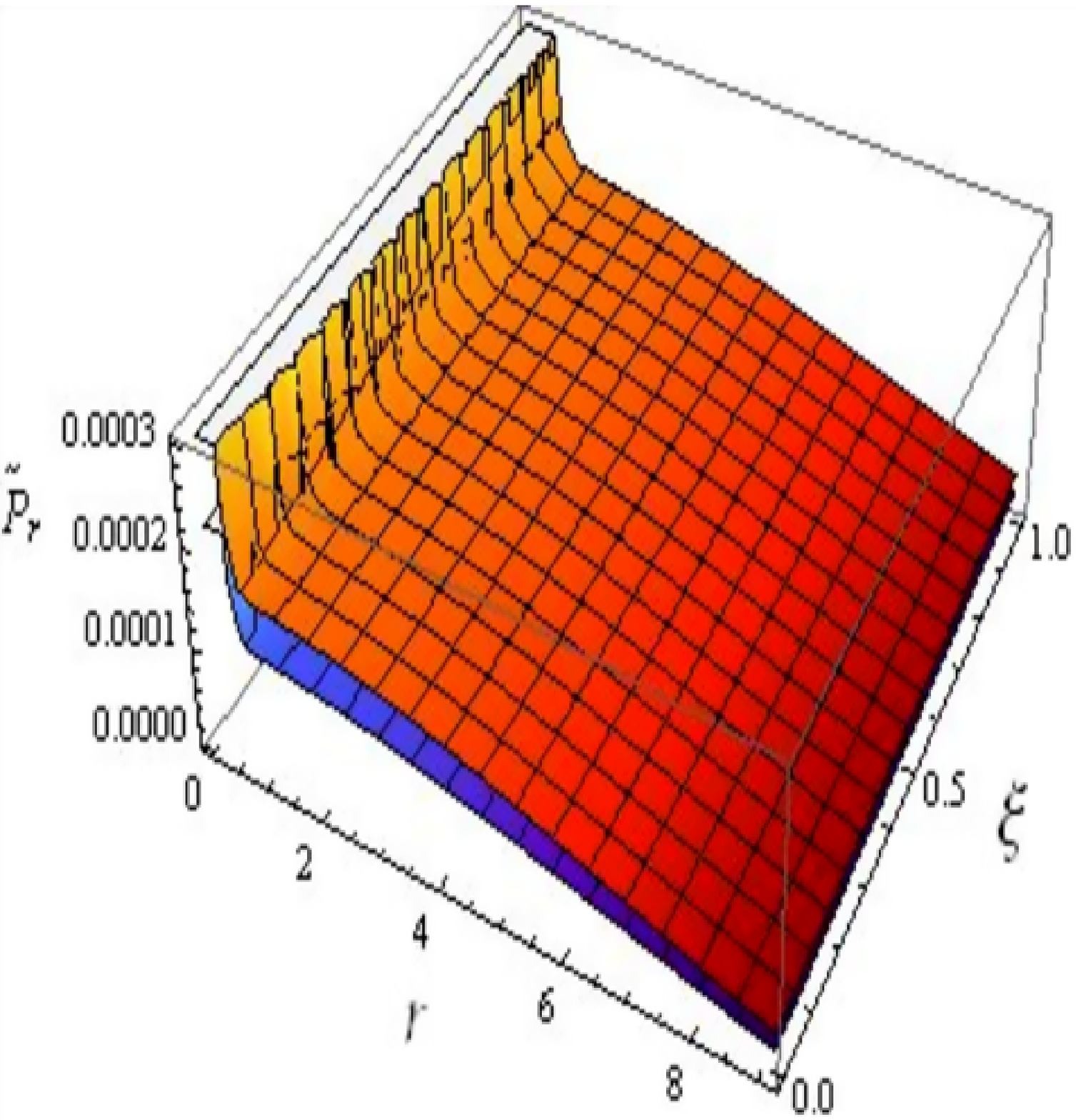,width=0.4\linewidth}
\epsfig{file=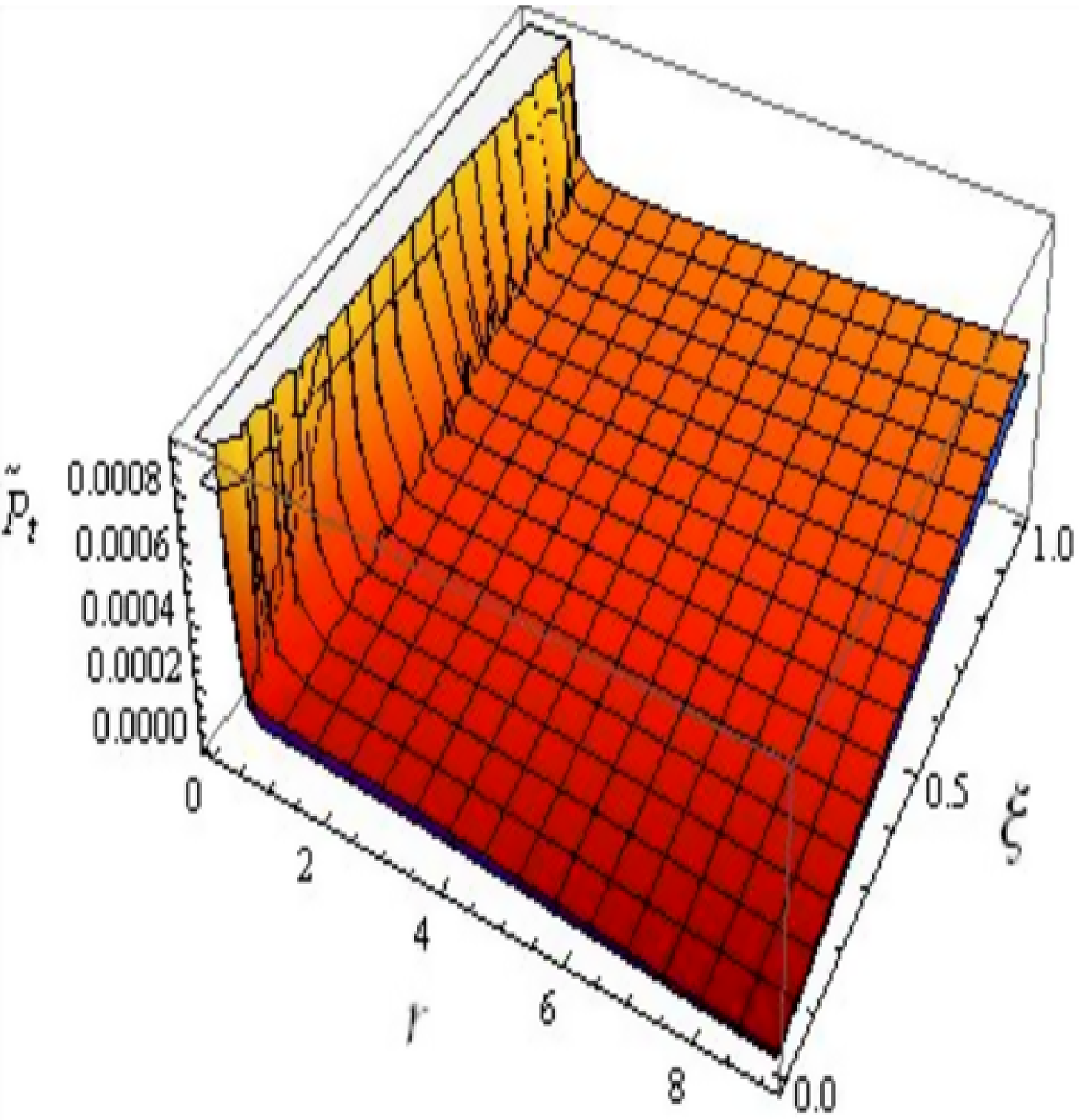,width=0.4\linewidth}\epsfig{file=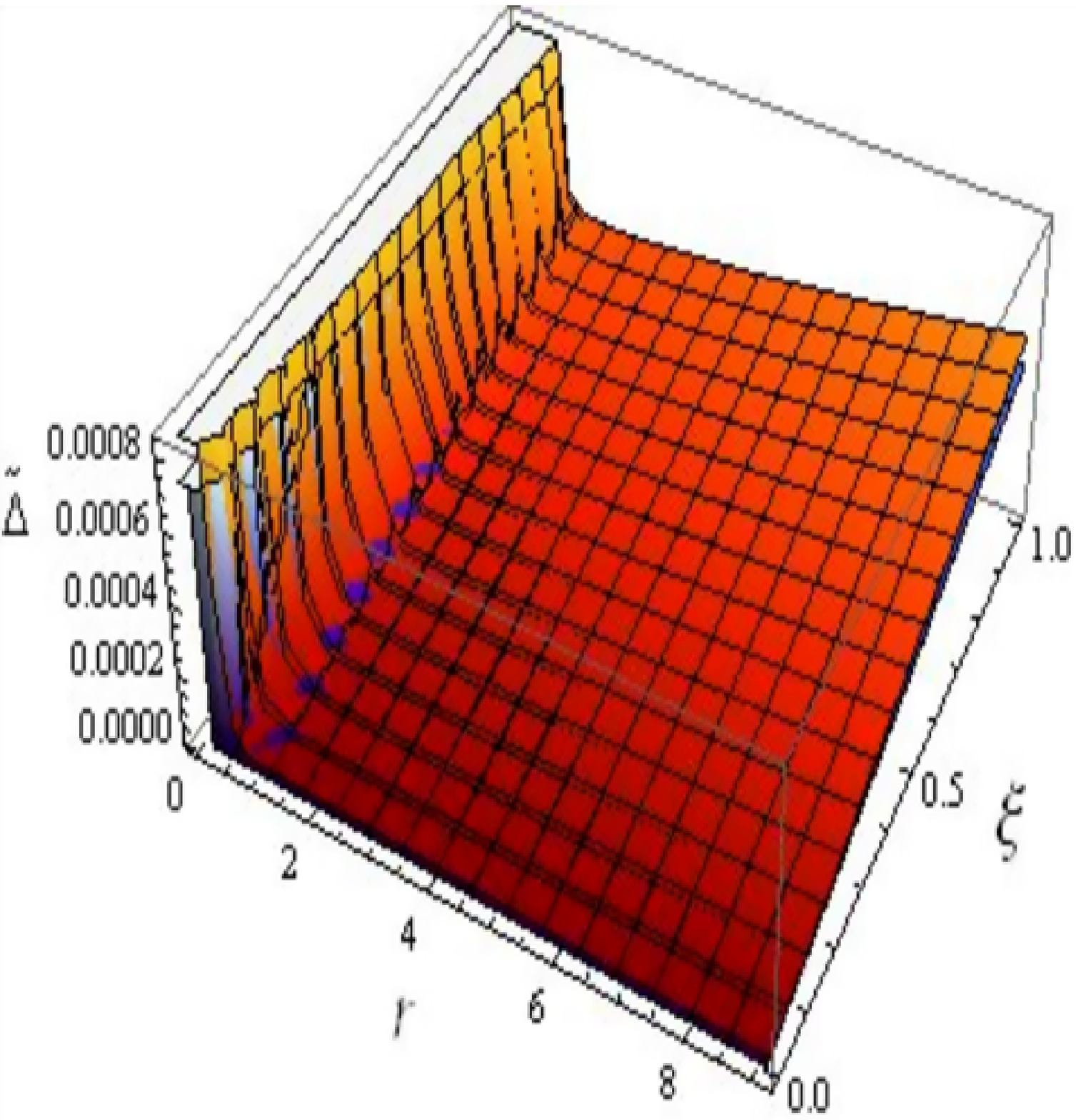,width=0.4\linewidth}
\caption{Plots of $\tilde{\mu},\tilde{P_r},\tilde{P_t}$ and
$\tilde{\Delta}$ versus $r$ and $\xi$ with $\mathbb{S}_o=0.1$
(Orange), $\mathbb{S}_o=0.9$ (Blue) for the solution II.}
\end{figure}
\begin{figure}\center
\epsfig{file=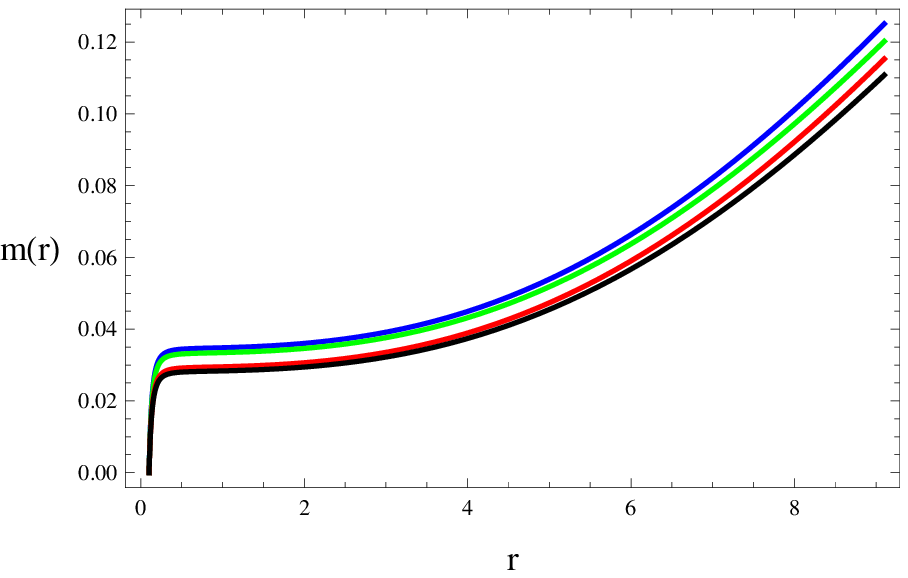,width=0.4\linewidth}\epsfig{file=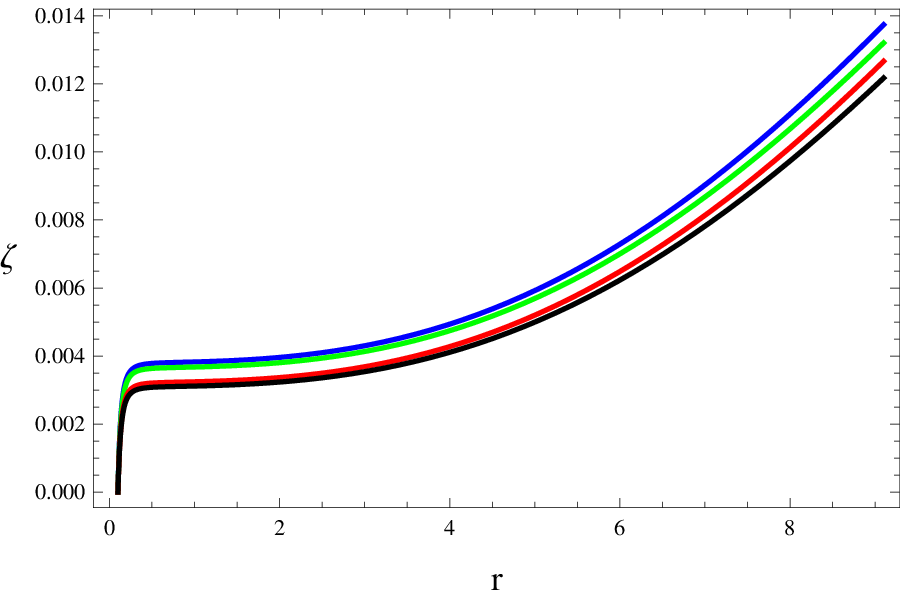,width=0.4\linewidth}
\epsfig{file=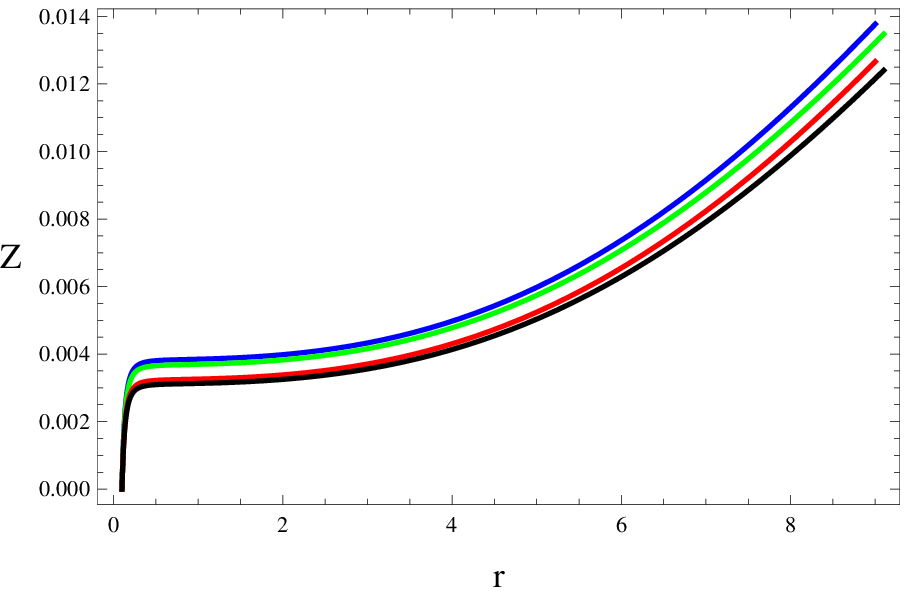,width=0.4\linewidth} \caption{Plots of mass,
compactness and redshift versus $r$ corresponding to
$\mathbb{S}_o=0.1$, $\xi=0.01$ (Blue), $\xi=0.9$ (Green) and
$\mathbb{S}_o=0.9$, $\xi=0.01$ (Red), $\xi=0.9$ (Black) for solution
II.}
\end{figure}
\begin{figure}\center
\epsfig{file=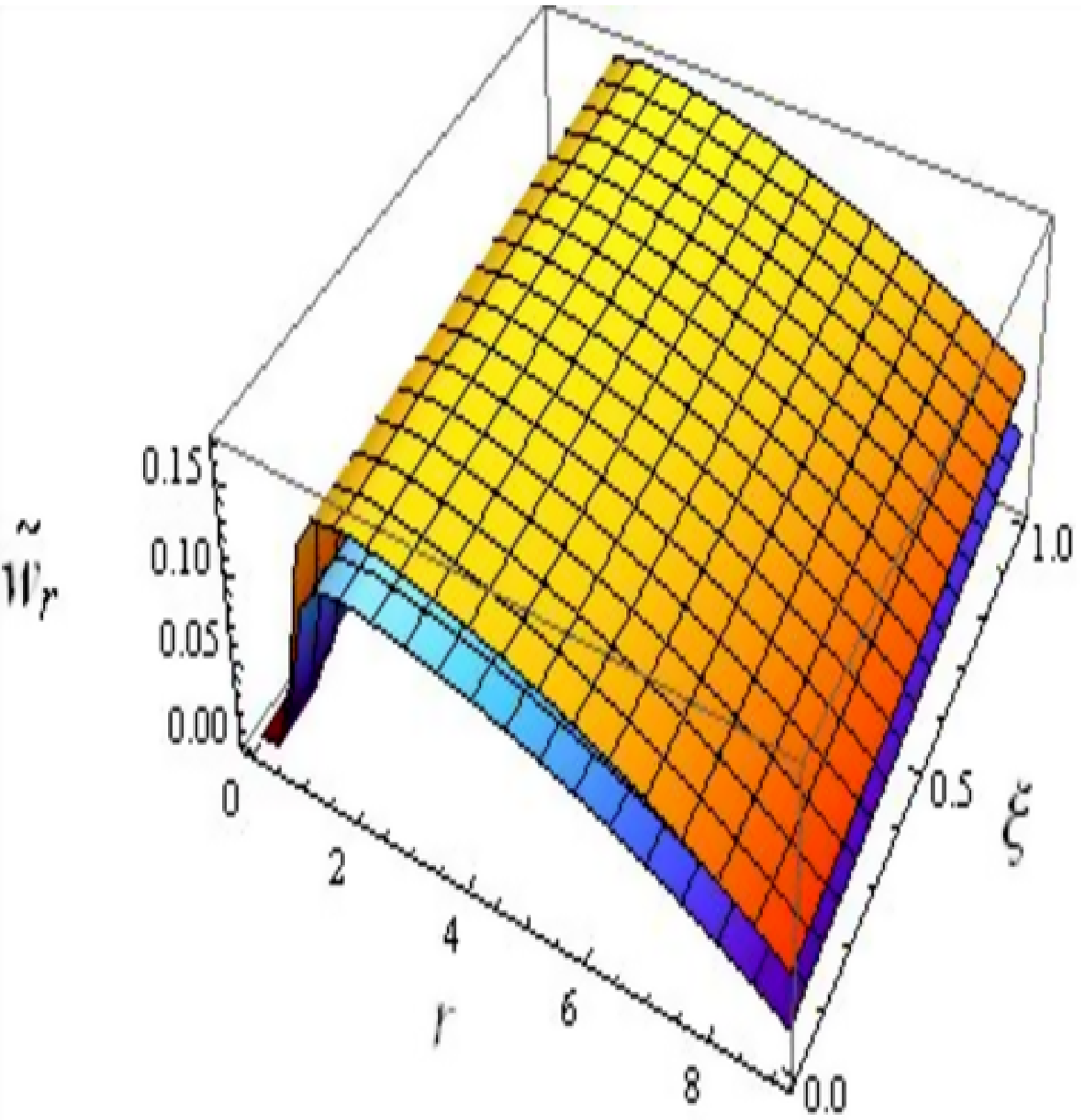,width=0.4\linewidth}
\epsfig{file=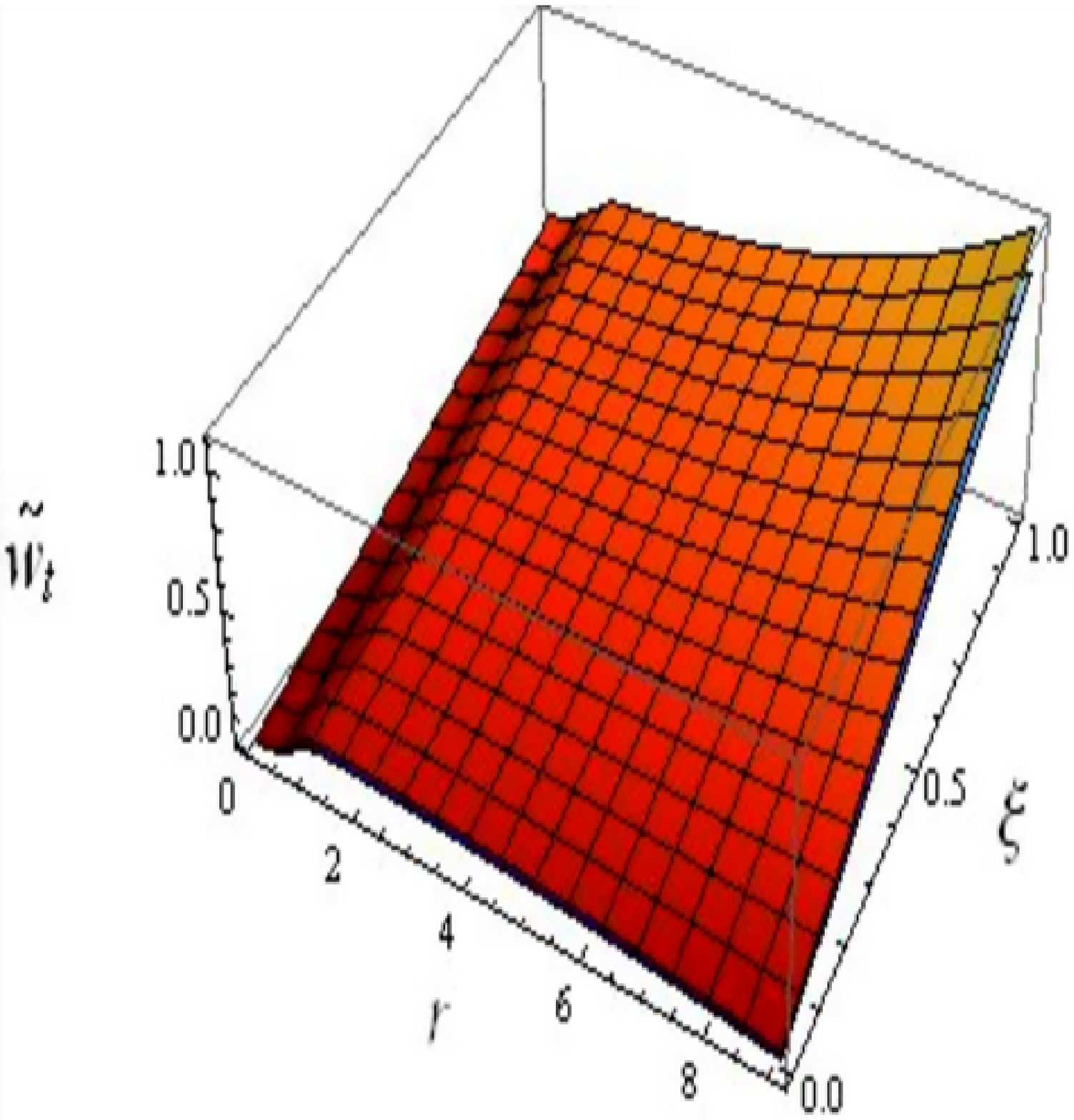,width=0.4\linewidth} \caption{Plots of EoS
parameters versus $r$ and $\xi$ with $\mathbb{S}_o=0.1$ (Orange),
$\mathbb{S}_o=0.9$ (Blue) for the solution II.}
\end{figure}
For mass, compactness and redshift parameter, we use distinct values
of the decoupling parameter along with charge. Figure \textbf{2}
(1st plot) illustrates that the mass function decreases for an
increasing charge. Figure \textbf{2} (2nd and 3rd plots) and Figure
\textbf{3} indicate that the compactness as well as redshift
parameters and the EoS parameters, respectively, lie within the
required limit. The energy conditions \eqref{59} for the solution I
are satisfied as shown in Figure \textbf{4} which indicates physical
viability of the solution. Figure \textbf{5} shows that the first
solution fulfills both the stability criterion, hence the solution I
is stable.
\begin{figure}\center
\epsfig{file=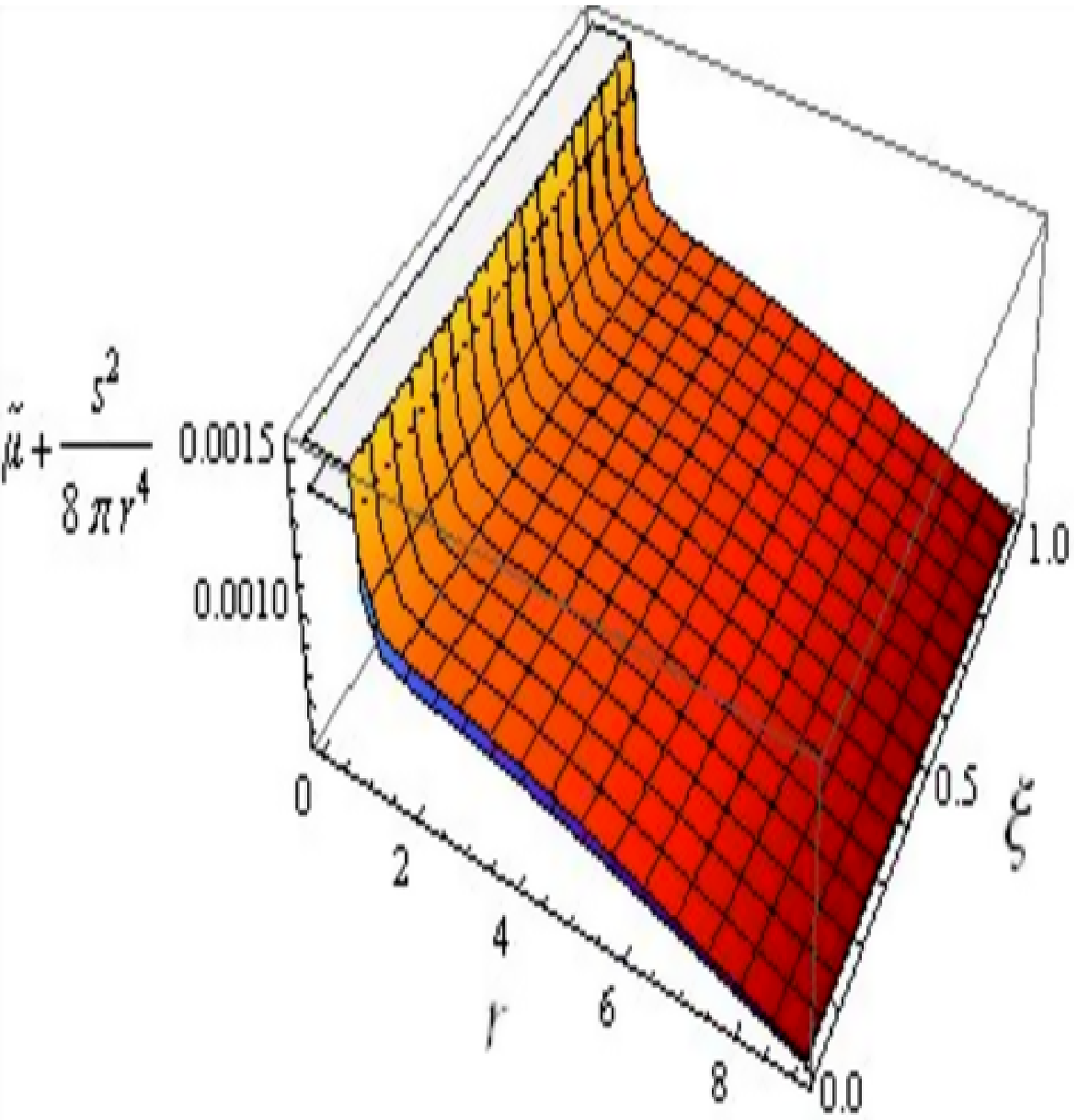,width=0.4\linewidth}\epsfig{file=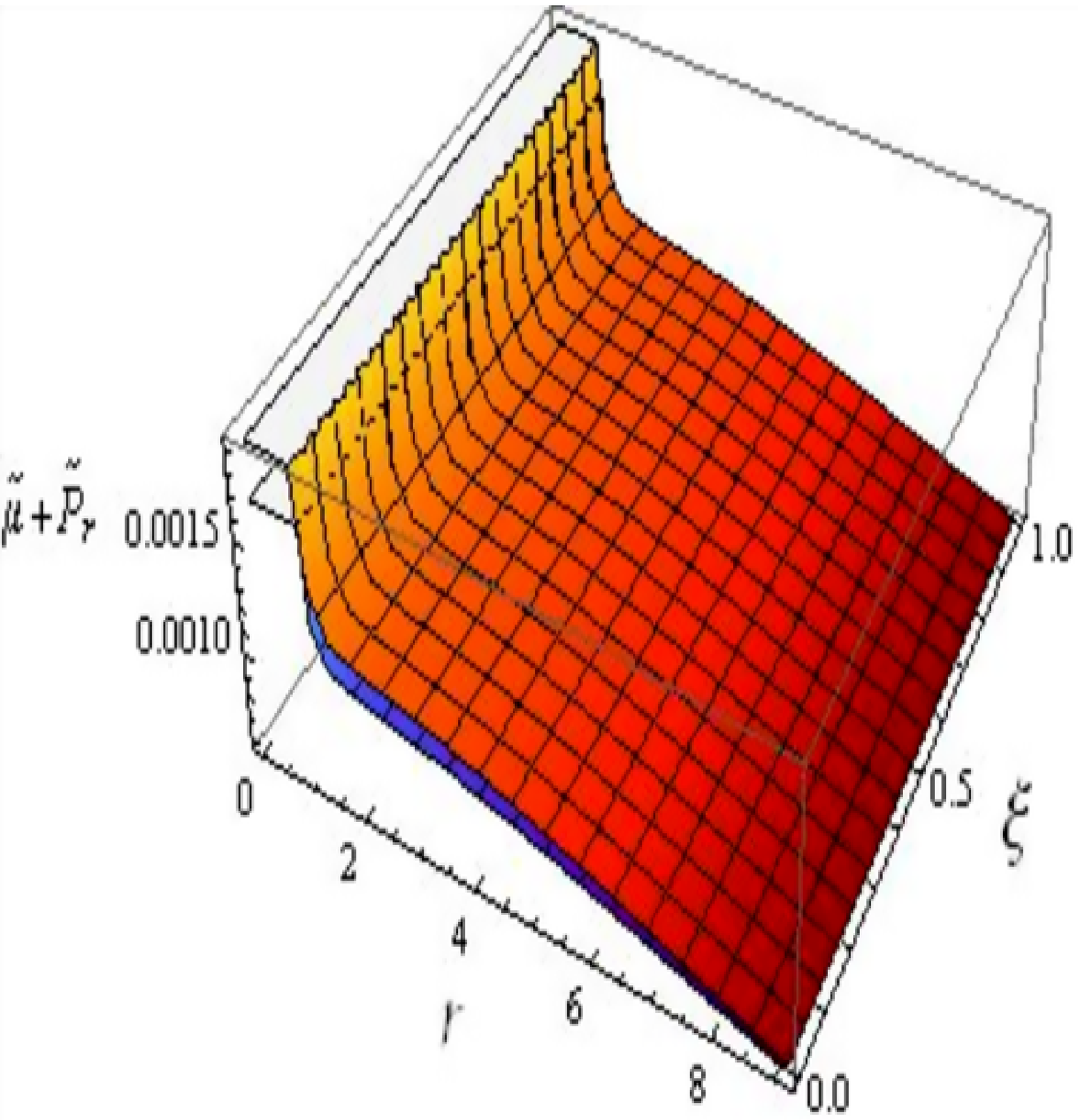,width=0.4\linewidth}
\epsfig{file=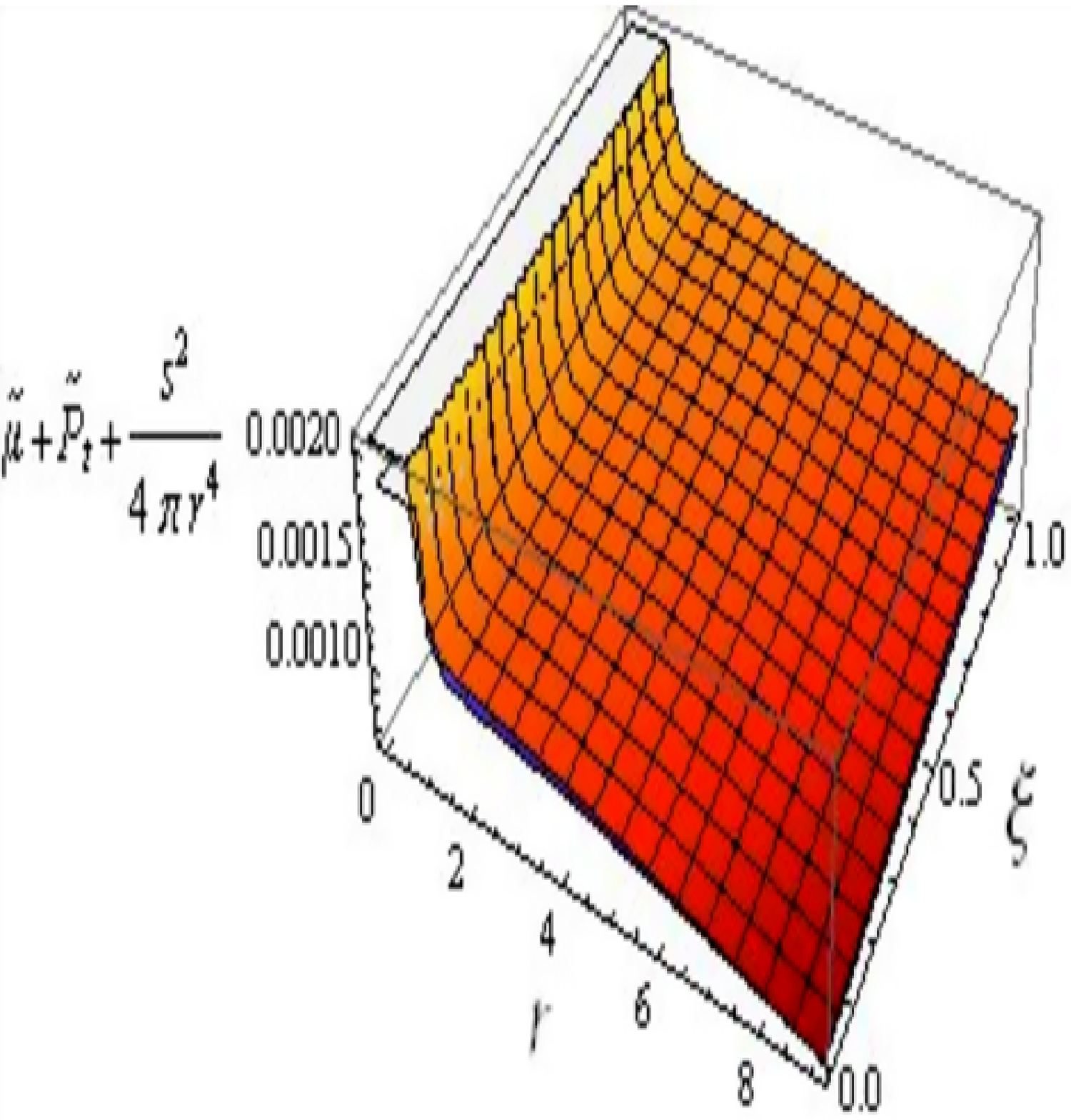,width=0.4\linewidth}\epsfig{file=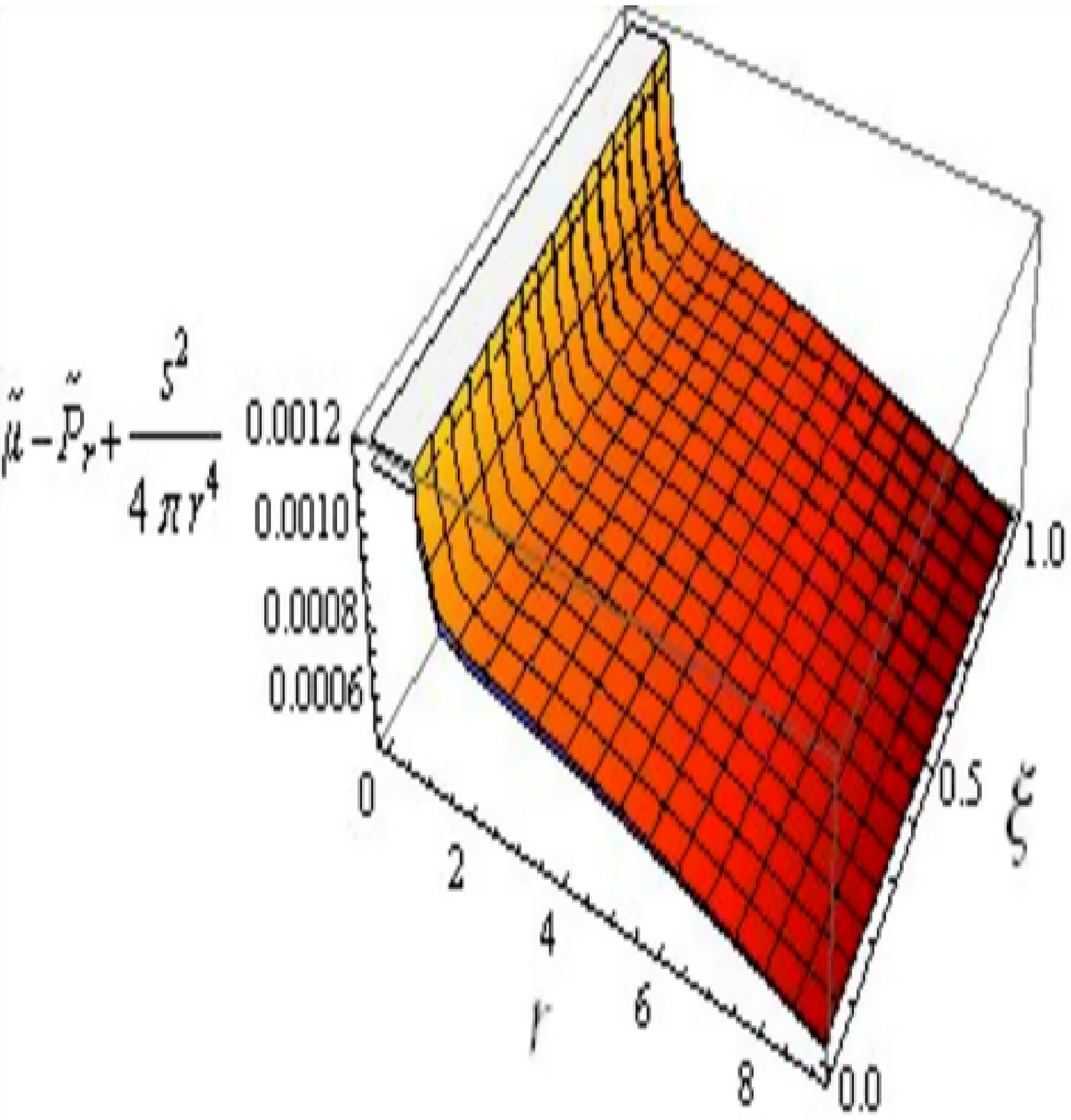,width=0.4\linewidth}
\epsfig{file=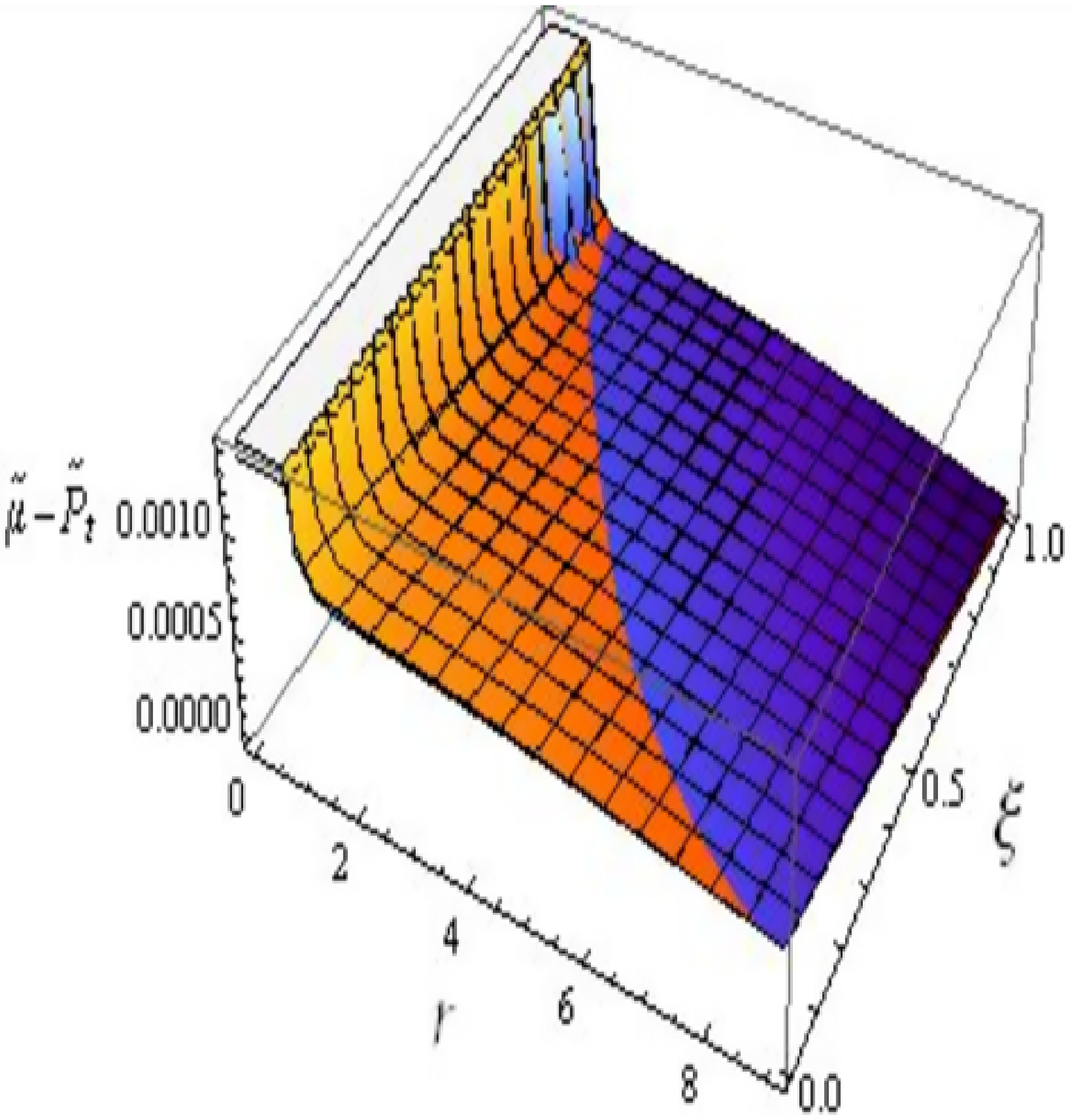,width=0.4\linewidth}\epsfig{file=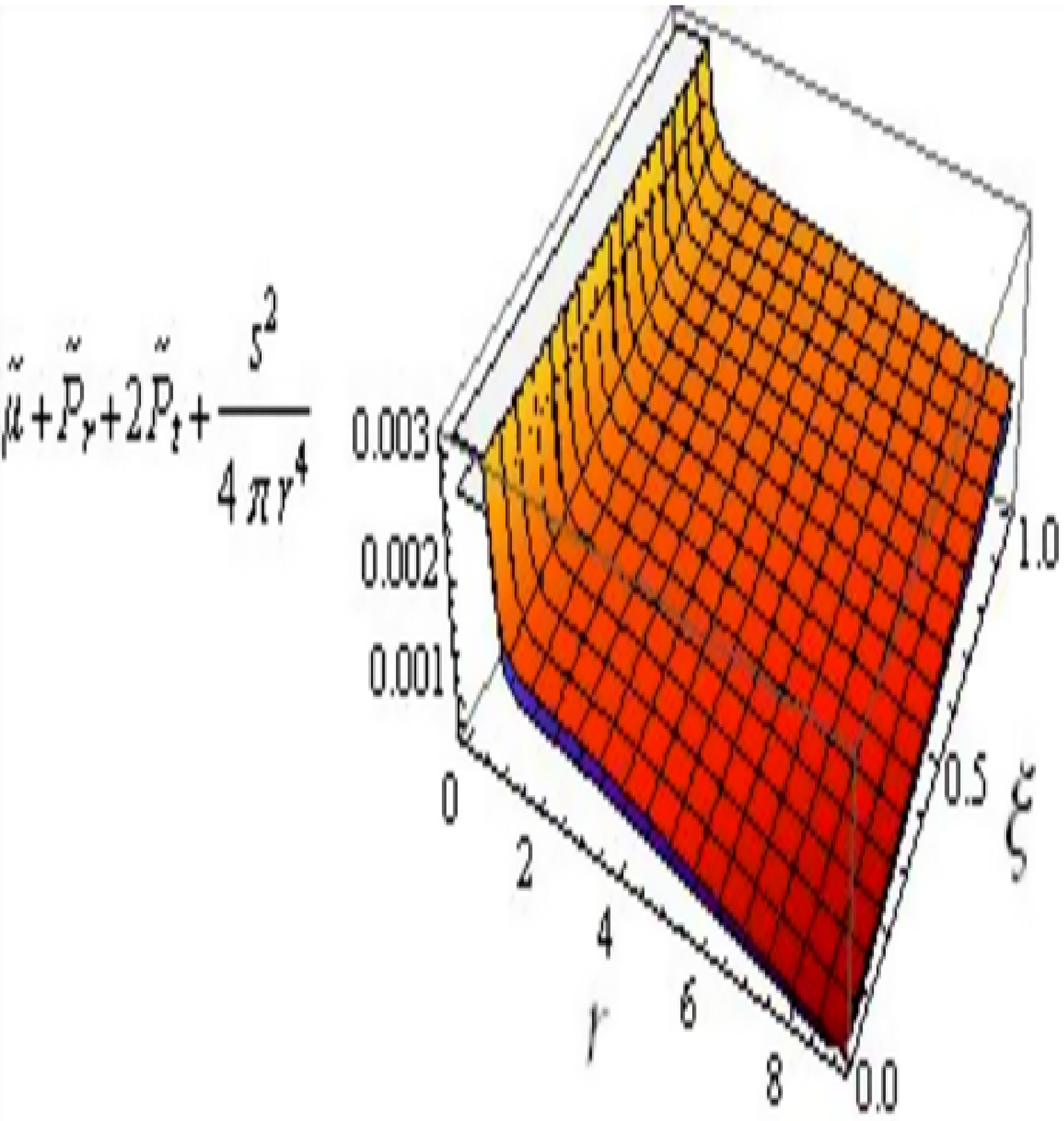,width=0.4\linewidth}
\caption{Plots of energy conditions versus $r$ and $\xi$ with
$\mathbb{S}_o=0.1$ (Orange), $\mathbb{S}_o=0.9$ (Blue) for the
solution II.}
\end{figure}
\begin{figure}\center
\epsfig{file=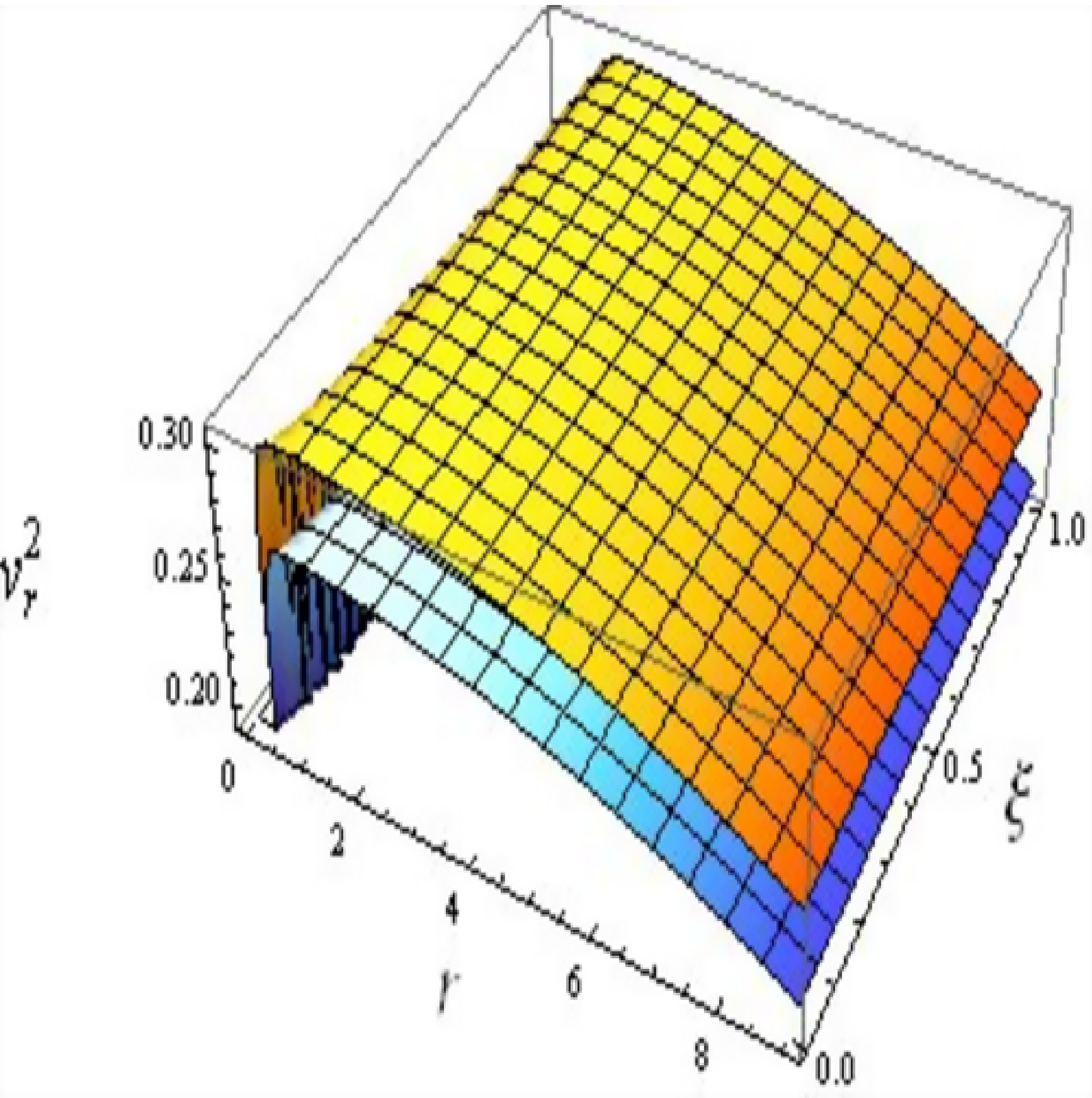,width=0.4\linewidth}\epsfig{file=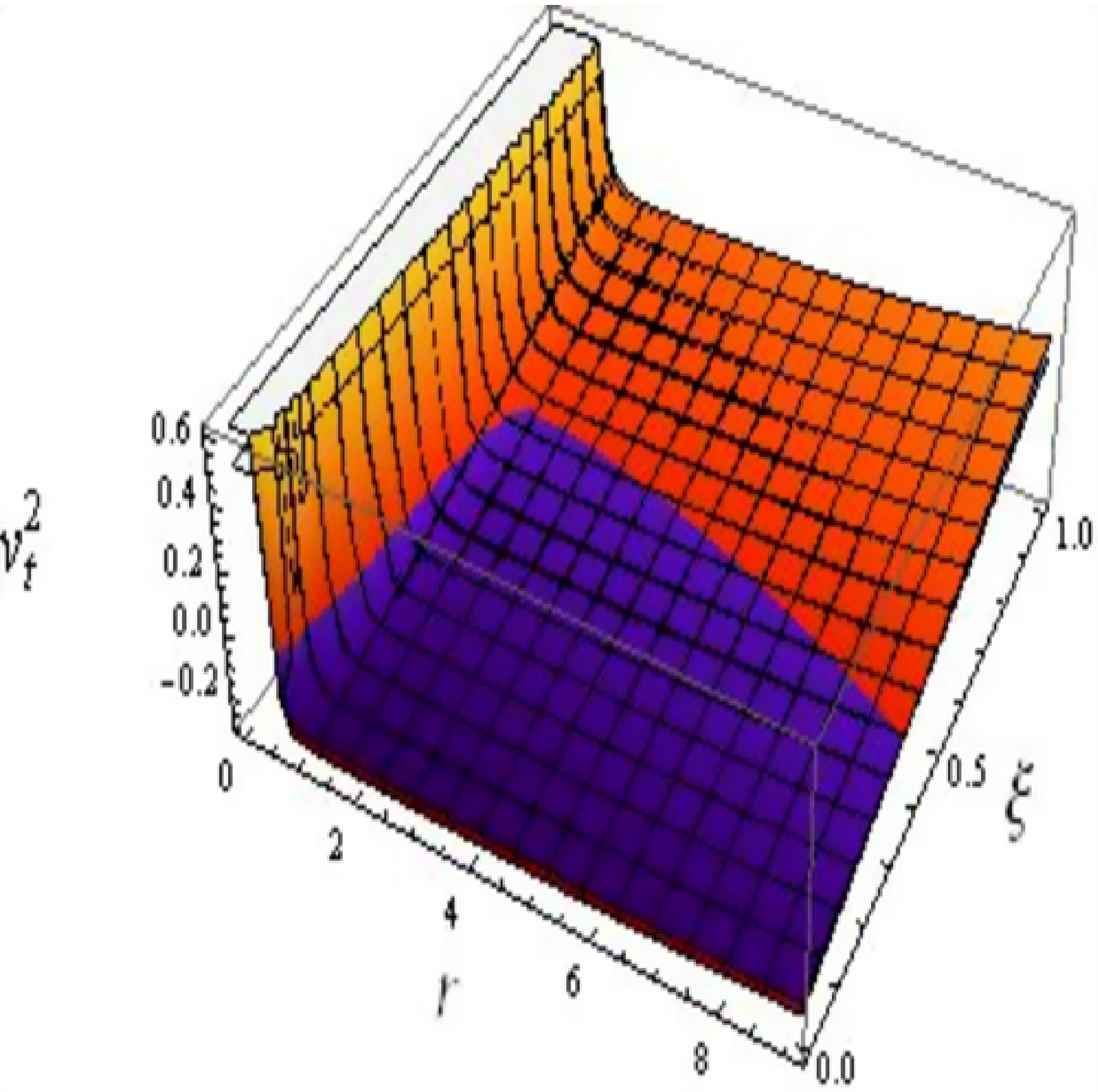,width=0.4\linewidth}
\epsfig{file=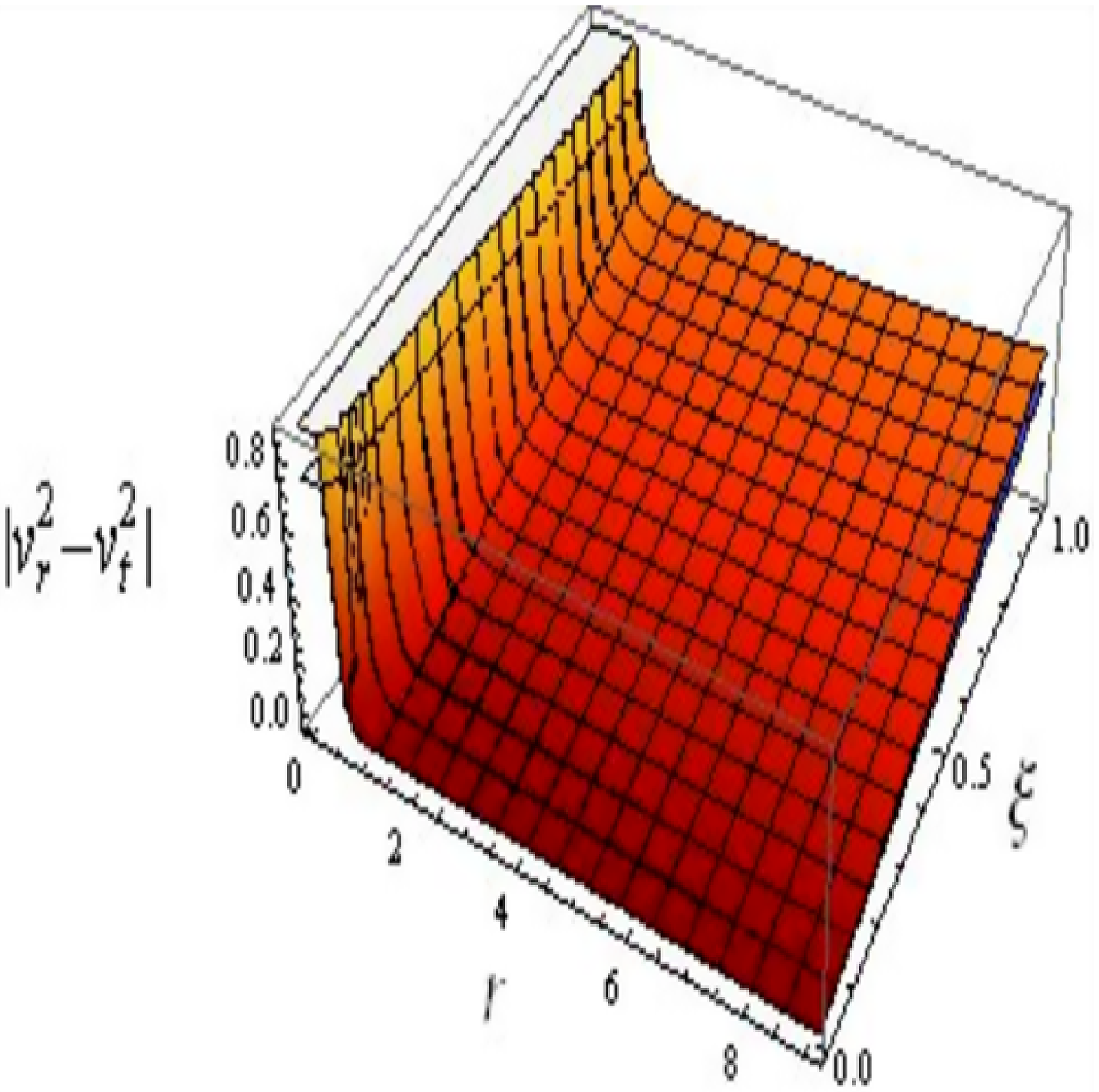,width=0.4\linewidth} \caption{Plots of
causality condition and Herrera cracking approach versus $r$ and
$\xi$ with $\mathbb{S}_o=0.1$ (Orange), $\mathbb{S}_o=0.9$ (Blue)
for the solution II.}
\end{figure}
The graphical analysis of the solution II is given for the same
values of the decoupling parameter and charge as for the solution I.
The constant terms $\mathcal{A}$ and $\mathcal{B}$ are taken from
Eqs.\eqref{36b} and \eqref{53a}. The effective energy density,
radial as well as tangential pressures of the second solution
exhibit the decreasing behavior with increase in $r$ (Figure
\textbf{6}) but for the larger charge, the effective energy density
of the star decreases. Figure \textbf{6} demonstrates that the
anisotropy increases with the decoupling parameter. Figure
\textbf{7} shows mass, compactness and redshift parameter meet their
desired ranges. Moreover, these factors decrease for higher $\xi$
and charge. Likewise, the EoS parameters for the second solution
meet the required limits (Figure \textbf{8}). Figure \textbf{9}
shows that all the energy conditions are satisfied ensuring the
viability of the system. The radial component of the sound velocity
obeys the stability criterion (Figure \textbf{10}) whereas the
tangential component shows unstable behavior at the core but becomes
stable for larger values of the decoupling parameter. However,
according to cracking approach, the second solution shows consistent
behavior.

\section{Conclusions}

Astrophysicists have made various efforts in developing physically
viable and stable solutions for compacts objects. A recently
developed technique termed as gravitational decoupling through MGD
has been found useful to understand the internal configurations of
the stellar structures through their anisotropic solutions. In this
paper, we have used this method to evaluate anisotropic solutions
for $f(G,T)= G^2+\beta T$ gravity model with charged spherically
symmetric geometry. The charged isotropic static sphere is filled
with the new source to induce anisotropy in it. This scheme
basically deforms the field equations into two sets representing the
isotropic and anisotropic systems. We have considered the
Krori-Barua ansatz to solve the system related to isotropic source
in which the values of unknowns are extracted through junction
conditions. For the second set \eqref{21}-\eqref{23}, we have
employed some extra constraints on $\omega_{\psi\chi}$ to evaluate
the unknown quantities. Finally, we have checked viable and stable
behavior of the obtained anisotropic solutions through graphical
analysis.

We have investigated physical properties of the effective state
variables $(\tilde{\mu}, \tilde{P_r}, \tilde{P_t})$, anisotropy
$(\tilde{\Delta})$ and energy conditions \eqref{59} with
$\beta=0.01$ to check physical viability of the developed solutions.
It is found that our both solutions satisfy the required limits for
mass, compactness, redshift as well as EoS parameters. For both
solutions, the self-gravitating object becomes less dense as charge
increases. Both solutions satisfy all the energy conditions, hence
they are physically viable. We have found that both the stability
criteria are satisfied for the first solution, while the second
solution is shown consistent only with the cracking approach.
Solution II shows stable behavior for higher values of the
decoupling parameter with causality condition. It is mentioned here
that two anisotropic solutions were found in GR \cite{15}, one was
viable but unstable, while the other was neither viable nor stable.
However, in $f(G)$ gravity all the developed solutions were
well-behaved \cite{16}. Here we have found that this theory provides
more better solutions than GR but similar to $f(G)$ gravity. It is
worth mentioning here that our results reduce to GR for $f(G,T)=0$.

\section*{Appendix A}
\renewcommand{\theequation}{A\arabic{equation}}
\setcounter{equation}{0}

The values of Gauss-Bonnet invariant and its derivatives are given
as
\begin{eqnarray}\label{63a}
G&=&\frac{1}{r^2}\bigg[2 e^{-2 \lambda } \left(\left(e^{\lambda
}-3\right) \phi ' \lambda '-\left(e^{\lambda }-1\right) \left(2
\lambda ''+\phi '^2\right)\right)\bigg],\\\nonumber
G'&=&\frac{-1}{r^3}\bigg[2 e^{-2 \lambda } \bigg(-r \phi '
\left(\left(e^{\lambda }-3\right) \lambda ''-2 \left(e^{\lambda
}-1\right) \phi ''\right)+r \left(e^{\lambda }-6\right) \phi '
\lambda '^2\\\nonumber &+&\lambda' \left(-r \left(3 e^{\lambda
}-7\right) \phi ''+r \left(-\left(e^{\lambda }-2\right)\right) \phi
'^2+2 \left(e^{\lambda }-3\right) \phi 'r\right)-2
\big(e^{\lambda}\\\label{64} &-&1\big) \phi '^2-2
\left(e^{\lambda(}-1\right) \left(2 \phi ''-r \phi
^{(3)}\right)\bigg)\bigg],
\end{eqnarray}
\begin{eqnarray}\nonumber
G''&=&\frac{1}{r^4}\bigg[2 e^{-2 \lambda } \bigg(\phi '^2 \big(r^2
\big(e^{\lambda }-2\big) \lambda ''-6 e^{\lambda }+6\big)-2
\bigg(\phi '' \big(6 \big(e^{\lambda }-1\big)-r^2 \big(2 e^{\lambda
}\\\nonumber&-&5\big) \lambda ''\big)+r^2 \big(e^{\lambda }-1\big)
\phi ''^2+r \big(r \phi ^{(4)}-4 \phi ^{(3)}\big) \big(e^{\lambda
}-1\big)\bigg)+r^2 \big(e^{\lambda }-12\big)\\\nonumber&\times& \phi
' \lambda '^3+\lambda ' \bigg(\phi ' \big(-3 r^2 \big(e^{\lambda
}-6\big) \lambda ''+4 r^2 \big(e^{\lambda }-2\big) \phi ''+6
\big(e^{\lambda }-3\big)\big)-4\\\nonumber&\times& r \big(e^{\lambda
}-2\big) \phi '^2+r \big(r \phi ^{(3)} \big(5 e^{\lambda }-11\big)-4
\big(3 e^{\lambda }-7\big) \phi ''\big)\bigg)-r \lambda '^2 \bigg(4
r \big(e^{\lambda }\\\nonumber&-5&\big) \phi ''+r \big(e^{\lambda
}-4\big) \phi '^2-4 \big(e^{\lambda }-6\big) \phi '\bigg)+r \phi '
\bigg(r \big(\big(e^{\lambda }-3\big) \lambda ^{(3)}-2 \phi
^{(3)}\\\label{65}&\times& \big(e^{\lambda }-1\big)\big)-4
\big(e^{\lambda }-3\big) \lambda ''+8 \big(e^{\lambda }-1\big) \phi
''\bigg)\bigg)\bigg].
\end{eqnarray}
The correction terms corresponding to $f(G,T)$ gravity become
\begin{eqnarray}\nonumber
T^{0(D)}_{0}&=&\frac{1}{8\pi}\bigg[-\frac{1}{2}\zeta
G^2+\bigg(\frac{4 e^{-2 \lambda} \phi ''}{r^2}-\frac{4 e^{-\lambda }
\phi ''}{r^2}-\frac{2 e^{-\lambda} \phi'^2}{r^2}+\frac{2 e^{-\lambda
} \phi ' \lambda '}{r^2}\\\nonumber &+&\frac{2 e^{-2 \lambda } \phi
'^2}{r^2}-\frac{6 e^{-2 \lambda } \phi ' \lambda
'}{r^2}\bigg)G+\bigg (\frac{12 e^{-2 \lambda } \lambda
'}{r^2}-\frac{4 e^{-\lambda }\lambda '}{r^2}\bigg)G'\\\label{61}
&&\bigg (-\frac{8 e^{-2 \lambda}}{r^2}+\frac{8 e^{-\lambda }
}{r^2}\bigg)G''\bigg],
\end{eqnarray}
\begin{eqnarray}\nonumber
T^{1(D)}_{1}&=&\frac{1}{8 \pi} \bigg[\frac{1}{2} \zeta G^2+
\bigg(-\frac{4 e^{-2 \lambda } \phi ''}{r^2}+\frac{4 e^{-\lambda }
\phi ''}{r^2}+\frac{6 e^{-2 \lambda } \phi ' \lambda '}{r^2}-\frac{2
e^{-\lambda } \phi ' \lambda '}{r^2}\\\nonumber &-&\frac{2 e^{-2
\lambda } \phi '^2}{r^2}+\frac{2 e^{-\lambda } \phi
'^2}{r^2}\bigg)G+ \bigg(\frac{12 e^{-2\lambda} \phi '}{r^2}-\frac{4
e^{-\lambda } \phi '}{r^2}\bigg)G'\bigg],\\\label{62}
\end{eqnarray}
\begin{eqnarray}\nonumber
T^{2(D)}_{2}&=&\frac{1}{8 \pi}\bigg[\frac{1}{2} G^2+ \bigg(-\frac{4
e^{-2\lambda} \phi ''}{r^2}+\frac{4 e^{-\lambda } \phi
''}{r^2}+\frac{2 e^{-\lambda } \phi '^2}{r^2}-\frac{2 e^{-2 \lambda
} \phi '^2}{r^2}\\\nonumber &-&\frac{2 e^{-\lambda } \phi ' \lambda
'}{r^2}+\frac{6 e^{-2 \lambda } \phi ' \lambda '}{r^2}\bigg)G+
\bigg(-\frac{6 e^{-2 \lambda } \phi ' \lambda '}{r}+\frac{4 e^{-2
\lambda } \phi ''}{r}\\\label{63} &+&\frac{2 e^{-2 \lambda } \phi
'^2}{r}\bigg)G'+ \frac{4 e^{-2 \lambda } \phi '}{r}G''\bigg].
\end{eqnarray}
The term $\Omega$ comes out to be
\begin{equation}\label{66}
\Omega=\frac{\beta}{8\pi-\beta}\bigg[-\frac{(-\mu+3P+\xi
\omega)'}{2}-\xi\omega^{1}_{1}(\ln
f_T)'+(-2P-2\xi\omega^{1}_{1})'-\frac{\mathrm{s}\mathrm{s}^{'}}{4\pi
r^4}\bigg].
\end{equation}

\subsection*{Data Availability Statement}

This manuscript has no associated data.


\begin{thebibliography}{40}

\bibitem{1} Van Albada, T.S. and Sancisi, R.: Philos. Trans. Royal Soc. A \textbf{320}(1986)447; Swaters, R.A., Madore, B.F. and Trewhella, M.: Astrophys. J.
\textbf{531}(2000)L107.

\bibitem{2} Barrow, J.D., Maartens, R. and Tsagas, C.G.: Phys. Rep. \textbf{449}(2007)131; Neveu, J. et al.: Astron. Astrophys. \textbf{600}(2017)A40.

\bibitem{2a} Lovelock, D.: J. Math. Phys. \textbf{12}(1971)498.

\bibitem{3} Nojiri, S. and Odintsov, S.D.: Phys. Lett. B \textbf{631}(2005)1.

\bibitem{8} Sharif, M. and Ikram, A.: Eur. Phys. J. C \textbf{76}(2016)640.

\bibitem{8a} Sharif, M. and Ikram, A.: Int. J. Mod. Phys. D \textbf{27}(2018)1750182.

\bibitem{8c} Yousaf, Z., Bhatti, M.Z. and Hassan, K.: Eur. Phys. J. Plus
\textbf{135}(2020)397; Yousaf, Z. et al.: New Astron.
\textbf{84}(2021)101541.

\bibitem{8d} Sharif, M. and Hassan, K.: Chin. J. Phys \textbf{77}(2022)1479; Pramana \textbf{96}(2022)50; Mod. Phys. Lett. A \textbf{37}(2022)2250027.

\bibitem{8e} Xingxiang, W.: Gen. Relativ. Gravit. \textbf{19}(1978)729.

\bibitem{8f} Das, B. et al.: Int. J. Mod. Phys. D \textbf{20}(2011)1675.

\bibitem{8g} Sharif, M. and Bhatti, M.Z.: Astrophys. Space Sci. \textbf{347}(2013)337.

\bibitem{8h} Murad, M.H.: Astrophys. Space Sci. \textbf{361}(2016)20.

\bibitem{8i} Singh, K.N. and Pant, N.: Astrophys. Space Sci. \textbf{358}(2015)1; Sharif, M. and Zeeshan Gul, M.: Eur. Phys. J. Plus \textbf{133}(2018)345; Sharif, M. and Naz, S.: Mod. Phys. Lett. A
\textbf{35}(2020)1950340.

\bibitem{9} Ruderman, M.: Annu. Rev. Astron. Astrophys. \textbf{10}(1972)427.

\bibitem{9a} Sokolov, A.I.: J. Exp. Theor. Phys. \textbf{49}(1980)1137.

\bibitem{9b} Kippenhahn, R., Weigert, A. and Weiss, A.: \emph{Stellar Structure and Evolution} (Springer, 1990).

\bibitem{9c} Herrera, L. and Santos, N.O.: Phys. Rep. \textbf{286}(1997)53.

\bibitem{10} Harko, T. and Mak, M.K.: Ann. Phys. \textbf{11}(2002)3.

\bibitem{10c} Paul, B.C. and Deb, R.: Astrophys. Space Sci. \textbf{354}(2014)421.

\bibitem{11} Ovalle, J.: Mod. Phys. Lett. A \textbf{23}(2008)3247.

\bibitem{13} Ovalle, J. et al.: Eur. Phys. J. C \textbf{78}(2018)960.

\bibitem{14} Gabbanelli, L., Rinc{\'o}n, {\'A}. and Rubio, C.: Eur. Phys. J. C \textbf{78}(2018)370.

\bibitem{15} Sharif, M. and Sadiq, S.: Eur. Phys. J. C
\textbf{78}(2018)410.

\bibitem{17} Estrada, M. and Tello-Ortiz, F.: Eur. Phys. J. Plus
\textbf{133}(2018)453.

\bibitem{18} Singh, K. et al.: Eur. Phys. J. C \textbf{79}(2019)851.

\bibitem{19} Hensh, S. and Stuchl{\'\i}k, Z.: Eur. Phys. J. C \textbf{79}(2019)834.

\bibitem{20} Zubair, M. and Azmat, H.:  Ann. Phys. \textbf{420}(2020)168248.

\bibitem{20a} Maurya, S.K. et al.: Phys. Dark Universe
\textbf{30}(2020)100640; Maurya, S.K. and Tello-Ortiz, F.: Phys.
Dark Universe \textbf{27}(2020)100442; ibid.
\textbf{29}(2020)100577; Maurya, S.K. et al.: Eur. Phys. J. C
\textbf{81}(2021)848; Maurya, S.K., Tello-Ortiz, F. and Govender,
M.: Fortsch. Phys. \textbf{69}(2021)2100099; Maurya, S.K.,
Tello-Ortiz, F. and Ray, S.: Phys. Dark Universe
\textbf{31}(2021)100753.

\bibitem{16} Sharif, M. and Saba, S.: Chin. J. Phys. \textbf{59}(2019)481; ibid. \textbf{63}(2020)348; Int. J. Mod. Phys. D \textbf{29}(2020)2050041.

\bibitem{21} Sharif, M. and Waseem, A.: Chin. J. Phys. \textbf{60}(2019)426; Ann. Phys. \textbf{405}(2019)14; Sharif, M. and Majid, A.: Chin. J. Phys. \textbf{68}(2020)406; Phys. Dark
Universe \textbf{30}(2020)100610; Sharif, M. and Naseer, T.: Chin.
J. Phys. \textbf{73}(2021)179; Naseer, T and Sharif, M.: Universe
\textbf{8}(2022)62.

\bibitem{21a} Sharif, M. and Hassan, K.: \emph{Anisotropic Decoupled Spheres in $f(G,T)$ Gravity} Int. J. Geom. Methods Mod. Phys. (2022, to appear).

\bibitem{23a} Shamir, M.F. and Ahmad, M.: Eur. Phys. J. C \textbf{77}(2017)674; Sharif, M. and Naeem, A.: Int. J. Mod. Phys. A \textbf{35}(2020)2050121.

\bibitem{23b} Ilyas, M.: Eur. Phys. J. C \textbf{78}(2018)757; Maurya, S.K., Singh, K.N. and Nag, R.: Chin. J. Phys.
\textbf{74}(2021)313; Sharif, M., Naeem, A. and Ramzan, A.:
Astrophys. Space Sci. \textbf{367}(2022)21.

\bibitem{22} Krori, K.D. and Barua, J.: J. Phys. A Math. Gen. \textbf{8}(1975)508.

\bibitem{22a} G{\"u}ver, T. et al.: Astrophys. J. \textbf{719}(2010)1807.

\bibitem{22b} Buchdahl, H.A.: Phys. Rev. \textbf{116}(1959)1027.

\bibitem{22bc} Andr{\'e}asson, H.: J. Differ. Equ. \textbf{245}(2008)2243.

\bibitem{22c} Ivanov, B.V: Phys. Rev. D \textbf{65}(2002)104011.

\bibitem{23} Mustafa, G. et al.: Chin. J. Phys. \textbf{67}(2020)576.

\bibitem{24} Herrera, L.: Phys. Lett. A \textbf{165}(1992)206.

\end{thebibliography}
\end{document}